\documentclass[english, 11pt, letter]{article}
\usepackage{amsmath,amssymb,amsthm}
\usepackage{natbib}
\usepackage[nodisplayskipstretch]{setspace}
\usepackage{bm}
\usepackage{threeparttable}
\usepackage{graphicx}
\usepackage{enumerate}
\usepackage{commath}			
\usepackage{bibentry}
\usepackage{booktabs}
\usepackage[titletoc,title]{appendix}
\usepackage{multirow}
\usepackage{tabularx}
\usepackage{dsfont}
\usepackage{rotating} 
\usepackage[font={small}]{subcaption}
\usepackage[left=1in, right=1in, top=1in, bottom=1in]{geometry}
\usepackage{float}
\usepackage{subcaption}
  
\usepackage[colorinlistoftodos,textsize=small]{todonotes}
\allowdisplaybreaks

\usepackage{url}
\usepackage{xr-hyper}  
\usepackage[
	breaklinks,
	colorlinks=true,
	pagebackref=false,
	hypertexnames=false,
	pdfauthor={},%
	pdftitle={},%
	citecolor=blue,
	linkcolor=blue,
	urlcolor=blue
	]{hyperref}      

\newtheorem{lem}{Lemma}
\newtheorem{assumption}{Assumption}

\newtheorem{cor}{Corollary}

\newtheorem{thm}{Theorem}

\newtheoremstyle{remark2}{1ex}{1ex}%
      {}
      {}
      {\bf}
      {.}
      {5pt}
      {\thmname{#1}\thmnumber{ #2}\thmnote{ \slshape{(#3)}}} 
\theoremstyle{remark2}

\newtheorem{example}{Example}

\newtheoremstyle{remark3}{1ex}{1ex}%
      {\it}
      {}
      {\bfseries}
      {.}
      {5pt}
      {\thmname{#1}\thmnumber{ #2*}\textnormal{\thmnote{ (#3)}}} 
\theoremstyle{remark3}
\newtheorem{assumptionA}{Assumption}

\newcommand{\lorder}{\preceq}
\newcommand{\lex}{\lorder_{\mathrm{lex}}}

\newcommand{\1}{\mathds{1}}

\usepackage{babel}
\input{ee.sty}

\@ifundefined{bibfont}{  } 
    {  }

\makeatletter
\renewenvironment{proof}[1][\bfseries\proofname]{\par
   \pushQED{\qed}%
   \normalfont \topsep6\p@\@plus6\p@\relax
   \trivlist
   \item[\hskip\labelsep
     #1\@addpunct{:}]\ignorespaces
}{%
   \popQED\endtrivlist\@endpefalse
}
\makeatother

\newcommand{\Comments}{1}
\newcommand{\mynote}[2]{\ifnum\Comments=1\textcolor{#1}{#2}\fi}
\newcommand{\mytodo}[2]{\ifnum\Comments=1%
  \todo[linecolor=#1!80!black,backgroundcolor=#1,bordercolor=#1!80!black]{#2}\fi}

\ifnum\Comments=1               
\fi

\ifnum\Comments=1               
\fi

\renewcommand\appendixpagename{Appendix}

\newcommand{\Q}{\operatorname{Q}}
\newcommand{\MES}{\operatorname{MES}}
\newcommand{\GDP}{\operatorname{GDP}}
\newcommand{\NFCI}{\operatorname{NFCI}}

\newcommand{\D}{\,\mathrm{d}}
\renewcommand{\E}{\mathbb{E}}
\renewcommand{\P}{\mathbb{P}}

\newcommand{\EL}{\operatorname{EL}}

\begin{document}

\baselineskip18pt
\renewcommand\floatpagefraction{.9}
\renewcommand\topfraction{.9}
\renewcommand\bottomfraction{.9}
\renewcommand\textfraction{.1}
\setcounter{totalnumber}{50}
\setcounter{topnumber}{50}
\setcounter{bottomnumber}{50}
\abovedisplayskip1.5ex plus1ex minus1ex
\belowdisplayskip1.5ex plus1ex minus1ex
\abovedisplayshortskip1.5ex plus1ex minus1ex
\belowdisplayshortskip1.5ex plus1ex minus1ex

\title{Self-Normalized Inference in (Quantile, Expected Shortfall) Regressions for Time Series\thanks{We are indebted to Timo Dimitriadis for insightful comments on an earlier version of this manuscript.
Both authors gratefully acknowledge support of the Deutsche Forschungsgemeinschaft (DFG, German Research Foundation) through grants 460479886 (first author) and 531866675 (first and second author).}
}

\author{
Yannick Hoga\thanks{Faculty of Economics and Business Administration, University of Duisburg-Essen, Universit\"atsstra\ss e 12, D--45117 Essen, Germany, \href{mailto:yannick.hoga@vwl.uni-due.de}{yannick.hoga@vwl.uni-due.de} and \href{mailto:christian.schulz@vwl.uni-due.de}{christian.schulz@vwl.uni-due.de}.}
\and
	Christian Schulz$^{\dagger}$
}

\date{\today}
\maketitle

\begin{abstract}
	\noindent 
	This paper proposes valid inference tools---based on self-normalization---in time series expected shortfall regressions and, as a corollary, also in quantile regressions.
	Extant methods for such time series regressions---based on a bootstrap or direct estimation of the long-run variance---are computationally more involved, require the choice of tuning parameters and have serious size distortions when the regression errors are strongly serially dependent.
	In contrast, our inference tools only require estimates of the (quantile, expected shortfall) regression parameters that are computed on an expanding window, and are correctly sized as we show in simulations.
	Two empirical applications to stock return predictability and to Growth-at-Risk demonstrate the practical usefulness of the developed inference tools.\\
	
\end{abstract}



\section{Motivation}

Much recent work has considered the expected shortfall (ES), i.e., the mean beyond a large quantile, as an appealing summary measure of the distribution of an outcome variable.
For instance, \citet{DB19} relate the ES of an outcome to covariates in their ES regressions for independent, identically distributed (i.i.d.) data.
Such ES regressions are more informative on the tail of the dependent variable than standard quantile regressions, because---by definition---the ES takes into account the complete tail beyond the quantile.
\citet{PZC19} develop predictive models for the ES, which are tailored to financial time series.
In financial modeling, a further reason the ES is preferred over quantiles (next to being informative on the complete tail) is that it is a coherent risk measure in the sense of \citet{Aea99}.
Since the early work of \citet{DB19} and \citet{PZC19}, the theoretical literature on ES regressions has increased rapidly in such diverse fields as microeconometrics \citep{WTH24,CY25,HLP25}, high-dimensional regressions \citep{HTZ23,Zea25}, and financial risk management \citep{BD22,Bar23}.

One applied field where ES regressions for time series have been used extensively is in forecasting macroeconomic risk in the Growth-at-Risk (GaR) literature \citep{ABG19,Pea20,Aea21,DDP24} and its offsprings \citep{LL24}.
However, the extant theoretical tools for analyzing such time series ES regressions suffer from several drawbacks that severely restrict their usefulness.
For instance, the i.i.d.~framework of \citet{DB19} is unsuitable as macroeconomic data typically exhibit strong serial dependence.
Similarly, the predictive ES models of \citet{PZC19} impose a martingale difference-type assumption on the errors, which---as we argue in more detail below---severely restricts their empirical applicability for time series regressions.
In particular, under their framework, the cross terms in the ``meat'' matrix of the ES estimator vanish.
Of course, it is those cross-terms in the long-run variance that have attracted decades worth of research in other contexts \citep[e.g.,][]{NW87,And91,JD00}.
Finally, in time series ES regressions, \citet{Bar23} proposes a heteroskedasticity and autocorrelation consistent (HAC) estimator of the long-run variance for his two-step parameter estimator.
However, as we show in simulations below, tests based on those HAC standard errors suffer from severe size distortions.

The main contribution of this paper is to propose valid inference methods in such time series ES regressions, where autocorrelated errors are a natural concern.
The only other paper we are aware of that deals with this issue is that of \citet{Bar23}.
Our method is different from his by being based on self-normalization (SN) as proposed by \citet{Sha10}.
As we show in simulations, SN almost completely eliminates the size distortions for the HAC-based tests of \citet{Bar23}.
Also, we use a slightly different two-step estimator for ES regressions than \citet{Bar23}.
This two-step estimation procedure first requires estimating a quantile regression (QR), after which the ES regression is estimated based only on data with a quantile exceedance.

Due to this two-step procedure, we obtain as a corollary that our SN-based inference methods are also valid for time series QR.
Of course, the need for robust inference in time series QR is no less than that for ES regressions.
For instance, QR methods are applied for time series in climate science \citep{Yea18,Mea21}, finance \citep{AB16,AGS22}, risk management \citep{EM04,Gea11} and macroeconomics \citep{ABG19,LL24}.

There already exist two main approaches for valid inference in time series QR.
The first, proposed by \citet{GY24}, is based on consistently estimating the asymptotic variance-covariance matrix of the QR estimator---much like the proposal of \citet{Bar23} for ES regressions.
This variance-covariance matrix has the usual sandwich form with two (identical) ``bread'' matrices and a ``meat'' matrix.
The bread matrix, which contains a scalar sparsity, can be consistently estimated following \citet{Pow91} and \citet{Kat12}.
The meat matrix requires HAC estimation techniques.
\citet{GY24} show consistency of such a HAC estimator of the meat matrix.

The second valid inference approach in time series QR is to use the smooth block bootstrap (SBB) of \citet{GLN18}.
This bootstrap procedure is quite general, encompassing in particular the popular moving block bootstrap (MBB).
In quantile regressions for dependent data, \citet{Fit97} was the first to propose such a MBB.
Yet, as \citet[Sec.~3.3]{GLN18} point out, his proof of bootstrap validity contains flaws.

The advantages of our self-normalized approach vis-\`{a}-vis the other two approaches---i.e., direct HAC estimates \`{a} la \citet{Bar23} and \citet{GY24}, or the bootstrap of \citet{GLN18}---are as follows.
First, our suggested inference method essentially only requires parameter estimates computed over an expanding window of the total sample.
This implementational convenience contrasts with the direct estimation of the asymptotic covariance and the bootstrap approach.
The former requires the choice of several tuning parameters:
the estimation of the bread matrix in the long-run variance depends on a bandwidth choice for estimating the scalar sparsity \citep{Kat12} and the meat matrix requires a bandwidth to be used in the kernel-weighted estimators of the cross terms \citep{Bar23,GY24}.
Similarly, the bootstrap approach of \citet{GLN18} depends on the choice of a bandwidth for resampling individual observations via kernel smoothing techniques and, moreover, the choice of the moving block length.
Additionally, the usual computational burden of bootstrap standard errors also applies in time series regressions.

The second advantage of our proposed self-normalized inference is the very good finite-sample performance.
Our simulations show that---in stark contrast to the other methods---our SN-based tests have excellent size.
In fact, even in large samples size distortions of the extant methods can exceed 20 percentage points, leading to many spurious rejections of the null.
To correct for the differences in size, we also compute size-corrected power and find somewhat lower power for SN.
This is typical of SN, where size improves at the expense of some slight loss in power---the well-known ``better size but less power'' phenomenon \citep[see, e.g.,][]{Sha15,WS20,Hea24}.
In this sense, SN provides a desirable robustness-efficiency tradeoff, where more reliable size (robustness) is traded off for some mild loss in power (efficiency).


Although our inference procedures for time series QR are not our main contribution, we nonetheless apply them to real data for a comparison with the other two methods.
Specifically, our first empirical application considers predictive quantile regressions for the equity premium \citep{CT08,Lee16,FL19}.
As popular stationary predictors, we use---among others---the stock variance and the long-term rate of return \citep{GW08,GW24}.
Since our approach proves to be most robust against autocorrelation in the errors, we do not find predictability for some predictors for some quantiles of the distribution of the equity premium.
In contrast, the (extremely liberal) HAC and SBB approaches---perhaps misleadingly---reject the null of no predictability.

Our second application, which concerns ES regressions, revisits the influential work on Growth-at-Risk by \cite{ABG19}. 
While \cite{ABG19} cannot test the statistical significance of financial conditions on the conditional distribution of US GDP growth, our SN-based inference tools for ES regressions allow us to do so. 
We show that national financial conditions play a significant role for future GDP growth only in the lower conditional tail of the GDP growth distribution. 
In the upper conditional tail, the coefficients of an ES regression of future GDP growth on the national financial conditions index (NFCI) are not distinguishable from 0.

The rest of the paper proceeds as follows. 
As a preliminary step, Section~\ref{sec:Quantile Regressions} introduces linear quantile regressions and the appertaining SN-based inference tools. 
The same follows for the linear ES model in Section~\ref{sec:Expected Shortfall Regressions}. 
The Monte Carlo simulations in Section~\ref{sec:Simulations} demonstrate the good finite-sample performance of our SN-based approach to inference. 
Section~\ref{sec:Empirical Application} presents two empirical applications and the final Section~\ref{sec:Conclusion} concludes.
The proofs of the main theoretical results are relegated to the Appendix.

\section{Quantile Regressions}\label{Quantile Regressions}
\label{sec:Quantile Regressions}

Since our two-step approach to estimating ES regressions (to be introduced in Section~\ref{sec:Expected Shortfall Regressions}) relies on an initial QR estimate, we first consider time series QRs in this section.
The theory will then be used later on for ES regressions, which we are mainly interested in.
However, we emphasize again that our SN-based tests for time series QR possess excellent size compared with the other two extant methods of \citet{GLN18} and \citet{GY24}, as our simulations in Section~\ref{sim:Quantile Regressions} demonstrate.
Therefore, this section's results are also of interest for their own sake.

\subsection{The Linear Quantile Regression Model}
\label{sec:The Linear Model}

We denote the real-valued dependent variable by $Y_t$ and the $(k\times1)$-vector of covariates by $\mX_t$. 
These variables are related via the QR
\begin{equation}\label{eq:QR model}
	Y_t = \mX_t^\prime\valpha_0 + \varepsilon_t,\qquad Q_{\tau}(\varepsilon_t\mid\mX_t)=0,
\end{equation}
where $\tau\in(0,1)$ is the probability level of interest, $\valpha_0=\valpha_0(\tau)$ is the true parameter, and $Q_{\tau}(\varepsilon_t\mid\mX_t):=F_{\varepsilon_t\mid\mX_t}^{-1}(\tau)$ is the $\tau$-quantile of the cumulative distribution function of $\varepsilon_t\mid\mX_t$, i.e., $F_{\varepsilon_t\mid\mX_t}(\cdot)=\P\{\varepsilon_t\leq\cdot\mid\mX_t\}$.
The assumption on the errors ensures that $Q_{\tau}(Y_t\mid\mX_t)=\mX_t^\prime\valpha_0$, such that $\valpha_0$ measures the impact of $\mX_t$ on the $\tau$-th conditional quantile of $Y_t$.
To ease the notational burden, we suppress the dependence of $\varepsilon_t$ and $\valpha_0$ on $\tau$.
As is already implicit in the subscript $t$ in \eqref{eq:QR model}, we allow $(Y_t, \mX_t^\prime)^\prime$ to be serially dependent.

Given a sample $\{(Y_t,\mX_t^\prime)^\prime\}_{t=1,\ldots,n}$, the QR parameter $\valpha_0$ is typically estimated via
\[
	\widehat{\valpha}=\argmin_{\valpha\in\mathbb{R}^{k}}\sum_{t=1}^{n}\rho(Y_t - \mX_t^\prime\valpha),
\]
where $\rho(u):=u(\tau-\1_{\{u\leq 0\}})$ is the usual tick loss function.
Under strict stationarity of $(Y_t, \mX_t^\prime)^\prime$ and some further regularity conditions, \citet[Theorem~2.2]{Fit97} and \citet[Theorem~1]{GLN18} show that $\sqrt{n}(\widehat{\valpha} - \valpha_0)\overset{d}{\longrightarrow}N(\vzero, \mD^{-1}\mJ\mD^{-1})$, as $n\to\infty$.
Here, $\mD=\E\big[f_{\varepsilon_t\mid\mX_t}(0)\mX_t\mX_t^\prime\big]$ is the ``bread'' matrix and the ``meat'' matrix $\mJ$ equals the long-run variance of $n^{-1/2}\sum_{t=1}^{n}\mX_t\psi(\varepsilon_t)$ with $\psi(u)=\tau-\1_{\{u\leq 0\}}$.
Note that $\psi(u)$ the subgradient of $\rho(u)$, which is equal to the standard gradient of $\rho(u)$ for $u\neq0$.
The $\psi(\varepsilon_t)$ are known as \textit{generalized errors} in QR, since they have mean zero conditional on covariates, i.e., $\E[\psi(\varepsilon_t)\mid\mX_t]=\tau - \P\big\{\varepsilon_t\leq0\mid\mX_t\big\}=0$.

Valid inference in quantile regressions can build on estimates of $\mD$ and $\mJ$. 
\citet{Pow91} and \citet{Kat12} propose a kernel-type estimator for $\mD$.
The meat matrix $\mJ$, also called the HAC component of the asymptotic variance-covariance matrix, is harder to estimate in time series QR. 
This difficulty arises because for serially dependent $(Y_t,\mX_t^\prime)^\prime$ cross terms generally appear in the long-run variance of $n^{-1/2}\sum_{t=1}^{n}\mX_t\psi(\varepsilon_t)$.
Note that for i.i.d.~data, this long-run variance simplifies to 
\[
	\mJ=\Var\big(\mX_1\psi(\varepsilon_1)\big),
\]
which is easy to estimate via its sample counterpart.
While for standard \textit{mean} regressions, the literature on HAC estimates fills volumes \citep[e.g.,][]{NW87,And91,JD00}, for \textit{quantile} regressions there is, to the best of our knowledge, only the work of \citet{GY24}.
The latter authors propose a HAC-type estimator for $\mJ$ and prove its consistency.
Of course, both the kernel estimator of $\mD$ and the HAC estimator of $\mJ$ require a bandwidth choice.
As usual, choosing an optimal bandwidth requires the estimation of unknown quantities.

The second available approach to inference in time series QR---the bootstrap approach of \citet{GLN18}---likewise involves the choice of tuning parameters, viz.~the bandwidth for the smoothing kernel estimator and the block length. 
To sidestep such problems, we propose an essentially tuning parameter-free inference method in time series QR in the next section. 
This method is based on SN and obviates the need to bootstrap or to estimate $\mD$ and $\mJ$.

\subsection{Self-Normalized Inference in QR}
\label{sec:Self-Normalized Inference in QR}

Our self-normalized approach to inference in time series QR requires parameter estimates that are computed sequentially over the sample, i.e., estimates of $\valpha_0$ based on the first $\lfloor ns\rfloor$ observations, where $s\in(0,1]$ and $\lfloor \cdot\rfloor$ denotes the floor function.
Therefore, we introduce the estimator
\[
	\widehat{\valpha}(s):=\argmin_{\valpha\in\mathbb{R}^{k}}\sum_{t=1}^{\lfloor ns\rfloor}\rho(Y_t - \mX_t^\prime\valpha).
\]
Obviously, it holds for the full-sample estimate that $\widehat{\valpha}(1)=\widehat{\valpha}$.
Following \citet{Sha10}, SN-based inference requires \textit{functional} central limit theory for $\widehat{\valpha}(s)$.
For this, we have to impose some regularity conditions.
To introduce these, we define $\Vert\cdot\Vert$ to be the Euclidean norm when applied to a vector or a matrix.

\begin{assumption}[Conditional Density of QR Errors]\label{ass:innov}
\begin{enumerate}[(i)]
	\item\label{it:dens} There exists some $d>0$, such that the distribution of $\varepsilon_t$ conditional on $\mX_t$ has a Lebesgue density $f_{\varepsilon_t\mid\mX_t}(\cdot)$ on $[-d,d]$.
	
	\item\label{it:dens bound} $\sup_{x\in[-d,d]}f_{\varepsilon_t\mid\mX_t}(x)\leq \overline{f}<\infty$ for all $t\geq 1$.
	
	\item\label{it:Lipschitz} $\big|f_{\varepsilon_t\mid\mX_t}(x) - f_{\varepsilon_t\mid\mX_t}(y)\big|\leq L |x-y|$ for all $-d\leq x,y\leq d$ and all $t\geq1$.

\end{enumerate}
\end{assumption}

\begin{assumption}[Serial Dependence]\label{ass:Serial Dependence}
The sequence $\big\{( \varepsilon_t, \mX_t^\prime)^\prime\big\}$ is $\alpha$-mixing with mixing coefficients $\alpha(\cdot)$ of size $-r/(r-2)$ for some $r>2$.
\end{assumption}

\begin{assumption}[Heterogeneity]\label{ass:Heterogeneity}
\begin{enumerate}[(i)]
	
	\item\label{it:(i)} $\E\big[\Vert\mX_t\Vert^{2r}\big]\leq\Delta<\infty$ for all $t\geq1$ and $r$ from Assumption~\ref{ass:Serial Dependence}.
	
	\item\label{it:(ii)} $\frac{1}{n}\E\Big[\big\{\sum_{t=1}^{n}\psi(\varepsilon_t)\mX_t\big\}\big\{\sum_{t=1}^{n}\psi(\varepsilon_t)\mX_t\big\}^\prime\Big]\longrightarrow \mJ$, as $n\to\infty$, for some positive definite $\mJ\in\mathbb{R}^{k\times k}$.
	
	\item\label{it:(iii)} $\frac{1}{n}\sum_{t=1}^{n}\E\big[ f_{\varepsilon_t\mid\mX_t}(0)\mX_t\mX_t^\prime \big]\longrightarrow \mD$, as $n\to\infty$, for some positive definite $\mD\in\mathbb{R}^{k\times k}$.
	
\end{enumerate}
\end{assumption}

Assumption~\ref{ass:innov} is a smoothness condition on the distribution of $\varepsilon_t$ conditional on the covariates $\mX_t$.
It ensures that in a neighborhood of $0$, a Lebesgue density exists that is uniformly bounded and Lipschitz continuous.
Note that outside of the interval $[-d,d]$, the distribution of $\varepsilon_t\mid\mX_t$ is not restricted at all.
The mixing condition in Assumption~\ref{ass:Serial Dependence} is standard, and ensures that suitable functional central limit theorems (FCLTs) hold.
Assumption~\ref{ass:Heterogeneity} restricts the distributional heterogeneity in $\big( \varepsilon_t, \mX_t^\prime, f_{\varepsilon_{t}\mid\mX_t}(0) \big)^\prime$.
Under strict stationarity the limiting matrix in Assumption~\ref{ass:Heterogeneity}~\eqref{it:(ii)} simplifies to $\mJ=\sum_{t=-\infty}^{\infty}\Cov\big(\psi(\varepsilon_0)\mX_0,\, \psi(\varepsilon_t)\mX_t\big)$ and that in Assumption~\ref{ass:Heterogeneity}~\eqref{it:(iii)} to $\mD=\E[f_{\varepsilon_t\mid\mX_t}(0)\mX_t\mX_t^\prime]$.

Overall, our assumptions are quite similar to those of \citet[Theorem~2.2]{Fit97} and \citet[Theorem~1]{GLN18}.
However, \citet{Fit97} and \citet{GLN18} assume strict stationarity of $\{(\varepsilon_t,\mX_t^\prime)^\prime\}$ \citep[as do][]{GY24}, while the distributional heterogeneity in our case is only restricted by Assumption~\ref{ass:Heterogeneity}. 
This allows for wider applicability in practice.
Note, however, that Assumption~\ref{ass:Heterogeneity}~\eqref{it:(iii)} does not cover deviations from stationarity, such as linear time trends or even unit root settings as those considered by \citet{KX04} and \citet{Xia09}.

To obtain the weak limit of $\widehat{\valpha}(s)$, we utilize a functional version of the well-known Convexity lemma of \citet{Kni89}; see Lemma~\ref{lem:Kato2009} in Appendix~\ref{sec:thm1}.
Our lemma builds on \citet[Theorem~1]{Kat09} and may be of independent interest, since it allows to develop functional limit theory for extremum estimators based on convex objective functions.
Our first main theoretical result, Theorem~\ref{thm:std est}, exploits this lemma to derive the functional weak limit of $\widehat{\valpha}(s)$.
In the following, we denote by $D_k[a,b]$ the space of $\mathbb{R}^k$-valued componentwise c\`{a}dl\`{a}g functions on the interval $[a,b]$ ($0\leq a<b<\infty$), which is endowed with the Skorohod topology \citep{Bil99}.

\begin{thm}\label{thm:std est}
Under Assumptions~\ref{ass:innov}--\ref{ass:Heterogeneity} it holds for any $\epsilon\in(0,1)$ that, as $n\to\infty$,
\begin{equation*}
s\sqrt{n}\big[\widehat{\valpha}(s)-\valpha_0\big]\overset{d}{\longrightarrow}\mOmega^{1/2}\mW_k(s)\qquad\text{in}\ D_{k}[\epsilon,1],
\end{equation*}
where $\mW_k(\cdot)$ is a $k$-variate standard Brownian motion and $\mOmega=\mD^{-1}\mJ\mD^{-1}$.
\end{thm}

\begin{proof}
See Appendix~\ref{sec:thm1}.
\end{proof}

For the non-functional case with $s=1$, Theorem~\ref{thm:std est} is similar to Theorem~2.2 of \citet{Fit97} and Theorem~1 of \citet{GLN18}. 
The aforementioned authors build on classic results in the extremum estimation literature, such as \citet{Hub67} and \citet{Wei91}.
This contrasts with our approach that uses a variant of Theorem~1 in \citet{Kat09}.
Despite this difference in the method of proof, the assumptions entertained here and in their work are remarkably similar.
Of course, our conclusion being a functional result for the QR estimator is much stronger.
This functional convergence allows the variability of the QR estimator on subsamples to be used for our self-normalized inference below.

Functional results for the QR estimator similar to those of Theorem~\ref{thm:std est} are available from \citet{Qu08}, \citet{SX08} and \citet{OQ11}.
However, under their assumptions (which include strict stationarity), the tricky cross terms of $\mJ$ vanish, such that simply $\mJ=\lim_{n\to\infty}\E\big[\frac{1}{n}\sum_{t=1}^{n}\psi^2(\varepsilon_t)\mX_t\mX_t^\prime\big]=\E\big[\psi^2(\varepsilon_1)\mX_1\mX_1^\prime\big]$.
To see this exemplarily in the work of \citet{OQ11}, note that their Assumption~1 requires, in our notation, $\psi(\varepsilon_t)$ to be a martingale difference sequence (MDS) with respect to $\mathcal{F}_{t-1}=\sigma(\mX_t,Y_{t-1},\mX_{t-1},Y_{t-2},\mX_{t-2},\ldots)$, such that $\E\big[\psi(\varepsilon_t)\mid\mathcal{F}_{t-1}\big]=0$.
Yet, in this case,
\begin{align*}
\E&\bigg[\Big\{\sum_{t=1}^{n}\psi(\varepsilon_t)\mX_t\Big\}\Big\{\sum_{t=1}^{n}\psi(\varepsilon_t)\mX_t\Big\}^\prime\bigg]\\
& = \sum_{t=1}^{n}\E\big[\psi^2(\varepsilon_t)\mX_t\mX_t^\prime\big] + \sum_{j=1}^{n-1}\sum_{t=j+1}^{n}\Big\{\E\big[\psi(\varepsilon_t)\mX_t\mX_{t-j}^\prime\psi(\varepsilon_{t-j})\big] + \E\big[\psi(\varepsilon_{t-j})\mX_{t-j}\mX_t^\prime\psi(\varepsilon_t)\big]\Big\}\\
& = \sum_{t=1}^{n}\E\big[\psi^2(\varepsilon_t)\mX_t\mX_t^\prime\big],
\end{align*}
because, by the law of iterated expectations for any $j>0$ and $t>0$,
\begin{equation*}
	\E\big[\psi(\varepsilon_t)\mX_t\mX_{t-j}^\prime\psi(\varepsilon_{t-j})\big] = \E\Big[\E\big\{\psi(\varepsilon_t)\mid \mathcal{F}_{t-1}\big\}\mX_t\mX_{t-j}^\prime\psi(\varepsilon_{t-j})\Big]=\vzero.
\end{equation*}
Therefore, imposing MDS QR errors $\psi(\varepsilon_t)$ implies that $\mJ=\E\big[\psi^2(\varepsilon_1)\mX_1\mX_1^\prime\big]$, such that the hard-to-estimate parts of $\mJ$ vanish and the problem of HAC inference is ``assumed away''.

Of course, only assuming that $\E[\psi(\varepsilon_t)\mid\mX_t]=0$ (as is equivalent to our assumption that $Q_{\tau}(\varepsilon_t\mid\mX_t)=0$) instead of the MDS assumption $\E[\psi(\varepsilon_t)\mid\mX_t,Y_{t-1},\mX_{t-1},\ldots]=0$ is a much weaker requirement. 
We stress that for cross-sectional data with i.i.d.~$\big\{(Y_t,\mX_t^\prime)^\prime\big\}$,
\[
	\E[\psi(\varepsilon_t)\mid\mX_t]=0\qquad \Longleftrightarrow \qquad\E[\psi(\varepsilon_t)\mid\mX_t,Y_{t-1},\mX_{t-1},\ldots]=0.
\]
However, in time-series contexts the left-hand side condition (``exogeneity with respect to the covariates'') allows for much more serial dependence in $\psi(\varepsilon_t)$ than does the right-hand side condition (``exogeneity with respect to the infinite past'').
The following example further illustrates this point.

\begin{example}
Suppose that $\mX_t=1$, such that the quantile regression $Y_t=\alpha_0+\varepsilon_t$ only contains an intercept.
In this case, Assumption~1 of \citet{OQ11} boils down to
\[
	\E\big[\psi(Y_t-\alpha_0)\mid Y_{t-1},Y_{t-2},\ldots\big] = 0,
\]
which imposes a MDS-type dependence structure for the $Y_t$.
In contrast, our assumption $\E\big[\psi(\varepsilon_t)\mid\mX_t\big]=0$ reduces to $\E\big[\psi(\varepsilon_t)\big]=0$ for $\mX_t=1$.
This does not restrict the dependence structure of the $Y_t$ \textit{at all} and, hence, is much more innocuous in time series applications.
\end{example}

Applying the continuous mapping theorem \citep[e.g.,][Theorem 26.13]{Dav94} to Theorem~\ref{thm:std est} essentially implies the following corollary.

\begin{cor}\label{cor:SN}
Let $\mA$ be a full rank $(\ell\times k)$-matrix with $\ell\leq k$, and define $\valpha_{0,\mA}=\mA\valpha_0$ and $\widehat{\valpha}_{\mA}(s) = \mA\widehat{\valpha}(s)$.
Under the assumptions of Theorem~\ref{thm:std est} it holds that, as $n\to\infty$,
\[
	\mathcal{T}_n:=n\big[\widehat{\valpha}_{\mA}(1) - \valpha_{0,\mA}\big]^\prime \mS_{n,\widehat{\valpha}_{\mA}(\cdot)}^{-1} \big[\widehat{\valpha}_{\mA}(1) - \valpha_{0,\mA}\big]\overset{d}{\longrightarrow}\mW_k^\prime(1)\mA^\prime\mV_k^{-1}\mA\mW_k(1)=:W(\mA),
\]
where
\begin{align*}
	\mV_k & = \mA\int_\epsilon^1 \big[\mW_k(s) - s\mW_k(1)\big] \big[\mW_k(s) - s\mW_k(1)\big]^\prime \D s \mA^\prime,\\
	\mS_{n,\widehat{\valpha}_{\mA}(\cdot)} & = \frac{1}{n^2}\sum_{j=\lfloor n\epsilon\rfloor+1}^{n}j^2\big[\widehat{\valpha}_{\mA}(j/n) - \widehat{\valpha}_{\mA}(1)\big]\big[\widehat{\valpha}_{\mA}(j/n) - \widehat{\valpha}_{\mA}(1)\big]^\prime.
\end{align*}
\end{cor}

\begin{proof}
See Appendix~\ref{sec:thm1}.
\end{proof}

Suppose one is interested in testing $H_0\colon \valpha_{0,\mA}=\valpha_{0,\mA}^{\circ}$, where $\valpha_{0,\mA}^{\circ}$ is some null-hypothetical value.
Then, Corollary~\ref{cor:SN} suggests to reject $H_0$ at significance level $\nu\in(0,1)$ if $\mathcal{T}_n$ (with $\valpha_{0,\mA}$ replaced by $\valpha_{0,\mA}^{\circ}$) exceeds the $(1-\nu)$-quantile of $W(\mA)$.
Note that these quantiles can be approximated to any desired level of accuracy by simulating sufficiently often from the nuisance parameter-free limiting distribution $W(\mA)$

The form of SN used in Corollary~\ref{cor:SN} with the specific form of $\mS_{n,\widehat{\valpha}_{\mA}(\cdot)}$ is due to \citet{Sha10}.
We also refer to Section~2.1 of \citet{Sha10} for a connection with fixed-$b$ approaches to inference that are popular in econometrics \citep{KV05}. 
The result in the literature closest to Corollary~\ref{cor:SN} is Proposition~2 in \citet{ZS13}.
However, that theorem only holds for $M$-estimators with differentiable loss function $\rho(\cdot)$ [see their Assumption A1] and for regressions with \textit{deterministic} covariates.
Regarding the latter restriction, \citet{ZS13} state that: ``As pointed out by a referee, the fixed regressor assumption is a limitation in many applications, which we acknowledge.''
Of course, the time series applications in economics and finance that we have in mind need to accommodate for stochastic regressors $\mX_t$.

We also point out that the assumptions Corollary~\ref{cor:SN} imposes for feasible inference seem to be rather mild.
In comparison, consistency of the estimator of the HAC matrix $\mJ$ requires more involved conditions, as shown in \citet[Theorem~4.2]{GY24}.
Similarly, consistent estimation of $\mD$ requires more stringent assumptions; see Assumptions~8--13 in \citet{Kat12}.
In particular, \citet[Assumption~8]{Kat12} and \citet[Assumption~1]{GY24} require strictly stationary $\{(\varepsilon_t,\mX_t^\prime)^\prime\}$, allowing for no distributional heterogeneity, which contrasts with our Assumption~\ref{ass:Heterogeneity}.
Corollary~\ref{cor:SN} therefore provides feasible inference for time series QR under quite general conditions.
These are similar to those entertained by \citet[Theorem~2]{GLN18} for their SBB.


\section{Expected Shortfall Regressions}
\label{sec:Expected Shortfall Regressions}

This section's main goal is to present our SN-based inference methods for time series ES regressions (Section~\ref{Self-Normalized ES Inference}).
Before doing so, Section~\ref{sec:The Linear ES Model} first introduces linear (quantile, ES) regressions.

\subsection{The Linear ES Regression Model}
\label{sec:The Linear ES Model}

As ES models can only be estimated together with quantile regressions \citep{FZ16a,DB19}, we consider the following \textit{joint} (quantile, ES) regression
\begin{align}
Y_t &=\mX_t^\prime\valpha_0+\varepsilon_t,&& Q_{\tau}(\varepsilon_t\mid\mX_t)=0,\label{eq:(1.0)}\\
Y_t &=\mX_t^\prime\vbeta_0+\xi_t,&& \ES_{\tau}(\xi_t\mid\mX_t)=0,\label{eq:(1.1)}
\end{align}
where $\ES_{\tau}(\xi_t\mid\mX_t)=\E_{t}\big[\xi_t\mid\xi_t\geq Q_{\tau}(\xi_t\mid\mX_t)\big]$ and $\E_{t}[\cdot]:=\E[\,\cdot\mid\mX_t]$.
The assumption in \eqref{eq:(1.1)} on the ES regression error $\xi_t$ ensures that $\ES_{\tau}(Y_t\mid\mX_t)=\mX_t^\prime \vbeta_0$, such that $\vbeta_0$ is the parameter of interest when relating the covariates $\mX_t$ to the ES of $Y_t$.
In this context, the QR in \eqref{eq:(1.0)} may be seen as an auxiliary regression, which is only of interest in so far as it allows the ES regression in \eqref{eq:(1.1)} to be estimated.

Often, the parameters in joint (quantile, ES) models are estimated by exploiting elicitability of the pair (quantile, ES) with appertaining loss functions given in \citet{FZ16a}; see, e.g., \citet{DB19} and \citet{PZC19}.
However, in general, these loss functions fail to be convex \citep[Proposition~4.2.31]{Fissler2017}, rendering estimation computationally difficult.

Therefore, in this paper, we take a different tack by drawing on the multi-objective elicitability of the pair (quantile, ES) in the sense of \citet{FH24}. 
Following the reasoning of their Theorem~4.2~(iv), it may be shown that the pair (quantile, ES) is multi-objective elicitable with multi-objective loss function
\[
	\mL^{(\Q,\ES)}\big((v,\mu), y\big) = \begin{pmatrix}L^{\Q}(v,y)\\ L^{\ES}\big((v,\mu),y\big) \end{pmatrix}=\begin{pmatrix}\rho(y-v)\\ \1_{\{y>v\}}(y-\mu)^2 \end{pmatrix},
\]
where we choose $\phi(y)=y^2$, $a(y)=y^2$ and $a^{\MES}(x,y)\equiv0$ in the notation of \citet[Theorem~4.2~(iv)]{FH24}.
Specifically, this means for some random variable $Y$ with $\tau$-quantile $Q_Y$ and $\tau$-expected shortfall $\ES_Y$ that
\[
	\E\Big[\mL^{(\Q,\ES)}\big((Q_Y,\ES_Y), Y\big)\Big]\lex \E\Big[\mL^{(\Q,\ES)}\big((v,\mu), Y\big)\Big]\qquad \text{for all }v\in\mathbb{R},\ \mu\in\mathbb{R},
\]
where $(x_1,x_2)\lex (y_1, y_2)$ in the lexicographic order if $x_1<y_1$ \textit{or} ($x_1=y_1$ and $x_2<y_2$).
In other words, the ($\mathbb{R}^2$-valued) expected loss is minimized with respect to the lexicographic order for the true quantile and the true ES.
This is akin to the expected ($\mathbb{R}$-valued) squared error loss being minimized by the true mean with respect to the canonical order $\leq$ on $\mathbb{R}$.

This suggests the QR estimator $\widehat{\valpha}$ from Section~\ref{Quantile Regressions} for $\valpha_0$ and the following estimator for the ES regression parameters $\vbeta_0$:
\[
	\widehat{\vbeta} = \argmin_{\vbeta\in\mathbb{R}^{k}}\sum_{t=1}^{n}\1_{\{Y_t>\mX_t^\prime\widehat{\valpha}\}}(Y_t- \mX_t^\prime\vbeta)^2.
\]
Therefore, $\widehat{\vbeta}$ is a least squares estimator for the observations with a quantile exceedance.
Note that this estimator is similar in spirit, but different from the two-step ES estimator of \citet{Bar23}.
His and our estimator share the desirable property that the minimization problems involved in computing $\widehat{\valpha}$ and $\widehat{\vbeta}$ are convex in the parameters.
This contrasts with estimators based on the loss functions of \citet{FZ16a}; see also Appendix~C of \citet{PZC19} for more on the intricate computational aspects of minimizing \citet{FZ16a} losses.
We show in the next section that the asymptotic variance of $\widehat{\vbeta}$ is even more complicated than that of $\widehat{\valpha}$.

\subsection{Self-Normalized ES Inference}\label{Self-Normalized ES Inference}

Once again, our approach requires estimates of $\vbeta_0$ that are computed over subsamples.
Formally, for $s\in(0,1]$ we define
\[
	\widehat{\vbeta}(s) = \argmin_{\vbeta\in\mathbb{R}^{k}}\sum_{t=1}^{\lfloor ns\rfloor}\1_{\{Y_t>\mX_t^\prime\widehat{\valpha}(s)\}}(Y_t- \mX_t^\prime\vbeta)^2.
\]
The proof of Theorem~\ref{thm:std est} reveals that the asymptotic limit of $\widehat{\valpha}(s)$ is (up to pre-multiplication with the matrix $\mD^{-1}$) driven by the asymptotic limit of $n^{-1/2}\sum_{t=1}^{\lfloor ns\rfloor}\psi(\varepsilon_t)\mX_t$, i.e., the suitably scaled partial sum process of the generalized errors multiplied by the covariates.
These generalized errors arise as
\begin{equation}\label{eq:(p.10)}
	\psi(\varepsilon_t)=\partial_{v}L^{\Q}(v,y)\big\vert_{(v,y)=(Q_{\tau}(Y_t\mid\mX_t),Y_t)}=\tau-\1_{\{Y_t\leq Q_{\tau}(Y_t\mid\mX_t)\}}=\tau-\1_{\{\varepsilon_t\leq 0\}}.
\end{equation}
Similarly, the asymptotic limit of $\widehat{\vbeta}(s)$ is also driven by a partial sum process of the generalized errors in ES regressions.
Following the construction in \eqref{eq:(p.10)}, up to pre-multiplication by (the inessential) $-1/2$, the ES generalized errors are
\begin{align*}
	\psi_{\ast}(\varepsilon_t,\xi_t)&=-(1/2)\partial_{\mu} L^{\ES}\big((v,\mu),y\big)\big\vert_{(v,\mu,y)=(Q_{\tau}(Y_t\mid\mX_t),\ES_{\tau}(Y_t\mid\mX_t), Y_t)}\\
	&=\1_{\{y>v\}}(y-\mu)\big\vert_{(v,\mu,y)=(Q_{\tau}(Y_t\mid\mX_t),\ES_{\tau}(Y_t\mid\mX_t), Y_t)}\\
	&=\1_{\{\varepsilon_t>0\}}\xi_t.
\end{align*}
Similarly as the QR generalized errors satisfy $\E_{t}\big[\psi(\varepsilon_t)\big]=0$, we also have that $\E_t\big[\psi_{\ast}(\varepsilon_t,\xi_t)\big]=0$. 
Moreover, the QR and the ES generalized errors are (conditionally) uncorrelated because
\[
	\E_{t}\big[\psi_{\ast}(\varepsilon_t,\xi_t)\psi(\varepsilon_t)\big] =\E_{t}\Big[\1_{\{\varepsilon_t> 0\}}\xi_t\big(\tau-\1_{\{\varepsilon_t\leq 0\}}\big)\Big]=\tau\E_{t}\big[\psi_{\ast}(\varepsilon_t,\xi_t)\big]-\E_{t}\big[\1_{\{\varepsilon_t> 0\}}\xi_t\1_{\{\varepsilon_t\leq 0\}}\big]=0.
\]

Deriving functional central limit theory for $\widehat{\vbeta}(s)$ requires extending Assumption~\ref{ass:Heterogeneity} by, among others, imposing conditions on the ES generalized errors.

\setcounter{assumptionA}{2}

\begin{assumptionA}[Heterogeneity]\label{ass:Heterogeneity ES}
Assumption~\ref{ass:Heterogeneity} holds and
\begin{enumerate}[(i)]

	\item\label{it:(i) ES} $\E\big[\Vert\mX_t\Vert^{3r}\big]\leq\Delta<\infty$ and $\E\big[(|\xi_t|\cdot\Vert\mX_t\Vert)^r\big]\leq \Delta<\infty$ for all $t\geq1$ and $r$ from Assumption~\ref{ass:Serial Dependence}.
	
	\item\label{it:(ii) ES} As $n\to\infty$,
	\[
	\frac{1}{n}\E\Bigg[\bigg\{\sum_{t=1}^{n}
		\begin{pmatrix}\psi_{\ast}(\varepsilon_t,\xi_t)\mX_t\\ \psi(\varepsilon_t)\mX_t\end{pmatrix}\bigg\}\Big\{\sum_{t=1}^{n}
		\big(\psi_{\ast}(\varepsilon_t,\xi_t)\mX_t,\psi(\varepsilon_t)\mX_t\big)\Big\}^\prime\Bigg]\longrightarrow\begin{pmatrix}\mJ_{\ast} & \mC^\prime \\ \mC & \mJ\end{pmatrix}
		\]
		for some positive definite limiting matrix.
		
		\item\label{it:(iii) ES} $\frac{1}{n}\sum_{t=1}^{n}\E\big[ \mX_t\mX_t^\prime \big]\longrightarrow \mD_{\ast}$, as $n\to\infty$, for some positive definite $\mD_{\ast}\in\mathbb{R}^{k\times k}$.
		
		\item\label{it:(iv) ES} $\frac{1}{n}\sum_{t=1}^{n}\E\big[ (\xi_t-\varepsilon_t)f_{\varepsilon_t\mid\mX_t}(0)\mX_t\mX_t^\prime \big]\longrightarrow \mK$, as $n\to\infty$, for some positive definite $\mK\in\mathbb{R}^{k\times k}$.
	
\end{enumerate}

\end{assumptionA}

Note that we do not need to strengthen Assumption~\ref{ass:Serial Dependence} by also imposing $\alpha$-mixing for the ES errors $\xi_t$.
To see why, equating \eqref{eq:(1.0)} and \eqref{eq:(1.1)} gives that
\begin{equation}\label{eq:QR and ES errors}
	\xi_t=\mX_{t}^\prime(\valpha_0 - \vbeta_0)+\varepsilon_t.
\end{equation}
Due to this and, e.g., \citet[Theorem~14.1]{Dav94}, it already follows from Assumption~\ref{ass:Serial Dependence} that $\big\{(\varepsilon_t, \xi_t, \mX_t^\prime)^\prime\big\}$ is $\alpha$-mixing with mixing coefficients of size $-r/(r-2)$.

\begin{thm}\label{thm:ES est}
Under Assumptions~\ref{ass:innov}, \ref{ass:Serial Dependence} and \ref{ass:Heterogeneity ES}* it holds for any $\epsilon\in(0,1)$ that, as $n\to\infty$,
\begin{equation*}
s\sqrt{n}\big[\widehat{\vbeta}(s)-\vbeta_0\big]\overset{d}{\longrightarrow}\mOmega_{\ast}^{1/2}\mW_k(s)\qquad\text{in }D_{k}[\epsilon,1],
\end{equation*}
where $\mW_k(\cdot)$ is a $k$-variate standard Brownian motion and $\mOmega_{\ast}=\frac{1}{(1-\tau)^2}\mD_{\ast}^{-1}\mSigma_{\ast}\mD_{\ast}^{-1}$ with positive definite $\mSigma_{\ast} = \mJ_{\ast} + \mK\mD^{-1}\mC + (\mK\mD^{-1}\mC)^\prime + \mK\mOmega\mK$.
\end{thm}

\begin{proof}
See Appendix~\ref{sec:thm2}.
\end{proof}

A careful reading of the proof of Theorem~\ref{thm:ES est} shows that if the QR parameters $\valpha_0$ are known, then the asymptotic variance of $\widehat{\vbeta}(s)$ simplifies to
\[
	\mOmega_{\ast} =  \frac{1}{(1-\tau)^2}\mD_{\ast}^{-1}\mJ_{\ast}\mD_{\ast}^{-1}.
\]
Yet, in practice, the QR parameters are not known, leading to the unwieldy form of the asymptotic variance $\mOmega_{\ast}$.
Estimating it would require an estimate of the meat matrix $\mJ$ from the QR part \textit{and} an estimate of the ES meat matrix $\mSigma_{\ast}$.
However, for this latter matrix, no estimation theory exists.
In sum, directly estimating the asymptotic variance-covariance seems a daunting task.

Once again, this problem is sidestepped by our self-normalized approach to inference.
The analog of Corollary~\ref{cor:SN} for ES regressions is the following

\begin{cor}\label{cor:SN ES}
Let $\mB$ be a full rank $(\ell\times k)$ matrix with $\ell\leq k$, and define $\vbeta_{0,\mB}=\mB\vbeta_0$ and $\widehat{\vbeta}_{\mB}(s) = \mB\widehat{\vbeta}(s)$.
Under the assumptions of Theorem~\ref{thm:ES est} it holds that, as $n\to\infty$,
\[
	n\big[\widehat{\vbeta}_{\mB}(1) - \vbeta_{0,\mB}\big]^\prime \mS_{n,\widehat{\vbeta}_{\mB}(\cdot)}^{-1} \big[\widehat{\vbeta}_{\mB}(1) - \vbeta_{0,\mB}\big]\overset{d}{\longrightarrow}W(\mB),
\]
where $\mS_{n,\widehat{\vbeta}_{\mB}(\cdot)}$ is defined similarly as $\mS_{n,\widehat{\valpha}_{\mA}(\cdot)}$ with $\widehat{\valpha}_{\mA}(\cdot)$ replaced by $\widehat{\vbeta}_{\mB}(\cdot)$ at every occurrence.
\end{cor}

\begin{proof}
Analogous to that of Corollary~\ref{cor:SN}.
\end{proof}

In principle, the framework of \citet[Section~3]{PZC19} is sufficiently general to also cover ES regressions.
This simply requires the inclusion of $\mX_{t}$ in their information set $\mathcal{F}_{t-1}=\sigma\{Y_{t-1},\mX_{t-1},\ldots,Y_{1},\mX_{1}\}$, which does not change any of their arguments.
However, their predictive modeling framework assumes that $Q_{\tau}(Y_t\mid\mathcal{F}_{t-1})=\mX_t^\prime \valpha_0$ and $\ES_{\tau}(Y_t\mid\mathcal{F}_{t-1})=\mX_t^\prime \vbeta_0$, which is a much more stringent assumption than our $Q_{\tau}(Y_t\mid\mX_t)=\mX_t^\prime \valpha_0$ and $\ES_{\tau}(Y_t\mid\mX_t)=\mX_t^\prime \vbeta_0$.
Similarly as in the discussion below Theorem~\ref{thm:std est}, the assumption of \citet{PZC19} leads to vanishing cross terms in the meat matrix; see the matrix $\mA_{0}$ in their Theorem~2.
Therefore, their modeling approach is well-suited for predictive modeling of quantiles and ES, but may lead to invalid inference in time series ES regressions.

\section{Simulations}\label{sec:Simulations}	

\subsection{The Data-Generating Process}\label{DGP}
We consider a time series regression as a data-generating process (DGP), which exhibits heteroskedasticity and autocorrelation in the errors.
To ensure comparability, the set-up is identical to \cite{GY24}. The data $\big\{(Y_{t}, \mX_{t}^\prime)^\prime\big\}_{t=1,\ldots,n}$ for all simulations is generated as follows.

The stochastic regressor $x_t$ in $\mX_t = (1,x_t)^\prime$ follows the AR(1) model
\begin{equation*}
x_t = \rho_x x_{t-1} + \eps_t,\qquad |\rho_x| < 1, \qquad \eps_t\overset{\text{i.i.d.}}{\sim}N(0,1).
\end{equation*}
The dependent variable $Y_t$ is generated from
\begin{align*}
	Y_t &= \mX_t^\prime \boldsymbol{\delta}  + (\mX_t^\prime \boldsymbol{\eta}) e_t,  \qquad \boldsymbol{\delta} \in \mathbb{R}^2, \qquad \boldsymbol{\eta} \in \mathbb{R}^2,
\end{align*}
with autoregressive
\begin{equation}\label{eq:dgp}
e_t = \rho e_{t-1} + \nu_t,\qquad |\rho|<1,\qquad \nu_t\overset{\text{i.i.d.}}{\sim}N(0,1-\rho^2).
\end{equation}
For this time series regression, the true conditional (quantile, ES) are linear functions in $\mX_t$ given by
\begin{equation*}
Q_{\tau}(Y_t\mid\mX_t) = \mX_t' \boldsymbol{\delta} + (\mX_t^\prime \boldsymbol{\eta}) Q_{\tau}(e_t)  ,\qquad \ES_{\tau}(Y_t \mid \mX_t) = \mX_t' \boldsymbol{\delta} + (\mX_t^\prime \boldsymbol{\eta})\ES_{\tau}(e_t ),
\end{equation*} 
where $Q_{\tau}(e_t):=F_{e_t}^{-1}(\tau)$ with $F_{e_t}(\cdot)=\P\{e_t\leq\cdot\}$ denoting the cumulative distribution function of $e_t$, and $\ES_{\tau}(e_t)=\E\big[e_t \mid e_t\geq Q_{\tau}(e_t)\big]$.
This implies for the true coefficients $\valpha_0 = (\alpha_{0,1}, \ \alpha_{0,2})^\prime$ and $\vbeta_0 = (\beta_{0,1}, \ \beta_{0,2})^\prime$ in \eqref{eq:(1.0)}--\eqref{eq:(1.1)} that $\valpha_0 = \vdelta + \veta Q_{\tau}(e_t) $ and $\vbeta_0 = \vdelta + \veta \ES_{\tau}(e_t)$. 

Note that we choose $\Var(\nu_t) = 1-\rho^2$ in \eqref{eq:dgp} to guarantee that $\Var(e_t) =  \frac{\Var(\nu_t)}{1-\rho^2} = 1$ is independent of $\rho$. 
Therefore, $e_t \sim N(0,1)$ for the Gaussian autoregression in \eqref{eq:dgp} for all $|\rho| < 1$, such that we can change the serial dependence structure of $e_t$ without affecting its marginal distribution. In particular, $Q_\tau(e_t)$ does not change with $\rho$.

We set $\rho_x = 0.8$, $\vdelta = (\delta_1, \ \delta_2)^\prime=(0,\ 1)^\prime$, and $\veta=(\eta_1,\ \eta_2)^\prime = (2,\ 0.5)^\prime$. The sample sizes we consider are $n \in \{100,\ 200,\ 500,\ 1000\}$, the degree of autocorrelation is $\rho \in \{0,\ 0.5,\ 0.9\}$, and the quantile levels are $\tau \in \{0.5,\ 0.75,\ 0.9\}$. All these values are taken from Section 6 of \cite{GY24}. Note that we set $\epsilon = 0.1$ in our QR simulation settings as suggested by \cite{ZS13}. For the ES regression case, we require a higher $\epsilon = 0.25$ (see Table~\ref{tab:sizeES}). 
Additional simulations (that are available upon request from the authors) show that the results are qualitatively unchanged for different choices of $\epsilon$ as long as the estimator can be computed.

\subsection{Quantile Regressions}\label{sim:Quantile Regressions}

We examine size and power of our SN-based tests from Corollary~\ref{cor:SN} in the following. Although we are mainly interested in ES regressions, considering quantile regressions here allows for a comparison of our self-normalized approach with the methods of \cite{GLN18} and \cite{GY24}, thus, shedding light on the relative merits of SN. We focus on tests with nominal size equal to 5\%. In order to assess size, for each $\tau \in \{0.5,\ 0.75,\ 0.9\}$ we test 
\begin{equation}\label{eq:truenull}
H_0: \alpha_{0,2} = \delta_{2}^{\circ}  + \eta_2 Q_\tau(e_t).
\end{equation} 
By setting $\delta_{2}^{\circ} =\delta_2= 1$ and $\eta_2 = 0.5$, we ensure that the null is true.
Later on, when we investigate power, we vary the null-hypothetical value $\delta_{2}^{\circ}$, but keep $\eta_2$ fixed at its true value of $0.5$.
Therefore, $\eta_2$ does not carry a superscript.

We conduct four statistical tests of $H_0$ using our SN-based approach, the HAC procedure of \cite{GY24}, the SBB of \cite{GLN18}, and, as a reference, conventional i.i.d. errors (IID). 
The HAC test proposed by \cite{GY24} is a $t$-test, where the standard error is based on an estimate of $\mJ = \sum_{t=-\infty}^{\infty} \Cov\big(\psi(\varepsilon_0)\mX_0, \psi(\varepsilon_t)\mX_t\big)$. 
In contrast, the IID test estimates the standard error assuming $\mJ = \Var\big(\mX_1 \psi(\varepsilon_t)\big)$. The SBB approach, on the other hand, constructs a resampling distribution for the estimator $\widehat{\valpha}$, from which the respective standard errors and confidence intervals can be derived. The SBB accounts for the serial dependence in $\mJ$ through its moving block design, and mitigates issues related to the smooth conditional density of the errors in $\mD$ by employing a two-fold smoothing strategy: tapering for data blocks and kernel smoothing for individual observations. Therefore, the HAC and the SBB approaches rely on a data-driven estimation of the bandwidth, and bandwidth and block size choices, respectively. The SN-based approach, instead, requires neither bandwidth nor block-size choices. 

\begin{table}[t!]\centering
	\caption{Rejection frequencies for IID, HAC, SBB, and SN-based tests for the QR model.}
\begin{tabular}[t]{rlrrrrrrrrrrrr}
	\toprule
	\multicolumn{2}{c}{ } & \multicolumn{4}{c}{$\tau= 0.5$} & \multicolumn{4}{c}{$\tau = 0.75$} & \multicolumn{4}{c}{$\tau = 0.9$} \\
	\cmidrule(l{3pt}r{3pt}){3-6} \cmidrule(l{3pt}r{3pt}){7-10} \cmidrule(l{3pt}r{3pt}){11-14}
	$n$ & $\rho$ & IID & HAC & SBB & SN & IID & HAC & SBB & SN & IID & HAC & SBB & SN\\
	\midrule
	100 & 0.0 & 6.8 & 5.6 & 3.6 & 3.2 & 8.9 & 6.6 & 2.0 & 3.5 & 13.0 & 7.1 & 3.8 & 3.6\\
	& 0.5 & 16.2 & 9.5 & 10.4 & 3.5 & 18.6 & 10.2 & 8.4 & 3.8 & 19.7 & 10.8 & 5.6 & 4.0\\
	& 0.9 & 40.6 & 16.6 & 23.6 & 6.0 & 40.2 & 17.6 & 22.4 & 6.7 & 41.2 & 23.4 & 17.4 & 7.9\\
	\noalign{\vskip 2pt}\hline\noalign{\vskip 2pt}
	200 & 0.0 & 5.9 & 6.0 & 1.6 & 3.9 & 6.8 & 6.1 & 2.4 & 3.8 & 9.7 & 6.4 & 0.6 & 3.9\\
	& 0.5 & 15.8 & 8.6 & 7.6 & 4.2 & 16.3 & 8.9 & 6.0 & 4.2 & 17.0 & 9.9 & 4.4 & 4.3\\
	& 0.9 & 38.7 & 12.6 & 24.2 & 5.8 & 39.2 & 13.9 & 23.0 & 6.4 & 38.7 & 18.4 & 22.4 & 7.3\\
	\noalign{\vskip 2pt}\hline\noalign{\vskip 2pt}
	500 & 0.0 & 5.1 & 6.3 & 2.2 & 4.1 & 5.3 & 6.2 & 1.4 & 4.4 & 7.0 & 6.7 & 1.0 & 4.5\\
	& 0.5 & 14.8 & 7.9 & 9.8 & 4.5 & 14.5 & 8.2 & 6.6 & 4.2 & 14.0 & 8.5 & 5.4 & 4.5\\
	& 0.9 & 38.4 & 10.5 & 29.0 & 5.1 & 37.6 & 11.0 & 22.6 & 5.7 & 36.1 & 13.7 & 18.6 & 6.2\\
	\noalign{\vskip 2pt}\hline\noalign{\vskip 2pt}
	1000 & 0.0 & 4.1 & 5.5 & 1.4 & 4.4 & 5.0 & 6.0 & 1.2 & 4.5 & 5.9 & 6.7 & 1.2 & 4.4\\
	& 0.5 & 13.7 & 7.1 & 6.8 & 4.3 & 13.7 & 7.4 & 4.8 & 4.4 & 13.0 & 8.3 & 5.0 & 4.8\\
	& 0.9 & 38.7 & 9.2 & 26.2 & 4.5 & 36.8 & 9.3 & 23.4 & 5.0 & 34.6 & 11.2 & 20.2 & 5.5\\
	\bottomrule
\end{tabular}
	\captionsetup[table]{font=small} 
	\caption*{\textit{Notes:} We compute the i.i.d. errors using the \texttt{iid} option from the \texttt{rq} function in the \texttt{quantreg} package in \texttt{R}. 
		For the HAC standard errors, the data-driven optimal bandwidth, $\widehat{S}_T^*$, is calculated by the alternative autocorrelation estimator $\widehat{\phi}_z^*(\tau)$ using the quadratic spectrum kernel. For the SBB, we simulate based on only 500 replications due to computational constraints. We use $B=2500$ bootstrap resamples per iteration. The empirical sizes are obtained under data-based bandwidth and nonparametric plug-in block size selection. This amounts to computing the block length $l$ using the \texttt{getNPPIblksizesQR} function and the bandwidth $h$ 
		using the \texttt{bw.SJ} function from the \texttt{QregBB} package in \texttt{R}. The SN-based approach requires neither bandwidth nor block-size choices. We set $\epsilon = 0.1$ for our SN-based test statistic. Nominal size is equal to 5\%.}
	\label{tab1}
\end{table}

Table~\ref{tab1} reports the empirical sizes of the four tests based on 10,000 Monte Carlo replications. The ``HAC'' columns in Table~\ref{tab1} correspond to the ``$t_r$'' columns of Table 3 in \cite{GY24}. The results show that the rejection frequencies of the SN-based approach are generally close to the nominal significance level of 5\%. Two notable exceptions occur when the sample size is rather small, i.e., $n \in \{100,\ 200\}$: First, when the autocorrelation in the errors $e_t$ is low ($\rho = 0$), slight under-rejection is observed. Second, when the autocorrelation in the errors is high ($\rho = 0.9$), we obtain slight over-rejections of the null hypothesis. As the sample size increases, the distortions diminish.

In contrast, the tests based on HAC and SBB standard errors exhibit severe size distortions. While the HAC approach also improves with larger sample sizes, it still struggles to maintain the 5\%-significance level under strong autocorrelation. 
Indeed, for $\rho=0.9$, size ranges from 9.2\% to 23.4\%.
Such liberal tendencies are well-known in HAC-type tests for mean regressions \citep{And91,KV05}. 
For the SBB, we observe size distortions independent of $n$ with large over-rejections of between 17.4\% and 29\% for high autocorrelation $(\rho = 0.9)$ and sizable under-rejections for no autocorrelation $(\rho = 0)$. 
When there is no dependence to model ($\rho = 0$), the block structure imposed by SBB becomes unnecessary and may introduce additional noise into the procedure. 
For highly persistent errors $e_t$ ($\rho = 0.9$), we speculate that the block lengths selection may lead to blocks that are too small, such that the strong serial dependence is not sufficiently captured in the resamples.

\begin{figure}[t!]
	\centering
	
	\includegraphics[width=\textwidth]{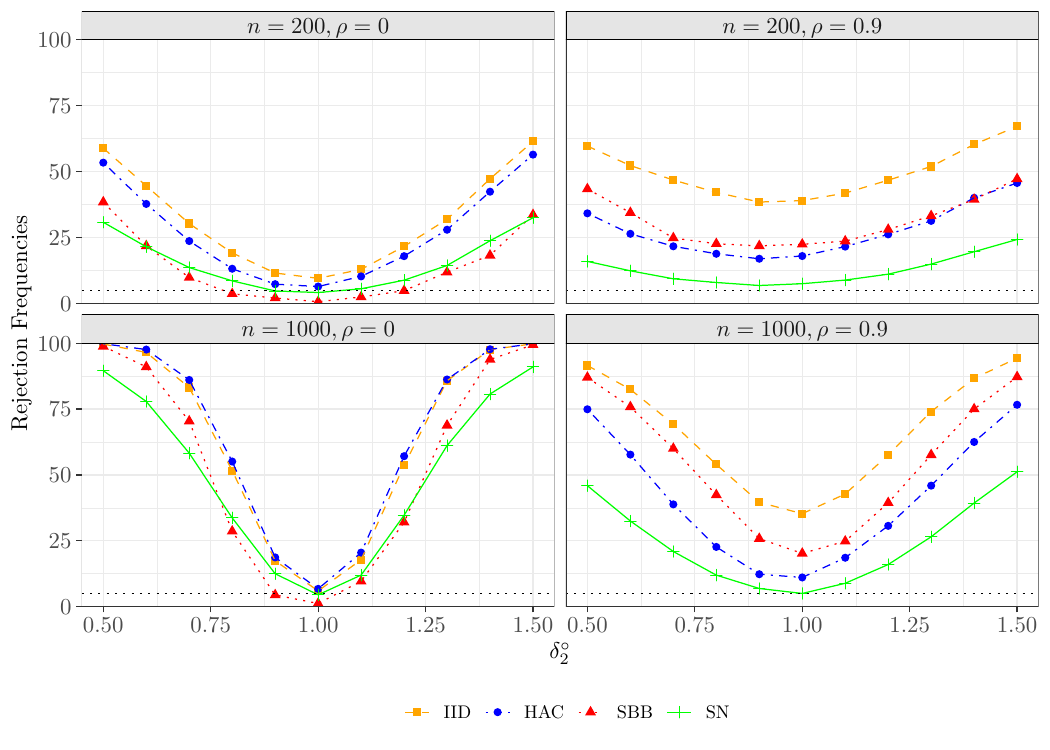}
	
	\caption{Power for IID, HAC, SBB, and SN-based tests for $\tau = 0.9$. Dotted horizontal lines indicate nominal level of 5\%.}
	\label{fig:powerQR}
\end{figure}

As is well-known in the SN literature \citep{Sha10,Sha15}, there is a trade-off between robustness and efficiency. Therefore, we now compute power to quantify this trade-off for our approach. Figure~\ref{fig:powerQR} illustrates the power of the four tests under the null hypothesis $H_0: \alpha_{0,2} = \delta_{2}^{\circ} + \eta_2 Q_\tau(e_t)$, where $\eta_2 = 0.5$ as for the true null in \eqref{eq:truenull}, and $\delta_{2}^{\circ}$ varies between $[0.5, 1.5]$. To conserve space, we only report results for $\tau = 0.9$. Results for the other values of $\tau$ are available from the authors upon request. 
We present the edge cases of low ($\rho = 0$) and high ($\rho =0.9$) autocorrelation in the errors $e_t$. 
As expected, power across all four tests is roughly symmetric in $\delta_{2}^{\circ}$ around the true value of $\delta_2=1$. 
It increases in $n$ and in the distance of the alternative from the null, i.e., in $|\delta_{2}^{\circ}-1|$. The analysis confirms that the strong control over size of the SN-based test comes at the cost of reduced power when testing against a false null hypothesis. 
Consider the upper right panel in Figure~\ref{fig:powerQR}. 
For $\delta_{2}^{\circ}=1$, the HAC approach exhibits a size of 18.4\%, more than double that of our SN-based method (7.3\%). 
In terms of power, however, the HAC approach outperforms our SN procedure, with a probability of rejecting $H_0$ based on, e.g., $\delta_{2}^{\circ} = 0.5$ of 34.1\% compared to 16.7\% for our approach. This result is, of course, not surprising given the substantial size distortion of the HAC procedure.

\begin{figure}[t!]
	\centering
	\includegraphics[width=\textwidth]{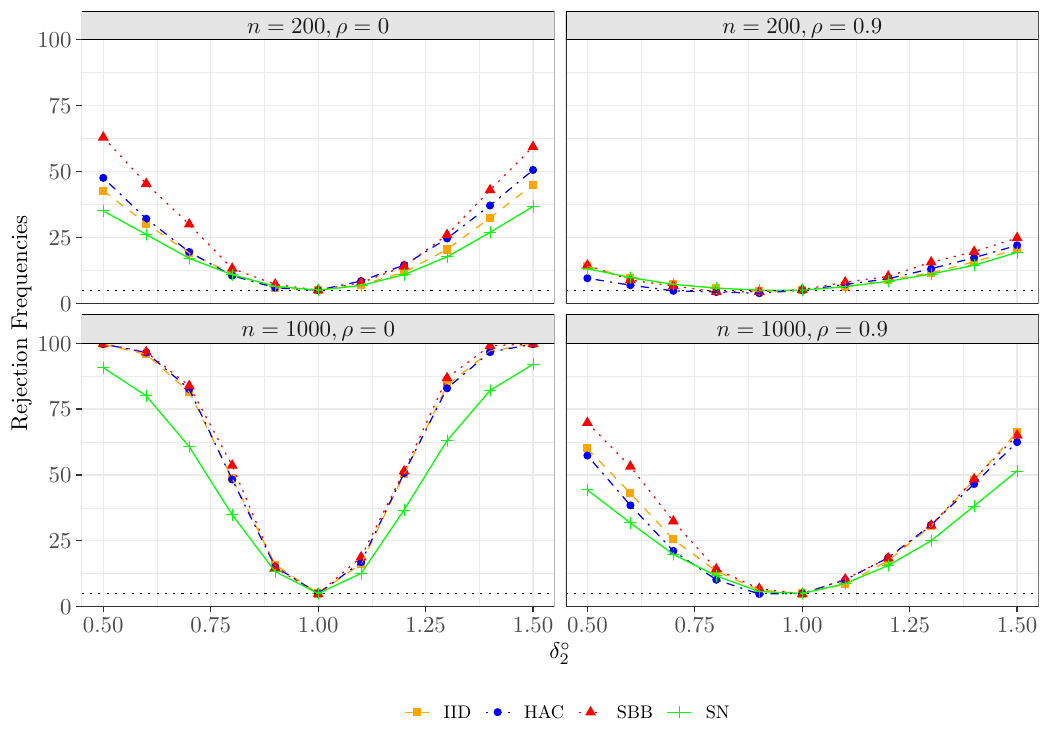}
	\caption{Size-adjusted power for IID, HAC, SBB, and SN-based tests for $\tau = 0.9$. Dotted horizontal lines indicate nominal level of 5\%.}
		\label{fig:spowerQR}
\end{figure}

For this reason, we compute size-adjusted power and present it in Figure~\ref{fig:spowerQR}. By construction, size-adjusted power for all four tests is equal to 5\% for the true value $\delta_{2}^{\circ} = 1$. The power advantage of the over-sized tests seen in Figure~\ref{fig:powerQR} shrinks after size-adjusting, but not entirely. This confirms the well-known ``better size, lower power'' property of SN-based tests.

\subsection{Expected Shortfall Regressions}

Now, we present size and power for the ES regressions. 
In order to assess size, for each $\tau \in \{0.5,\ 0.75,\ 0.9\}$ we evaluate the true null 
\begin{equation}\label{eq:truenullES}
 H_0: \beta_{0,2} = \delta_{2}^{\circ} + \eta_2 \ES_\tau(e_t),
\end{equation}
with $\delta_{2}^{\circ}  = 1$ and $\eta_2 = 0.5$.
We conduct three statistical tests of $H_0$ using our SN-based approach and the HAC procedure of \citet{Bar23} based on his two-stage ES regression parameter estimator in its unweighted version (HAC) and its weighted version (HAC-W).\footnote{We are grateful to Sander Barendse for sharing his computer code.}

\begin{table}[t!]\centering
	\caption{Rejection frequencies for the HAC, HAC-W, and SN-based tests for the ES regression.}

\begin{tabular}[t]{rlrrrrrrrrr}
	\toprule
	\multicolumn{2}{c}{ } & \multicolumn{3}{c}{$\tau= 0.5$} & \multicolumn{3}{c}{$\tau = 0.75$} & \multicolumn{3}{c}{$\tau = 0.9$} \\
	\cmidrule(l{3pt}r{3pt}){3-5} \cmidrule(l{3pt}r{3pt}){6-8} \cmidrule(l{3pt}r{3pt}){9-11}
	$n$ & $\rho$ & HAC & HAC-W & SN & HAC & HAC-W & SN & HAC & HAC-W & SN\\
	\midrule
	100 & 0.0 & 9.3 & 20.7 & 2.8 & 11.6 & 23.2 & 3.4 & 18.2 & 26.7 & 2.5\\
	& 0.5 & 13.8 & 23.4 & 4.9 & 17.3 & 27.5 & 4.6 & 25.2 & 32.9 & 3.3\\
	& 0.9 & 27.9 & 35.2 & 8.9 & 34.5 & 41.2 & 10.2 & 46.3 & 50.4 & 6.7\\
	\noalign{\vskip 2pt}\hline\noalign{\vskip 2pt}
	200 & 0.0 & 7.3 & 18.2 & 4.0 & 9.0 & 19.2 & 3.9 & 12.8 & 20.2 & 3.1\\
	& 0.5 & 11.4 & 19.7 & 5.1 & 13.3 & 22.1 & 5.0 & 18.1 & 24.8 & 4.3\\
	& 0.9 & 23.5 & 31.0 & 8.2 & 27.4 & 34.0 & 8.8 & 36.0 & 41.6 & 9.1\\
	\noalign{\vskip 2pt}\hline\noalign{\vskip 2pt}
	500 & 0.0 & 6.2 & 15.9 & 4.4 & 6.9 & 16.7 & 4.2 & 8.9 & 17.3 & 4.0\\
	& 0.5 & 9.2 & 17.2 & 5.3 & 10.1 & 18.4 & 5.3 & 12.1 & 19.8 & 5.5\\
	& 0.9 & 19.0 & 25.9 & 6.9 & 19.3 & 27.1 & 7.2 & 25.4 & 31.8 & 9.8\\
	\noalign{\vskip 2pt}\hline\noalign{\vskip 2pt}
	1000 & 0.0 & 5.6 & 14.8 & 4.4 & 6.2 & 15.0 & 4.4 & 7.7 & 15.9 & 4.4\\
	& 0.5 & 7.8 & 15.5 & 5.1 & 8.1 & 16.9 & 5.0 & 9.9 & 18.0 & 5.2\\
	& 0.9 & 14.8 & 23.4 & 5.7 & 16.3 & 25.4 & 7.1 & 18.9 & 26.7 & 8.0\\
	\bottomrule
\end{tabular}
	\label{tab:sizeES}
	\captionsetup[table]{font=small} 
	\caption*{\textit{Notes:}  We compute HAC standard errors (HAC) based on the unweighted two-stage ES regression parameter estimator of \cite{Bar23}. In its efficiently weighted (HAC-W) version, we additionally apply location-scale weights, $\overline{w}_t(\hat{\omega}_T)$, and set them to 0.4 and 0.6 as in the original paper. For $n = 100$ and $\tau = 0.9$, we set $\epsilon = 0.3$ in our SN-based test statistic from Corollary~\ref{cor:SN ES}. For all other parameter combinations, we set $\epsilon = 0.25$. A higher $\epsilon$ is required for ES regressions, as they are more data-intensive than QR. Using for instance $\epsilon = 0.1$, the first sequential estimate of $\valpha_0$ is computed on only 10 observations. At $\tau = 0.9$, this would result in an insufficient number of observations for the first sequential estimate of $\vbeta_0$. To ensure comparability to the QR results in Table~\ref{tab1}, we show results for $\tau \in \{0.5, \ 0.75\}$ as well, even though the ES defined as  $\ES_{\tau}(Y_t\mid\mX_t)=\E_{t}\big[Y_t\mid Y_t \geq Q_{\tau}(Y_t\mid\mX_t)\big]$ is typically thought of as the expected value of the conditional tail of the distribution for a large quantile such as $\tau = 0.9$. Nominal size is equal to 5\%.}
\end{table}

We first compare the results of our SN-based approach for the ES regressions in Table~\ref{tab:sizeES} to the SN-based QR results in Table~\ref{tab1}. The empirical sizes for the ES regressions show slightly more distortions. This is as expected, since the sample size in ES regressions is reduced to approximately $\tau n$ when testing whether $x_t$ influences the conditional tail expectation of $Y_t$. Despite this reduction, the empirical sizes still appear very reasonable. In particular, as for the QR, distortions occur mostly when the sample size is rather small. When the autocorrelation in the errors $e_t$ is low, some mild under-rejection is observed. In contrast, when the autocorrelation in the errors is high, we see slight over-rejections of the null hypothesis. As the sample size increases, the size distortions become less severe or vanish completely.

As expected, the HAC standard errors computed based on the approach of \citet{Bar23} are not sufficient to account for highly autocorrelated error processes. The efficient weighting approach does not alleviate the problem as well with size between 14.8\% and 50.4\% across all setups.
 On the contrary, distortions only increase when efficient weights are used.
We mention that the size distortions seen here for the HAC approaches are not uncommon.
In fact, the liberal tendencies of HAC-based tests motivated the development of many alternative inference methods \citep{KV02,KV05}.


\begin{figure}[t!]
	\centering
	\includegraphics[width=\textwidth]{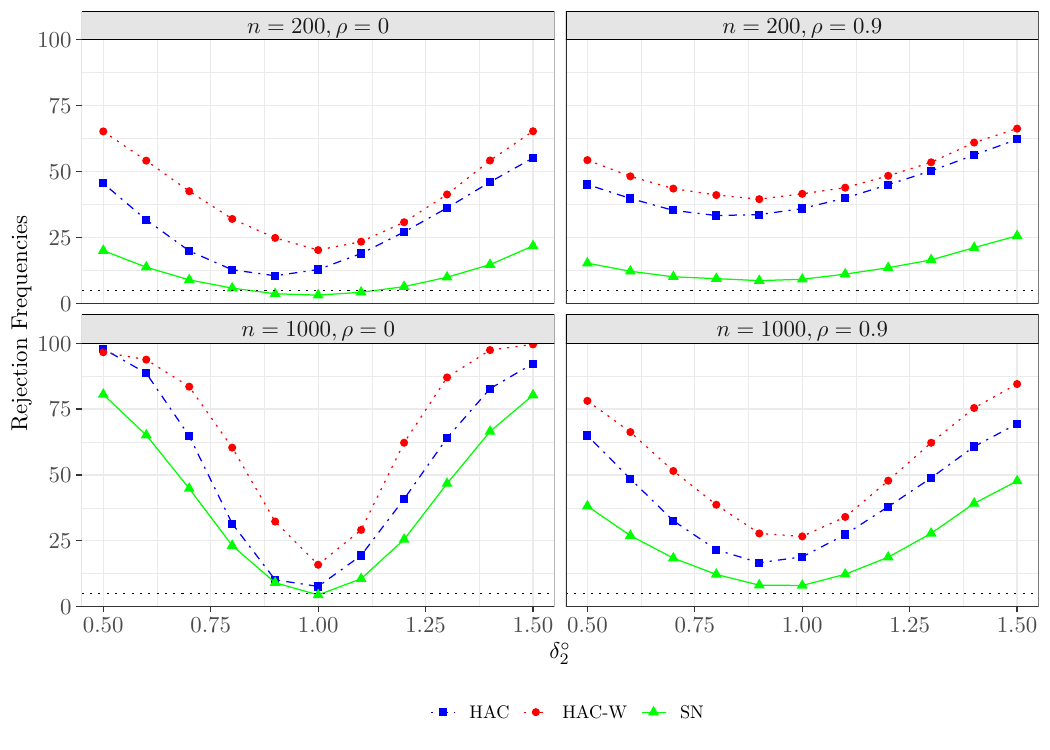}
	\caption{Power for HAC, HAC-W, and SN-based tests for the ES regressions at $\tau = 0.9$. Dotted horizontal lines indicate nominal level of 5\%.}
	\label{fig:powerES}
\end{figure}

\begin{figure}[t!]
	\centering
	\includegraphics[width=\textwidth]{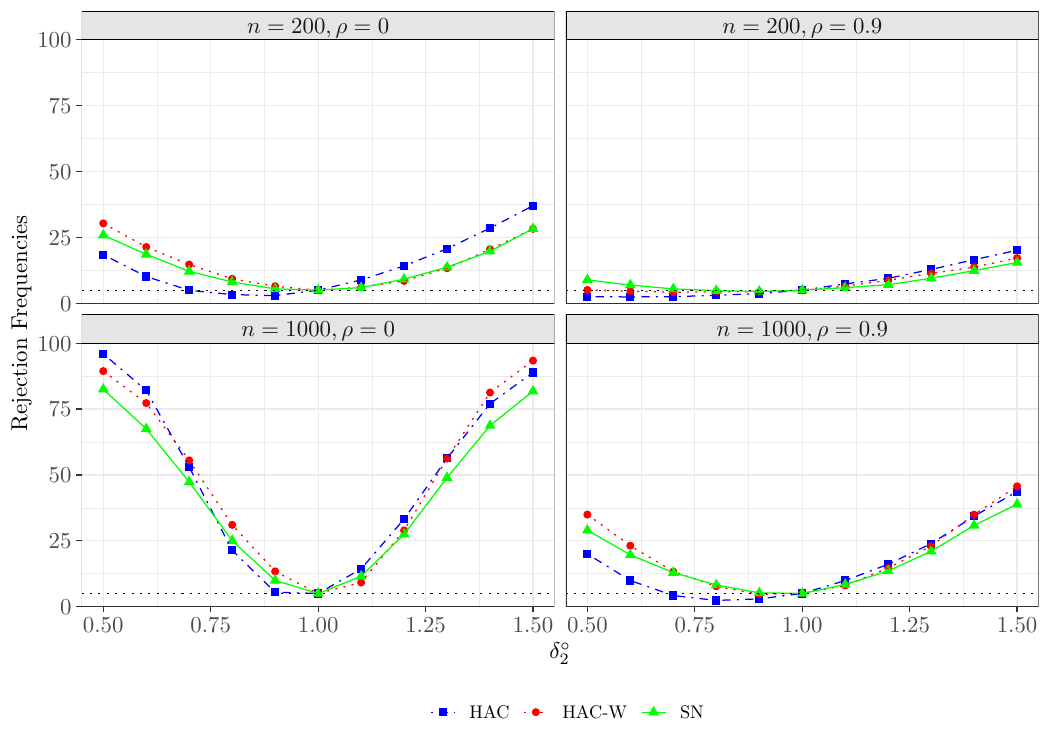}
	\caption{Size-adjusted power for for HAC, HAC-W, and SN-based tests for $\tau = 0.9$. Dotted horizontal lines indicate nominal level of 5\%.}
		\label{fig:spowerES}
\end{figure}

Figure~\ref{fig:powerES} illustrates the power results for the hypothesis $H_0: \beta_{0,2} = \delta_{2}^{\circ} + \eta_2 \ES_\tau(e_t)$, where $\eta_2 = 0.5$, consistent with the true null in \eqref{eq:truenullES}.
The power is evaluated for different values of $\delta_{2}^{\circ} \in [0.5, 1.5]$. Again, we present results only for $\tau = 0.9$ to save space. As for the quantile regressions, power is roughly symmetric in $\delta_{2}^{\circ}$ around the true value of 1. 
Moreover, it increases in $n$ and in the distance of the alternative from the null, i.e., in $|\delta_{2}^{\circ}-1|$. 

For completeness, we compute size-adjusted power and present it in Figure~\ref{fig:spowerES}. By construction, size-adjusted power for all three tests is equal to 5\% for the true value $\delta_{2}^{\circ} = 1$. The power advantage of the over-sized tests seen in Figure~\ref{fig:powerES} shrinks after size-adjusting. For some values of $\delta_{2}^{\circ}$, the size-adjusted power of the SN-based test even exceeds that of the HAC and HAC-W approaches.

Overall, we find that the tests based on the bootstrap or HAC estimators are extremely liberal, leading to many spurious rejections of the null.
In contrast, our SN-based tests deliver accurate size with little or---in the case of ES regressions---no price to pay in terms of (size-adjusted) power.

\section{Empirical Applications}\label{sec:Empirical Application}


\subsection{Predictive Quantile Regressions for the Equity Premium}\label{sec:Empirical Application I}

The question of whether the distribution of the equity premium is predictable has long intrigued academics and investors alike. While \citet{GW08} and \citet{GW24} find only limited predictive power both in-sample and out-of-sample for most predictors of the mean of the equity premium, for the whole distribution of the equity premium a different picture emerges. \cite{CT08} run quantile regressions using the predictors from \cite{GW08} and find that many variables have an asymmetric effect on the return distribution, affecting lower, central and upper quantiles very differently. 

A key issue in these predictive regressions is, however, the asymptotic distribution of estimators when predictors are highly persistent. In particular, the limiting distributions of slope coefficient estimates depend on the degree of persistence in the predictors; see, e.g., \citet{P15H} for an overview. Another concern in this literature is autocorrelation and heteroskedasticity in the errors, which---if not sufficiently accounted for---leads to over-rejections of the null hypothesis as demonstrated in our Monte Carlo simulations. 

In recent years, researchers have developed inference approaches allowing for persistent regressors \citep{Lee16} and various other stylized facts, such as heteroskedasticity \citep{ CCL23, LLPY24, MSK24}. These authors use different time spans of the data and find essentially less and less predictability across all quantile levels the more robust the approach gets. Since their approaches, however, are developed for near-unit root predictors, they do not consider stationary regressors in their empirical applications.


Thus, a particularly underexplored case is that of stationary predictors allowing for heteroskedastic and autocorrelated errors. Using our SN-based approach to inference in time series quantile regressions, we can shed light on this case. 
We examine four predictors investigated in \citet{GW08}: inflation in levels, stock variance, long-term rate of returns, and the default return spread. These four predictors are commonly regarded as stationary in the literature \citep[see, e.g.,][Table~1]{FLS23}. 
In studying predictability, we adopt a comprehensive approach by utilizing all available monthly data from 1927 to 2023 for all four regressors.\footnote{We use the updated data set on Amit Goyal's website (\url{https://sites.google.com/view/agoyal145}).}
This gives us a sample of size $n = 1,163$.

\begin{table}[t!]\centering
	\caption{$p$-values (in \%) of the Dynamic Quantile Test of \cite{EM04}}
\begin{tabular}[t]{lrrrrrrrrr}
	\toprule
	\multicolumn{1}{c}{ } & \multicolumn{9}{c}{Quantile Level} \\
	\cmidrule(l{3pt}r{3pt}){2-10}
	& 0.1 & 0.2 & 0.3 & 0.4 & 0.5 & 0.6 & 0.7 & 0.8 & 0.9\\
	\midrule
	Default Return Spread & \textbf{0.0} & \textbf{1.8} & \textbf{3.2} & \textbf{2.7} & \textbf{1.3} & 7.7 & \textbf{0.3} & \textbf{0.2} & \textbf{0.0}\\
	Inflation & \textbf{0.0} & \textbf{2.9} & 6.9 & 7.4 & \textbf{3.4} & 18.8 & \textbf{1.2} & \textbf{1.6} & \textbf{0.0}\\
	Long-term Rate of Return & \textbf{0.0} & \textbf{0.5} & 6.8 & \textbf{0.9} & \textbf{3.4} & 16.1 & \textbf{0.2} & \textbf{0.2} & \textbf{0.0}\\
	Stock Variance & 24.4 & 55.4 & \textbf{4.7} & \textbf{2.2} & \textbf{0.6} & 12.0 & \textbf{4.7} & 45.9 & \textbf{1.5}\\
	\bottomrule
\end{tabular}
	\captionsetup[table]{font=small} 
	\caption*{\textit{Notes:} We use 10 lags in the specification of the DQT. $p$-values $\le 5\%$  in bold.}
	\label{tab:DQT}
\end{table}

We estimate the parameters $\valpha_0 = (\alpha_{0,1}, \ \alpha_{0,2})^\prime$ of the linear predictive QR
\begin{equation}\label{eq:p19}
Y_t=\alpha_{0,1}+\alpha_{0,2} x_{t-1}+\varepsilon_t ,  \qquad Q_{\tau}(\varepsilon_t\mid x_{t-1})=0,
\end{equation}
for the monthly equity premium $Y_t=\log \left(\frac{P_t+D_t}{P_{t-1}}\right)-\log \left(1+R_t^f\right)$.
Here, $P_t$ denotes the index value of the S\&P 500, $D_t$ the dividends paid out by all S\&P 500 constituents over the current month, and $R_t^f$ the Treasury-bill rate. 
As mentioned above, the predictors $x_{t-1}$ are the default return spread (drs), inflation (infl), long-term rate of return (ltr), and stock variance (svar). Our simple linear QR specification in \eqref{eq:p19} follows \cite{Lee16}, \cite{FL19}, and \cite{LLPY24}. 

We emphasize again that, compared to the popular approaches of the aforementioned authors, we only need to assume $Q_{\tau}(\varepsilon_t\mid x_{t-1})=0$ instead of $Q_{\tau}(\varepsilon_t\mid Y_{t-1}, x_{t-1}, Y_{t-2}, x_{t-2},\ldots)=0$.
These two conditions are equivalent to $\E\big[\psi(\varepsilon_t)\mid x_{t-1}\big]=0$ and $\E\big[\psi(\varepsilon_t)\mid Y_{t-1}, x_{t-1}, Y_{t-2}, x_{t-2},\ldots\big]=0$, respectively.
Thus, our assumption allows for autocorrelation in the (generalized) errors, while the other does not. 
Of course, assuming non-autocorrelated generalized errors, implied by the latter assumption, is particularly suspect in such simple linear regressions as \eqref{eq:p19} with only \textit{one} predictor to capture all the dynamics.
As shown in our Monte Carlo simulations, our SN-based test statistic is most robust to autocorrelation in the errors compared with all other testing approaches. 

In order to corroborate the necessity of our autocorrelation-robust approach, we test the generalized errors $\psi(\varepsilon_t)$ for autocorrelation. 
Recall that in QR, $\sqrt{n}(\widehat{\valpha} - \valpha_0)\overset{d}{\longrightarrow}N(\vzero, \mD^{-1}\mJ\mD^{-1})$, as $n\to\infty$, with $\mJ$ being the long-run variance of $n^{-1/2}\sum_{t=1}^{n}\mX_t\psi(\varepsilon_t)$. These generalized errors then have mean zero conditional on covariates, i.e., $\E[\psi(\varepsilon_t)\mid x_{t-1}]=\tau - \P\big\{\varepsilon_t\leq0\mid x_{t-1}\big\}=0$, compared to $\E[\varepsilon_t\mid x_{t-1}]= 0$ in a mean regression. 
We use the Dynamic Quantile test (DQT) of \citet{EM04} to test the null hypothesis whether $\big\{\psi(\varepsilon_t)=\tau-\1_{\{\varepsilon_t\leq0\}}=\tau-\1_{\{Y_t\leq Q_{\tau}(Y_t\mid x_{t-1})\}}\big\}_{t=1}^{n}$ is i.i.d., with QR errors $\varepsilon_{t} = Y_t - \alpha_{0,1}- \alpha_{0,2}x_{t-1}=Y_t - Q_{\tau}(Y_t\mid x_{t-1})$.
(The generalized errors $\psi(\varepsilon_t)$ correspond to what \citet{EM04} call the \textit{hit function}.)
In carrying out the DQT, we of course have to replace the QR generalized errors $\psi(\varepsilon_t)$ by the QR generalized residuals $\psi(\widehat{\varepsilon}_t)$, where $\widehat{\varepsilon}_t=Y_t-\widehat{\alpha}_{0,1}-\widehat{\alpha}_{0,2}x_{t-1}=Y_t-\widehat{Q}_{\tau}(Y_t\mid x_{t-1})$.
Table~\ref{tab:DQT} shows that the null hypothesis can be rejected across most quantile levels, indicating autocorrelation in the generalized errors.
This implies the need to use a robust testing procedure, such as our SN-based approach.

\begin{figure}[t!]
	\centering
	\includegraphics[width=\textwidth]{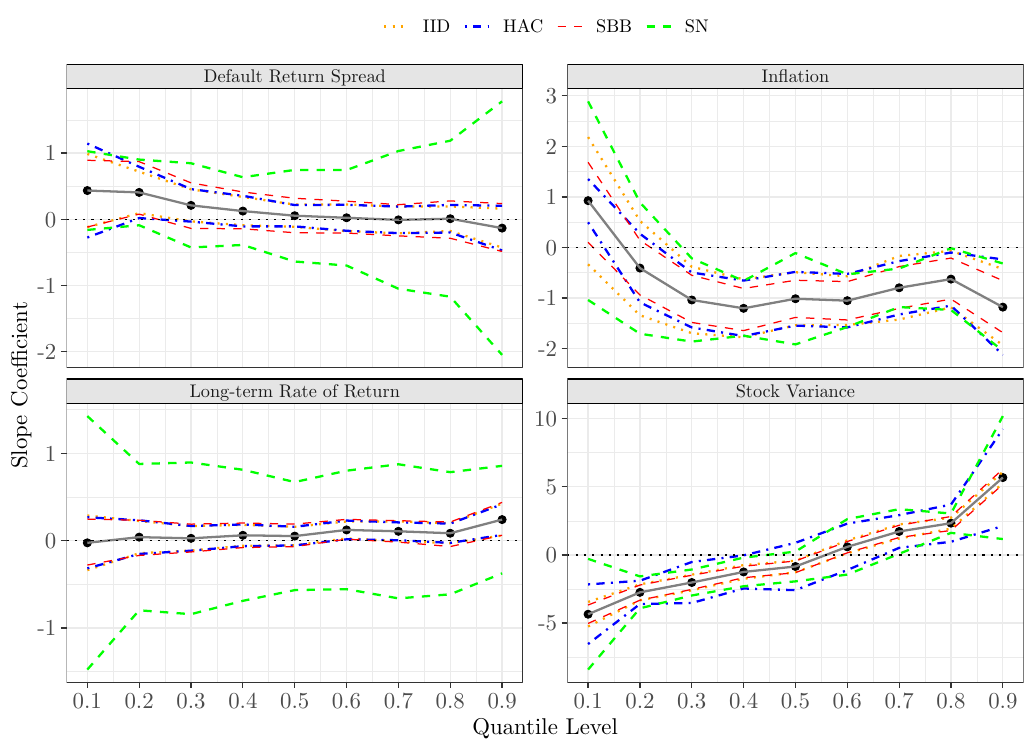} \\
	\caption{Estimates of the slope coefficients for the QR at relevant quantile levels. The colored lines represent pointwise 95\%-confidence intervals based on the IID, HAC, SBB, and self-normalized standard errors.}
	\label{fig:equity}
\end{figure}

Figure~\ref{fig:equity} presents the estimates of $\alpha_{0,2}$ from \eqref{eq:p19} for each predictor over different quantile levels ranging from 0.1 to 0.9. The colored lines around the estimates are confidence intervals based on the IID, HAC, SBB, and self-normalized approach. Since we only test one coefficient at a time, $S_{n,\widehat{\alpha}_{\mA}(\cdot)}$ and $\widehat{\alpha}_{\mA}$ are scalars and therefore the SN-based asymptotic $(1-\nu)$-confidence interval for $\alpha_{0,2}$ is simply
\begin{equation*}
\left[\widehat{\alpha}_{\mA}- \sqrt{S_{n,\widehat{\alpha}_{\mA}(\cdot)}W_{1-\nu}(\mA) /n }, \ \widehat{\alpha}_{\mA}+ \sqrt{S_{n,\widehat{\alpha}_{\mA}(\cdot)}W_{1-\nu}(\mA) / n} \right],
\end{equation*}
where $\mA =  (0, \ 1)\in\mathbb{R}^{1\times2}$ and $W_{1-\nu}(\mA)$ denotes the $(1-\nu)$-quantile of the limiting distribution $W(\mA)$; cf.~\citet[Sec.~2.1]{Sha15}.

Generally, the results differ strongly between the predictors and change with the quantile levels. As expected from our Monte Carlo simulations, the SN-based approach proves to be most robust against autocorrelation in the errors for the majority of the considered cases. We highlight two instances, where it is particularly apparent in our results. First, if we were to rely on autocorrelation-robust approaches other than our own, we might conclude that inflation (upper right panel in Figure~\ref{fig:equity}) has predictive power for the lowest quantile at $\tau = 0.1$. Second, the long-term rate of return (lower left panel) proves to be insignificant using our approach, whereas all the others find predictability for large equity premium quantiles ($\tau = 0.9$).
Note that for both inflation (for $\tau=0.1$) and the long-term rate of return (for $\tau=0.9$), Table~\ref{tab:DQT} indicates significantly autocorrelated QR errors.

It is worth emphasizing that both \citet{GW08} and \citet{CT08} focus on conditional means and quantiles of the equity premium, typically using the full set of available predictors. However, to reduce estimation error, they often estimate regressions with a single predictor at a time. Our approach, which also analyzes one covariate at a time, can be seamlessly integrated into ensemble forecasting frameworks that aggregate results across multiple predictors, thereby leveraging the full information set while maintaining estimation precision.

\subsection{Growth-at-Risk Regressions}\label{sec:Empirical Application II}

In recent years, researchers and policymakers have increasingly turned their attention to measuring economic downside risks as part of monitoring the outlook of the economy.
One prominent example is the GaR approach of \citet{ABG19}. They use predictive quantile regressions and the ES to capture time-varying tail risks to US GDP growth.

 First, \citet{ABG19} estimate quantile regressions using lagged GDP growth and the lagged US NFCI\footnote{The NFCI is computed by the Federal Reserve Bank of Chicago and consists of 105 indicators capturing U.S. financial conditions in money markets, debt and equity markets, and the traditional and ``shadow'' banking systems.} as covariates. They find an association of the NFCI with the lower quantiles of the GDP growth distribution. Second, they match the predicted quantiles to the theoretical moments of the skewed-$t$ distribution developed by \citet{AZ13}. This conditional distribution is then used to calculate the ES of US GDP growth. They conclude that the dependence of future GDP growth on current financial conditions is significantly stronger for the lower quantiles of the distribution than for the upper quantiles.
 
 \begin{figure}[t!]\centering
	\includegraphics[width=.75\linewidth]{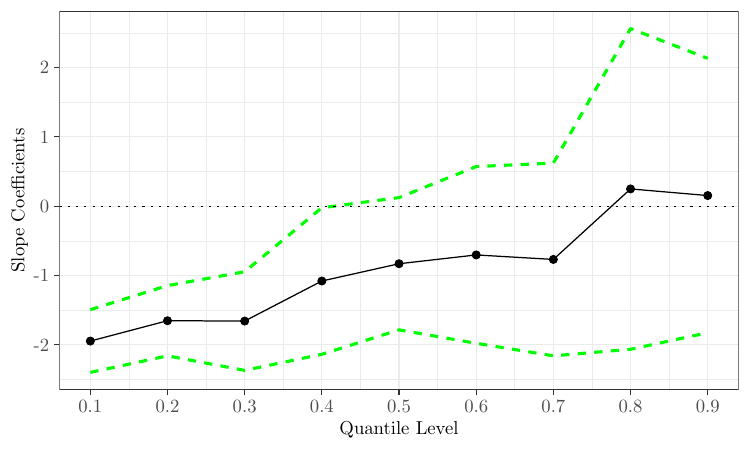}
	\caption{Estimates for the QR regression at relevant quantile levels for the lagged NFCI. Dashed lines represent pointwise 95\%-confidence intervals based on our self-normalized approach.}
	\label{fig:QR}
\end{figure}	

Their analysis has sparked a vast literature on capturing macroeconomic tail risks using Bayesian vector autoregressions \citep{CCM24}, several other predictors \citep{Pea20,Aea21}, a GARCH specification using a panel of countries \citep{BS21}, or a time-varying skewed-$t$ model for GDP Growth \citep{DDP24}, among others. Overall, these results highlight that both asymmetry and time-variation in higher moments crucially depend on the choice of the model and the conditioning information.

In our analysis, we revisit the parsimonious \cite{ABG19} QR specification and add a crucial aspect to their results---statistical inference. 
Another distinguishing feature in terms of the methodology is that in order to make statements about the conditional ES of GDP growth, we do not have to rely on the fitted skewed-$t$ distribution but can directly estimate the effect of the NFCI on the ES of the GDP growth distribution. 
We then test the statistical significance of the predictors using our SN-based approach.

We emphasize that, as widely recognized in financial risk management through the Basel Accords \citep{BCBS16}, the ES is the appropriate risk measure to consider in recession risk forecasting as well. Rather than focusing solely on a specific conditional quantile of the GDP growth distribution, economists should aim to understand the entire left tail of the distribution to accurately measure severe economic disruptions. The conditional quantile as a risk measure has additional limitations, such as its lack of coherence \citep{Aea99}. In the context of GaR, this means that aggregating risks across sectors or economies may fail to account for diversification effects, potentially overstating the total risk.


\begin{figure}[t!]\centering
	\includegraphics[width=0.8\linewidth]{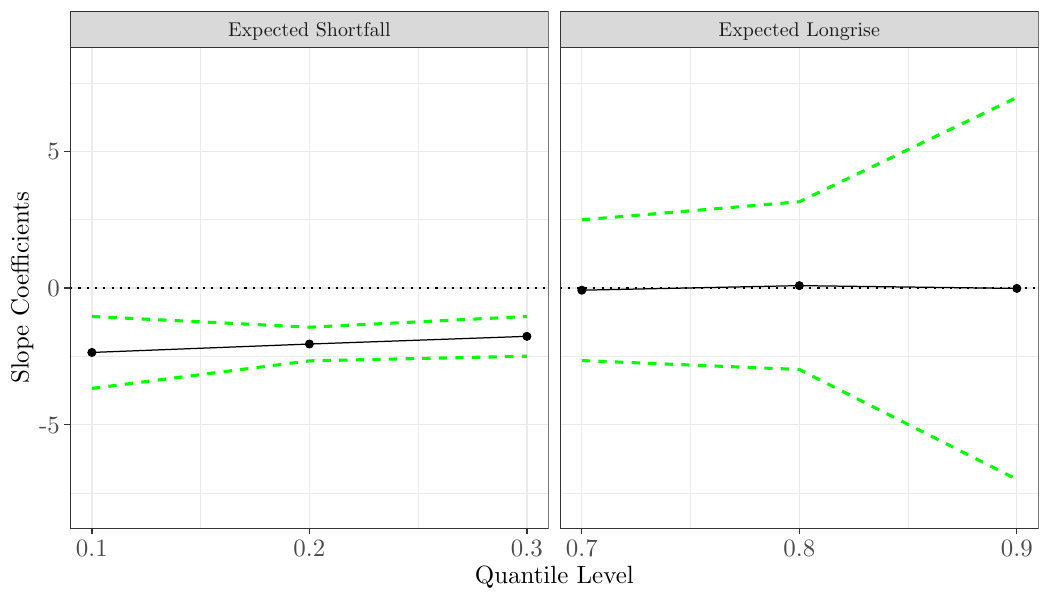}
	\caption{Estimates for the ES and the EL regression at relevant quantile levels for the lagged NFCI. Dashed lines represent pointwise 95\%-confidence intervals based on our self-normalized approach.}
	\label{fig:ES}
\end{figure}

 We use data on the NFCI and quarter-over-quarter annualized GDP growth from 1973Q1 up to 2019Q4 ($n=187$) to estimate the parameters $\boldsymbol{\alpha}_0 = (\alpha_{0,1}, \ \alpha_{0,2}, \ \alpha_{0,3})'$ and $\boldsymbol{\beta}_0 = (\beta_{0,1}, \ \beta_{0,2}, \ \beta_{0,3})'$ of the linear (quantile, ES) regression
\begin{align*}
	\GDP_t &=  \alpha_{0,1} + \alpha_{0,2} \NFCI_{t-1} + \alpha_{0,3} \GDP_{t-1} + \varepsilon_t,  \qquad Q_{\tau}(\varepsilon_t\mid \NFCI_{t-1}, \GDP_{t-1})=0,\\
	\GDP_t &= \beta_{0,1} + \beta_{0,2} \NFCI_{t-1} + \beta_{0,3} \GDP_{t-1} +  \xi_t,  \qquad \ES_{\tau}(\xi_t\mid \NFCI_{t-1}, \GDP_{t-1})=0.
\end{align*}
Note that since the GaR literature commonly defines the ES as the left conditional tail expectation of the GDP growth distribution, we slightly abuse notation here and re-define the ES as $\ES_{\tau}(\GDP_t\mid\mX_t)=\E_{t}\big[\GDP_t\mid\GDP_t\leq Q_{\tau}(\GDP_t\mid\mX_t)\big]$, where $\mX_{t}=(1,\GDP_{t-1},\NFCI_{t-1})^\prime$.

As introduced by \citet{ABG19}, the ES has a right-tail counterpart---the expected longrise (EL), which we define as $\EL_{\tau}(\GDP_t\mid\mX_t)=\E_{t}\big[\GDP_t\mid\GDP_t\geq Q_{\tau}(\GDP_t\mid\mX_t)\big]$. The EL, thus, measures the entire right tail of the GDP growth distribution beyond an upper conditional quantile, e.g., $\tau = 0.9$, and can be interpreted as an estimate for a best-case scenario outcome. We can now estimate the parameters $\boldsymbol{\alpha}_0 = (\alpha_{0,1}, \ \alpha_{0,2}, \ \alpha_{0,3})'$ and $\boldsymbol{\gamma}_0 = (\gamma_{0,1}, \ \gamma_{0,2}, \ \gamma_{0,3})'$ of the linear (quantile, EL) regression 
\begin{align*}
	\GDP_t &=  \alpha_{0,1} + \alpha_{0,2} \NFCI_{t-1} + \alpha_{0,3} \GDP_{t-1} + \varepsilon_t,\qquad Q_{\tau}(\varepsilon_t\mid \NFCI_{t-1}, \GDP_{t-1})=0, \\
	\GDP_t &= \gamma_{0,1} + \gamma_{0,2} \NFCI_{t-1} + \gamma_{0,3} \GDP_{t-1} +  \xi_t,  \qquad \EL_{\tau}(\xi_t\mid \NFCI_{t-1}, \GDP_{t-1})=0.
\end{align*}


\begin{figure}[t!]\centering
	\includegraphics[width  = 0.8\linewidth]{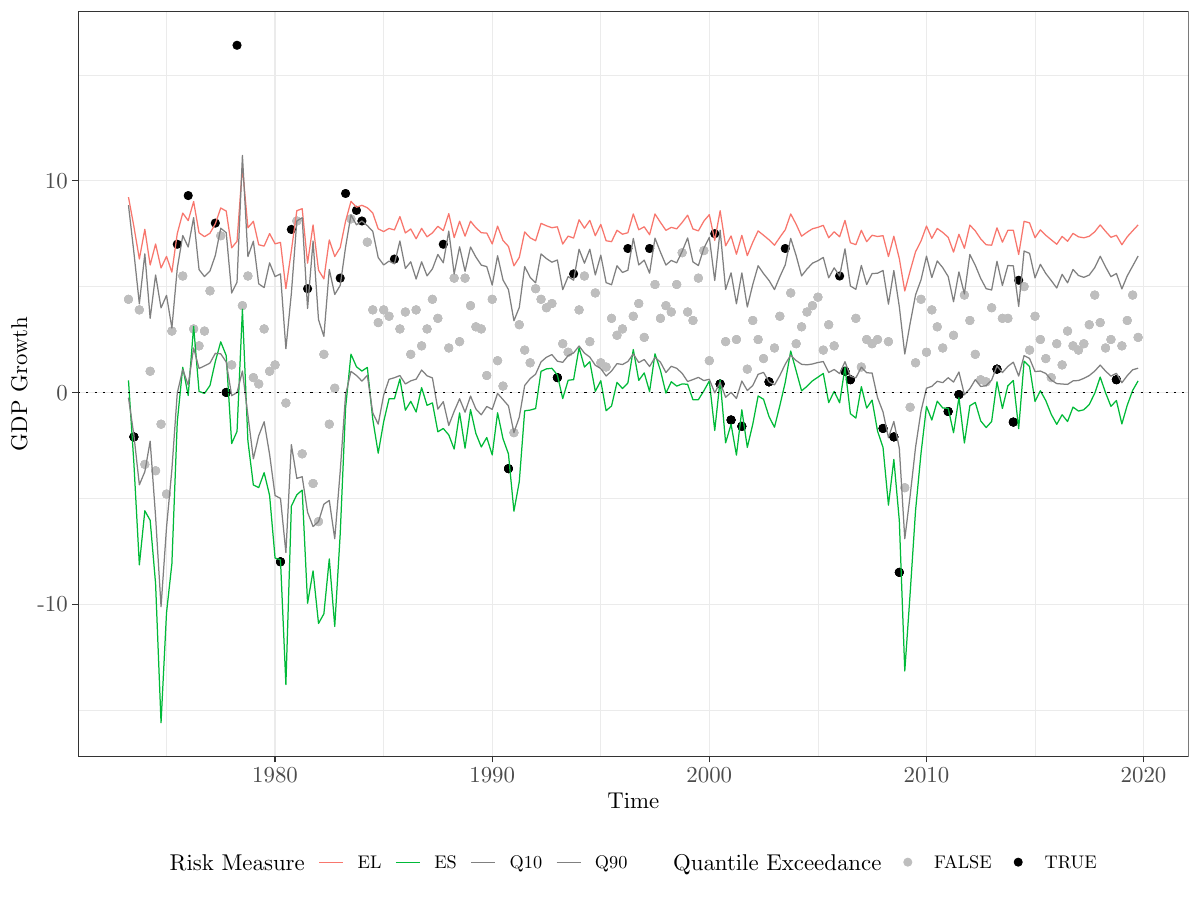}
	\caption{Estimated EL ($\tau=0.9$) and ES ($\tau=0.1$) sample paths for US GDP growth from 1973Q1 to 2019Q4. GDP growth rates with an exceedance of the estimated conditional quantile at $\tau = 0.1$ and $\tau = 0.9$ are displayed in black while all other growth rates are shown in gray color.}
	\label{fig}
\end{figure}

The estimates of the NFCI coefficients $\alpha_{0,2}$ are summarized in Figure~\ref{fig:QR}. The figure reproduces the quantile regression results from \citet[Figure 4, Panel C]{ABG19} and---as a novelty---adds SN-based confidence intervals to it. Note that \cite{ABG19} do not provide such or other autocorrelation-robust confidence intervals, e.g., those of \cite{GLN18} or \cite{GY24}. Figure~\ref{fig:QR} shows that the regression slopes change for the NFCI across the quantiles, with the lowest quantiles of GDP being significantly negatively related to last quarter's NFCI, while the highest quantiles of the GDP growth distribution are not significantly affected by the NFCI. 

Figure~\ref{fig:ES} shows how the ES regression estimates for $\beta_{0,2}$ and its EL counterpart for $\gamma_{0,2}$ vary over the quantile levels $\tau \in \{0.1,\ 0.2,\ 0.3\}$ and $\tau \in \{0.7,\ 0.8,\ 0.9\}$, respectively. The estimated effect of the lagged NFCI on the ES at $\tau = 0.1$ of GDP growth ranges from $-3.7$ to $-1.0$, and is significantly different from 0 at the 95\%-confidence level. 
The estimated effects for the EL of GDP growth at all relevant quantile levels vary slightly around 0 and come with wide confidence intervals. 
Note that this finding is much stronger than the results in \cite{ABG19}. While \cite{ABG19} only visually observe the stronger association of the ES with the NFCI compared to that of the NFCI with the EL, we can quantify and test the effect of the NFCI on the ES/EL of the GDP growth distribution.  

Finally, Figure~\ref{fig} plots the estimated sample paths of GDP growth for the 10\%- and 90\%-quantile and illustrates our results graphically. The ES of GDP growth varies more with its lagged covariates than its counterpart the EL, indicating skweness in the conditional distribution of GDP growth.

\section{Conclusion}\label{sec:Conclusion}	

In this paper, we propose feasible inference methods for time series ES regressions.
Our SN-based approach requires nothing more than parameter estimates computed over expanding windows.
This eliminates the need for data-driven estimates of a bandwidth \citep{Bar23,GY24} or block length \citep{GLN18}, making it both simple and computationally efficient. 
Simulations demonstrate that while size may be concerningly high for the other methods under strongly autocorrelated errors, it is adequate for our SN-based approach.
Our simulations also highlight a favorable robustness-efficiency tradeoff: more reliable size (robustness) is traded off for some mild---or, in the case of ES regressions, even non-existent---loss in power (efficiency), which is often desirable in practical applications.
We see this as the main advantage of our method, because it guards against spurious rejections of the null by practitioners.

We illustrate our inference methods in two examples. First, we test whether common stationary predictors \citep{GW08,GW24,CT08} have predictive power for the distribution of the equity premium using data from 1927 to 2023. The SN-based approach proves to be most robust for the majority of the considered cases.

Second, we revisit \citeauthor{ABG19}'s \citeyearpar{ABG19} seminal work in the Growth-at-Risk literature. \citet{ABG19} do not statistically test the effect of lagged financial conditions on the conditional tails of the GDP growth distribution. Using our SN-based approach to ES regressions, we are able to confirm a significant effect of the NFCI on the future GDP growth distribution in times of crises. In economically prosperous times, the NFCI plays no significant role, highlighting the relevance of macrofinancial linkages in the explanation of economic downturns.

\appendix
\renewcommand\appendixpagename{Appendix}
\appendixpage

First, we fix some notation.
The probability space we work on is $(\Omega, \mathcal{F}, \P)$.
We denote by $C_k[a,b]$ the space of $\mathbb{R}^k$-valued (componentwise) continuous functions on the interval $[a,b]$ ($0\leq a<b<\infty$), which is endowed with the uniform topology. 
We refer to \citet{PD86} and \citet{Bil99} for more detail on $C_k[a,b]$ and $D_k[a,b]$ (the latter being defined in the main text).
We denote by $\Vert\cdot\Vert_{p}$ the $L_p$-norm of a random variable, i.e., $\Vert X \Vert_p=\big\{\E|X|^p\big\}^{1/p}$.
The symbols $o_{\P}(1)$, $O_{\P}(1)$, $o(1)$ and $O(1)$ carry their usual meaning and, if not indicated otherwise, are to be understood with respect to $n\to\infty$.
The identity matrix is denoted by $\mI$, where the dimension will be clear from the context.
Finally, $K>0$ denotes a large universal constant, whose precise value may change from place to place.
Recall that $\Vert\cdot\Vert$ denotes the Euclidean norm.
We exploit in the following without further mention that the Euclidean norm is submultiplicative, such that $\Vert\mA\mB\Vert\leq \Vert\mA\Vert \cdot \Vert\mB\Vert$ for conformable matrices $\mA$ and $\mB$.

\section{Proofs of Theorem~\ref{thm:std est} and Corollary~\ref{cor:SN}}
\label{sec:thm1}
\renewcommand{\theequation}{A.\arabic{equation}}	
\setcounter{equation}{0}

The proof of Theorem~\ref{thm:std est} requires the following three lemmas.
We first present the most important lemma, which may be of independent interest.
To introduce it, we define $f_n(\vx,s,\omega):\mathbb{R}^k\times[a,b]\times\Omega\to\mathbb{R}$ ($n\geq1$) and $f_{\infty}(\vx,s,\omega):\mathbb{R}^k\times[a,b]\times\Omega\to\mathbb{R}$ to be random elements.
As usual, we suppress their dependence on $\omega\in\Omega$ and abbreviate them as $f_n(\vx,s)$ and $f_{\infty}(\vx,s)$, respectively.

\begin{lem}[{cf. \citealp[Theorem~1]{Kat09}}]\label{lem:Kato2009}
Suppose (i) the real-valued random elements $f_n(\vx,s)$ ($n\geq1$) and $f_{\infty}(\vx,s)$ are convex in $\vx$ for each $s$, $f_n(\vx,s)$ ($n\geq1$) is c\`{a}dl\`{a}g in $s$ for each $\vx$ and $f_{\infty}(\vx,s)$ is continuous in $s$ for each $\vx$; (ii) $\vx_{\infty}(s)$ is the unique minimum point of $f_{\infty}(\cdot,s)$ for each $s\in [a,b]$, and $\vx_{n}(s)$ is the minimum point of $f_{n}(\cdot,s)$ for each $s\in [a,b]$; (iii) $\vx_n(\cdot)\in D_{k}[a,b]$ $(n\geq1)$ and $\vx_{\infty}(\cdot)\in C_k[a,b]$.
If, as $n\to\infty$,
\begin{equation}\label{eq:fn finf}
	\big(f_n(\vx_1,\cdot), \ldots, f_n(\vx_{\ell},\cdot)\big)^\prime\overset{d}{\longrightarrow}\big(f_{\infty}(\vx_1,\cdot), \ldots, f_{\infty}(\vx_{\ell},\cdot)\big)^\prime\qquad\text{in}\ D_{\ell}[a,b]
\end{equation}
for each ${\ell}\geq1$, where $\{\vx_1,\vx_2,\ldots\}$ is a dense subset of $\mathbb{R}^k$, then we also have that, as $n\to\infty$,
\[
	\vx_n(\cdot)\overset{d}{\longrightarrow}\vx_{\infty}(\cdot)\qquad\text{in}\ D_k[a,b].
\]
\end{lem}

\begin{proof}
The proof is almost identical to that of Theorem~1 in \citet{Kat09}. 
The only difference is that one needs to resort to \citet[Section~15]{Bil99} to deduce that the convergence in \eqref{eq:fn finf} [the equivalent of \citeauthor{Kat09}'s equation (1)] nonetheless implies (on a Skorohod probability space) the uniform convergence $\sup_{s\in[a,b]}\big|f_n(\vx_i,s) - f_{\infty}(\vx_i,s)\big|\longrightarrow0$ almost surely for each $i$ [the equivalent of \citeauthor{Kat09}'s equation (4)]. 
This conclusion holds because \citet{Bil99} shows that when convergence in $D_{\ell}[a,b]$ is to a continuous limit [here: the $f_{\infty}(\vx_i,\cdot)$'s in \eqref{eq:fn finf}] convergence in the Skorohod metric (which we have equipped $D_{\ell}[a,b]$ in \eqref{eq:fn finf} with) implies convergence in the uniform metric.
The remainder of the proof then follows as in \citet{Kat09}.
\end{proof}

\begin{lem}\label{lem:c2 FS}
Under the assumptions of Theorem~\ref{thm:std est} it holds that, as $n\to\infty$,
\begin{equation*}
	\frac{1}{\sqrt{n}}\sum_{t=1}^{\lfloor ns\rfloor}\psi(\varepsilon_t)\mX_{t}
	\overset{d}{\longrightarrow} \mJ^{1/2}\mW(s)\qquad\text{in}\ D_k[0,1],
\end{equation*}
where $\mW(\cdot)$ is a $k$-variate standard Brownian motion.
\end{lem}

\begin{proof}
See Section~\ref{sec:QRest Lemmas}.
\end{proof}

\begin{lem}\label{lem:c3 FS}
Under the assumptions of Theorem~\ref{thm:std est} it holds for every fixed $\vw\in\mathbb{R}^{k}$ that, as $n\to\infty$,
\begin{equation*}
	\sup_{s\in[0,1]}\bigg| \sum_{t=1}^{\lfloor ns\rfloor}\big(\varepsilon_t - \vw^\prime n^{-1/2}\mX_{t}\big)\big(\1_{\{\vw^\prime n^{-1/2}\mX_{t}<\varepsilon_t<0\}} - \1_{\{0<\varepsilon_t<\vw^\prime n^{-1/2}\mX_{t}\}}\big)	-	\frac{1}{2}s\vw^\prime\mD\vw\bigg|=o_{\P}(1).
\end{equation*}
\end{lem}

\begin{proof}
See Section~\ref{sec:QRest Lemmas}.
\end{proof}

\begin{proof}[{\textbf{Proof of Theorem~\ref{thm:std est}:}}]
The estimator $\widehat{\valpha}(s)$ can equivalently be written as
\begin{align*}
\widehat{\valpha}(s) &= \argmin_{\valpha\in\mathbb{R}^{k}} \sum_{t=1}^{\lfloor ns\rfloor}\big[\rho(Y_t-\mX_{t}^\prime\valpha) - \rho(\varepsilon_t)\big]\\
&= \argmin_{\valpha\in\mathbb{R}^{k}} \sum_{t=1}^{\lfloor ns\rfloor}\big[\rho\big(Y_t-\mX_{t}^\prime\valpha_0 -\mX_t^\prime(\valpha-\valpha_0)\big) - \rho(\varepsilon_t)\big]\\
&\overset{\eqref{eq:QR model}}{=} \argmin_{\valpha\in\mathbb{R}^{k}} \sum_{t=1}^{\lfloor ns\rfloor}\big[\rho\big(\varepsilon_t -\mX_t^\prime(\valpha-\valpha_0)\big) - \rho(\varepsilon_t)\big].
\end{align*}
Thus, defining
\[
 W_n(\vw,s)=\sum_{t=1}^{\lfloor ns\rfloor}\big[\rho(\varepsilon_t - \vw^\prime n^{-1/2}\mX_{t}) - \rho(\varepsilon_t)\big],
\] 
the minimizer $\vw_n(s)$ of $W_n(\cdot,s)$ satisfies that 
\[
	\vw_n(s)=\sqrt{n}\big[\widehat{\valpha}(s) - \valpha_0\big].
\]

To derive the weak limit of $\vw_n(\cdot)$, we invoke Lemma~\ref{lem:Kato2009} (where $\vw_n(\cdot)$ plays the role of $\vx_n(\cdot)$ in Lemma~\ref{lem:Kato2009}, and $W_n(\vw,s)$ that of $f_n(\vx,s)$).
For this, we first derive the limit of the finite-dimensional distributions of $W_n(\cdot,s)$ to verify \eqref{eq:fn finf}.
Note that for $u\neq0$,
\begin{equation}\label{eq:rho eq}
	\rho(u-v) - \rho(u) = -v\psi(u) + (u-v)\big[\1_{\{v<u<0\}} - \1_{\{0<u<v\}}\big],
\end{equation}
where $\psi(u)=\tau-\1_{\{u\leq0\}}$ is defined in Section~\ref{sec:The Linear Model}.
Since
\[
	\P\big\{\exists\ t\in\mathbb{N}\colon \varepsilon_t=0\big\} = \P\Big\{\bigcup_{t\in\mathbb{N}}\big\{\varepsilon_t=0\big\}\Big\}\leq\sum_{t\in\mathbb{N}}\P\big\{\varepsilon_t=0\big\}=0
\]
by Assumption~\ref{ass:innov}~\eqref{it:dens}, we may use \eqref{eq:rho eq} to conclude that almost surely
\begin{align*}
	W_n(\vw,s) &= -\vw^\prime\frac{1}{\sqrt{n}}\sum_{t=1}^{\lfloor ns\rfloor}\mX_{t}\psi(\varepsilon_t)\\
	& \hspace{2cm} + \sum_{t=1}^{\lfloor ns\rfloor}\big(\varepsilon_t - \vw^\prime n^{-1/2}\mX_{t}\big)\big[\1_{\{\vw^\prime n^{-1/2}\mX_{t}<\varepsilon_t<0\}} - \1_{\{0<\varepsilon_t<\vw^\prime n^{-1/2}\mX_{t}\}}\big].
\end{align*}
We obtain from Lemmas~\ref{lem:c2 FS}--\ref{lem:c3 FS} that for each fixed $\vw\in\mathbb{R}^{k}$, as $n\to\infty$,
\begin{align}
	W_n(\vw,s)&\overset{d}{\longrightarrow}-\vw^\prime\mJ^{1/2}\mW(s) +\frac{1}{2}\vw^\prime s\mD\vw\notag\\
	& = -\vw^\prime \va(s)+ \frac{1}{2}\vw^\prime \mB(s)\vw\notag\\
	& =: W_{\infty}(\vw,s)\qquad\text{in}\ D[\epsilon,1],\label{eq:(W)}
\end{align}
where $\va(s) = \mJ^{1/2}\mW(s)$ and $\mB(s) = s\mD$.
A straightforward application of a Cram\'{e}r--Wold device \citep[e.g.,][Theorem~29.16]{Dav94} gives that
\[
	\big(W_n(\vw_{1},s),\ldots,W_n(\vw_{\ell},s)\big)^\prime\overset{d}{\longrightarrow}\big(W_{\infty}(\vw_{1},s), \ldots, W_{\infty}(\vw_{\ell},s)\big)^\prime\qquad\text{in}\ D_{\ell}[\epsilon,1]
\]
for any $\vw_{1},\ldots,\vw_{\ell}\in\mathbb{R}^{k}$.
Therefore, assumption \eqref{eq:fn finf} of Lemma~\ref{lem:Kato2009} is satisfied.

Next, we verify the remaining conditions \textit{(i)}--\textit{(iii)} of Lemma~\ref{lem:Kato2009}.
The only non-trivial part is to show that $W_{\infty}(\cdot,s)$ has a unique minimum.
A simple exercise in multivariable calculus shows that $W_{\infty}(\cdot,s)$ in \eqref{eq:(W)} has a minimum for $\vw_{\infty}(s)$ satisfying the first-order condition
\[
	-\va(s) + \mB(s)\vw_{\infty}(s) = \vzero.
\]
Since $\mB(s)=s\mD$ is positive definite by Assumption~\ref{ass:Heterogeneity}~\eqref{it:(iii)}, $\vw_{\infty}(s)$ is also the unique minimum of $W_{\infty}(\cdot,s)$ and it satisfies
\begin{align*}
	\vw_{\infty}(s)&= \mB^{-1}(s)\va(s)= \frac{1}{s}\mD^{-1}\mJ^{1/2}\mW(s).
\end{align*}
Therefore, $\vw_n(s)\overset{d}{\longrightarrow}\vw_{\infty}(s)$ in $D_k[\epsilon,1]$ by Lemma~\ref{lem:Kato2009}, such that the conclusion follows.
\end{proof}

\begin{proof}[{\textbf{Proof of Corollary~\ref{cor:SN}:}}]
Since $s\mapsto\valpha_{\mA}(s)$ is constant on intervals $\big[(j-1)/n, j/n\big)$ for $j=1,\ldots,n$, it follows that
\begin{align*}
	\mS_{n,\widehat{\valpha}_{\mA}(\cdot)}  &= \frac{1}{n^2}\sum_{j=\lfloor n\epsilon\rfloor+1}^{n}j^2\big[\widehat{\valpha}_{\mA}(j/n) - \widehat{\valpha}_{\mA}(1)\big]\big[\widehat{\valpha}_{\mA}(j/n) - \widehat{\valpha}_{\mA}(1)\big]^\prime \\
	&\leq n\int_\epsilon^1 s^2 \big[\widehat{\valpha}_{\mA}(s) - \widehat{\valpha}_{\mA}(1)\big]\big[\widehat{\valpha}_{\mA}(s) - \widehat{\valpha}_{\mA}(1)\big]^\prime \D s\\
	&\leq \frac{1}{n^2}\sum_{j=\lfloor n\epsilon\rfloor}^{n}(j+1)^2\big[\widehat{\valpha}_{\mA}(j/n) - \widehat{\valpha}_{\mA}(1)\big]\big[\widehat{\valpha}_{\mA}(j/n) - \widehat{\valpha}_{\mA}(1)\big]^\prime,
\end{align*}
where the inequalities are to be understood with respect to the Loewner order $\leq$ for matrices.
Thus, with $j_{\ast}=\lfloor n\epsilon\rfloor$,
\begin{align*}
	0&\leq \frac{1}{n^2}\sum_{j=\lfloor n\epsilon\rfloor}^{n}(j+1)^2\big[\widehat{\valpha}_{\mA}(j/n) - \widehat{\valpha}_{\mA}(1)\big]\big[\widehat{\valpha}_{\mA}(j/n) - \widehat{\valpha}_{\mA}(1)\big]^\prime - \mS_{n,\widehat{\valpha}_{\mA}(\cdot)}\\
	&= \frac{1}{n^2}\sum_{j=\lfloor n\epsilon\rfloor+1}^{n}(j+1)^2\big[\widehat{\valpha}_{\mA}(j/n) - \widehat{\valpha}_{\mA}(1)\big]\big[\widehat{\valpha}_{\mA}(j/n) - \widehat{\valpha}_{\mA}(1)\big]^\prime - \mS_{n,\widehat{\valpha}_{\mA}(\cdot)}\\
	&\hspace{1cm}+ \frac{1}{n^2}(j_{\ast}+1)^2\big[\widehat{\valpha}_{\mA}(j_{\ast}/n) - \widehat{\valpha}_{\mA}(1)\big]\big[\widehat{\valpha}_{\mA}(j_{\ast}/n) - \widehat{\valpha}_{\mA}(1)\big]^\prime\\	
	&= \frac{1}{n^2}\sum_{j=\lfloor n\epsilon\rfloor+1}^{n}(2j+1)\big[\widehat{\valpha}_{\mA}(j/n) - \widehat{\valpha}_{\mA}(1)\big]\big[\widehat{\valpha}_{\mA}(j/n) - \widehat{\valpha}_{\mA}(1)\big]^\prime + o_{\P}(1)\\
	&= \frac{1}{n} \bigg(\frac{1}{n}\sum_{j=\lfloor n\epsilon\rfloor+1}^{n}\big[\widehat{\valpha}_{\mA}(j/n) - \widehat{\valpha}_{\mA}(1)\big]\big[\widehat{\valpha}_{\mA}(j/n) - \widehat{\valpha}_{\mA}(1)\big]^\prime\bigg)\\
	&\hspace{1cm} +2\bigg(\frac{1}{n}\sum_{j=\lfloor n\epsilon\rfloor+1}^{n}\frac{j}{n}\big[\widehat{\valpha}_{\mA}(j/n) - \widehat{\valpha}_{\mA}(1)\big]\big[\widehat{\valpha}_{\mA}(j/n) - \widehat{\valpha}_{\mA}(1)\big]^\prime\bigg)+o_{\P}(1)\\
	&\leq \frac{1}{n} \bigg(\int_{\epsilon}^{1}\big[\widehat{\valpha}_{\mA}(s) - \widehat{\valpha}_{\mA}(1)\big] \big[\widehat{\valpha}_{\mA}(s) - \widehat{\valpha}_{\mA}(1)\big]^\prime\D s\bigg)\\
	&\hspace{1cm} + 2\bigg(\int_{\epsilon}^{1}s\big[\widehat{\valpha}_{\mA}(s) - \widehat{\valpha}_{\mA}(1)\big] \big[\widehat{\valpha}_{\mA}(s) - \widehat{\valpha}_{\mA}(1)\big]^\prime\D s\bigg)+o_{\P}(1)\\
	&= \frac{1}{n^2} \bigg(n\int_{\epsilon}^{1}\big[\widehat{\valpha}_{\mA}(s) - \widehat{\valpha}_{\mA}(1)\big] \big[\widehat{\valpha}_{\mA}(s) - \widehat{\valpha}_{\mA}(1)\big]^\prime\D s\bigg)\\
	&\hspace{1cm} + \frac{2}{n}\bigg(n\int_{\epsilon}^{1}s\big[\widehat{\valpha}_{\mA}(s) - \widehat{\valpha}_{\mA}(1)\big] \big[\widehat{\valpha}_{\mA}(s) - \widehat{\valpha}_{\mA}(1)\big]^\prime\D s\bigg)+o_{\P}(1)\\
	&=O_{\P}(n^{-2}) + O_{\P}(n^{-1})+o_{\P}(1)\\
	&=o_{\P}(1),
\end{align*}
where the second-to-last line follows from the continuous mapping theorem \citep[CMT; e.g.,][Theorem~26.13]{Dav94} in combination with Theorem~\ref{thm:std est}.
The above two displays imply that
\begin{equation}\label{eq:(p.26)}
	\mS_{n,\widehat{\valpha}_{\mA}(\cdot)}=n\int_\epsilon^1 s^2 \big[\widehat{\valpha}_{\mA}(s) - \widehat{\valpha}_{\mA}(1)\big]\big[\widehat{\valpha}_{\mA}(s) - \widehat{\valpha}_{\mA}(1)\big]^\prime \D s+ o_{\P}(1)=:\overline{\mS}_n + o_{\P}(1).
\end{equation}
The CMT and Theorem~\ref{thm:std est} imply that, as $n\to\infty$,
\begin{align*}
	n&\big[\widehat{\valpha}_{\mA}(1) - \valpha_{0,\mA}\big]^\prime \overline{\mS}_n^{-1} \big[\widehat{\valpha}_{\mA}(1) - \valpha_{0,\mA}\big]\\
	& =\sqrt{n}\big[\widehat{\valpha}_{\mA}(1) - \valpha_{0,\mA}\big]^\prime \overline{\mS}_n^{-1} \sqrt{n}\big[\widehat{\valpha}_{\mA}(1) - \valpha_{0,\mA}\big]\\
	&\overset{d}{\longrightarrow}\mW_k^\prime(1)(\mOmega^{1/2})^\prime\mA^\prime\Big\{\mA\mOmega^{1/2}\int_\epsilon^1 \big[\mW_k(s) - s\mW_k(1)\big] \big[\mW_k(s) - s\mW_k(1)\big]^\prime \D s\ (\mOmega^{1/2})^{\prime}\mA^\prime\Big\}^{-1}\times\\
	&\hspace{13cm}\times\mA\mOmega^{1/2}\mW_k(1)\\
	&\overset{d}{=}\mW_k^\prime(1)\mA^\prime\mV_k\mA\mW_k(1),
\end{align*}
where the final step follows from Proposition~2 of \citet{ZS13}.
Combining this and \eqref{eq:(p.26)}, the conclusion follows.
\end{proof}

\section{Proof of Theorem~\ref{thm:ES est}}
\label{sec:thm2}

\renewcommand{\theequation}{B.\arabic{equation}}	
\setcounter{equation}{0}

The proof of Theorem~\ref{thm:ES est} requires the four preliminary Lemmas~\ref{lem:ES 1}--\ref{lem:ES 4}.
Lemmas~\ref{lem:ES 1} and \ref{lem:ES 3} play a role similar to that of Lemma~\ref{lem:c2 FS} in the proof of Theorem~\ref{thm:std est}.
Similarly, Lemmas~\ref{lem:ES 2} and \ref{lem:ES 4} are the equivalents of Lemma~\ref{lem:c3 FS} in the proof of Theorem~\ref{thm:std est}.

\begin{lem}\label{lem:ES 1}
Under the assumptions of Theorem~\ref{thm:ES est} it holds that, as $n\to\infty$,
\begin{equation*}
	\frac{1}{\sqrt{n}}\sum_{t=1}^{\lfloor ns\rfloor}\1_{\{\varepsilon_t>0\}}\xi_t\mX_{t}
	\overset{d}{\longrightarrow} \mJ_{\ast}^{1/2}\mW_{\ast}(s)\qquad\text{in}\ D_k[0,1],
\end{equation*}
where $\mW_{\ast}(\cdot)$ is a $k$-variate standard Brownian motion.
\end{lem}

\begin{proof}
See Section~\ref{sec:ESest Lemmas}.
\end{proof}

\begin{lem}\label{lem:ES 2}
Under the assumptions of Theorem~\ref{thm:ES est} it holds that, as $n\to\infty$,
\begin{equation*}
	\sup_{s\in[0,1]}\bigg\Vert \frac{1}{n}\sum_{t=1}^{\lfloor ns\rfloor} \1_{\{\varepsilon_t>0\}}\mX_t\mX_t^\prime - s(1-\tau)\mD_{\ast}\bigg\Vert=o_{\P}(1).
\end{equation*}
\end{lem}

\begin{proof}
See Section~\ref{sec:ESest Lemmas}.
\end{proof}

\begin{lem}\label{lem:ES 3}
Under the assumptions of Theorem~\ref{thm:ES est} it holds that, as $n\to\infty$,
\begin{equation*}
	\frac{1}{\sqrt{n}}\sum_{t=1}^{\lfloor ns\rfloor} \big[\1_{\{\mX_t^\prime(\widehat{\valpha}(s) - \valpha_0)<\varepsilon_t\leq 0 \}} - \1_{\{0<\varepsilon_t\leq \mX_t^\prime(\widehat{\valpha}(s) - \valpha_0)\}}\big]\xi_t\mX_t\overset{d}{\longrightarrow} \mK\mOmega^{1/2}\mW(s)\qquad\text{in }D_k[\epsilon,1],
\end{equation*}
where $\mW(\cdot)$ is a $k$-variate standard Brownian motion.
\end{lem}

\begin{proof}
See Section~\ref{sec:ESest Lemmas}.
\end{proof}

\begin{lem}\label{lem:ES 4}
Under the assumptions of Theorem~\ref{thm:ES est} it holds that, as $n\to\infty$,
\begin{equation*}
	\sup_{s\in[\epsilon,1]}\bigg\Vert\frac{1}{n}\sum_{t=1}^{\lfloor ns\rfloor} \big[\1_{\{\mX_t^\prime(\widehat{\valpha}(s) - \valpha_0)<\varepsilon_t\leq 0 \}} - \1_{\{0<\varepsilon_t\leq \mX_t^\prime(\widehat{\valpha}(s) - \valpha_0)\}}\big]\mX_t\mX_t^\prime\bigg\Vert=o_{\P}(1).
\end{equation*}
\end{lem}

\begin{proof}
See Section~\ref{sec:ESest Lemmas}.
\end{proof}

We are now in a position to prove Theorem~\ref{thm:ES est}.

\begin{proof}[{\textbf{Proof of Theorem~\ref{thm:ES est}:}}]
The first part of the proof is similar to that of Theorem~\ref{thm:std est}.
To highlight the similarities, we often overload notation by redefining quantities that already appeared in the proof of Theorem~\ref{thm:std est}, such as $W_n(\vw, s)$ and $\vw_n(s)$.
Note that the estimator $\widehat{\vbeta}(s)$ can equivalently be written as
\begin{align*}
\widehat{\vbeta}(s) &= \argmin_{\vbeta\in\mathbb{R}^{k}}\sum_{t=1}^{\lfloor ns\rfloor}\1_{\{Y_t>\mX_t^\prime\widehat{\valpha}(s)\}}\big[ (Y_t- \mX_t^\prime\vbeta)^2 - \xi_t^2\big]\\
&= \argmin_{\vbeta\in\mathbb{R}^{k}} \sum_{t=1}^{\lfloor ns\rfloor}\1_{\{Y_t>\mX_t^\prime\widehat{\valpha}(s)\}}\Big[ \big(Y_t- \mX_t^\prime\vbeta_0 -\mX_t^\prime(\vbeta-\vbeta_0)\big)^2 - \xi_t^2\Big]\\
&\overset{\eqref{eq:(1.1)}}{=} \argmin_{\vbeta\in\mathbb{R}^{k}} \sum_{t=1}^{\lfloor ns\rfloor}\1_{\{Y_t>\mX_t^\prime\widehat{\valpha}(s)\}}\Big[ \big(\xi_t -(\vbeta-\vbeta_0)^\prime\mX_t\big)^2 - \xi_t^2\Big].
\end{align*}
Therefore, letting
\[
 W_n(\vw,s)=\sum_{t=1}^{\lfloor ns\rfloor}\1_{\{Y_t>\mX_t^\prime\widehat{\valpha}(s)\}}\big[ (\xi_t -\vw^\prime n^{-1/2}\mX_t)^2 - \xi_t^2\big],
\] 
the minimizer $\vw_n(s)$ of $W_n(\cdot,s)$ satisfies that 
\[
	\vw_n(s)=\sqrt{n}\big[\widehat{\vbeta}(s) - \vbeta_0\big].
\]

We proceed by invoking Lemma~\ref{lem:Kato2009} to derive the weak limit of $\vw_n(\cdot)$.
To do so, we again first prove convergence of the finite-dimensional distributions of $W_n(\cdot,s)$ to verify \eqref{eq:fn finf}.
To that end, we simplify $W_n(\vw, s)$ and write
\[
	 W_n(\vw,s)=\sum_{t=1}^{\lfloor ns\rfloor}\1_{\{Y_t>\mX_t^\prime\widehat{\valpha}(s)\}}\big[ -2\xi_t\vw^\prime n^{-1/2}\mX_t + (\vw^\prime n^{-1/2}\mX_t)^2\big].
\]
The indicator can be written as 
\begin{align*}
	\1_{\{Y_t>\mX_t^\prime\widehat{\valpha}(s)\}} &= \1_{\{Y_t>\mX_t^\prime\valpha_0 + \mX_t^\prime(\widehat{\valpha}(s)-\valpha_0)\}} - \1_{\{\varepsilon_t>0\}} + \1_{\{\varepsilon_t>0\}}\\
	&= \1_{\{\varepsilon_t> \mX_t^\prime(\widehat{\valpha}(s)-\valpha_0)\}} - \1_{\{\varepsilon_t>0\}} + \1_{\{\varepsilon_t>0\}}\\
	&= \1_{\{\mX_t^\prime(\widehat{\valpha}(s) - \valpha_0)<\varepsilon_t\leq 0\}} - \1_{\{0<\varepsilon_t\leq \mX_t^\prime(\widehat{\valpha}(s) - \valpha_0)\}} + \1_{\{\varepsilon_t>0\}},
\end{align*}
whence
\begin{multline*}
	W_n(\vw,s)=\sum_{t=1}^{\lfloor ns\rfloor}\1_{\{\varepsilon_t>0\}}\big( -2\xi_t\vw^\prime n^{-1/2}\mX_t + \frac{1}{n}\vw^\prime\mX_t\mX_t^\prime\vw\big)\\
	 + \sum_{t=1}^{\lfloor ns\rfloor}\big[\1_{\{\mX_t^\prime(\widehat{\valpha}(s) - \valpha_0)<\varepsilon_t\leq 0\}} - \1_{\{0<\varepsilon_t\leq \mX_t^\prime(\widehat{\valpha}(s) - \valpha_0)\}}\big] \big( -2\xi_t\vw^\prime n^{-1/2}\mX_t + \frac{1}{n}\vw^\prime\mX_t\mX_t^\prime\vw\big).
\end{multline*}
We obtain from Lemmas~\ref{lem:ES 1}--\ref{lem:ES 4} that for each fixed $\vw\in\mathbb{R}^{k}$, as $n\to\infty$,
\begin{align*}
	W_n(\vw,s)&\overset{d}{\longrightarrow}-2\vw^\prime\mJ_{\ast}^{1/2}\mW_{\ast}(s) + s(1-\tau)\vw^\prime\mD_{\ast}\vw - 2\vw^\prime\mK\mOmega^{1/2}\mW(s)\\
	& = -2\vw^\prime \va(s)+\vw^\prime \mB(s)\vw\\
	& =: W_{\infty}(\vw,s)\qquad\text{in}\ D[\epsilon,1]
\end{align*}
for $\va(s) = \mJ_{\ast}^{1/2}\mW_{\ast}(s) + \mK\mOmega^{1/2}\mW(s)$ and $\mB(s) = s(1-\tau)\mD_{\ast}$.
Similarly as in the proof of Theorem~\ref{thm:std est}, we deduce from this that
\[
	\big(W_n(\vw_{1},s),\ldots,W_n(\vw_{\ell},s)\big)^\prime\overset{d}{\longrightarrow}\big(W_{\infty}(\vw_{1},s), \ldots, W_{\infty}(\vw_{\ell},s)\big)^\prime\qquad\text{in}\ D_{\ell}[\epsilon,1]
\]
for any $\vw_{1},\ldots,\vw_{\ell}\in\mathbb{R}^{k}$, such that \eqref{eq:fn finf} of Lemma~\ref{lem:Kato2009} is met.

Once more, the only remaining condition of \textit{(i)}--\textit{(iii)} in Lemma~\ref{lem:Kato2009} that is not immediately obvious, is to show that $W_{\infty}(\cdot,s)$ has a unique minimum.
Because $\mB(s)$ is positive definite by Assumption~\ref{ass:Heterogeneity ES}*~\eqref{it:(iii) ES} for $s\in[\epsilon,1]$, it is easy to deduce that $W_{\infty}(\cdot,s)$ has a unique minimum for $\vw_{\infty}(s)$ satisfying the first-order condition
\[
	-\va(s) + \mB(s)\vw_{\infty}(s) = \vzero,
\]
such that
\begin{equation}\label{eq:vw char}
	\vw_{\infty}(s)= \mB^{-1}(s)\va(s)= \frac{1}{s}\frac{1}{1-\tau}\mD_{\ast}^{-1}\big[\mJ_{\ast}^{1/2}\mW_{\ast}(s) + \mK\mOmega^{1/2}\mW(s)\big].
\end{equation}
Therefore, as $n\to\infty$,
\begin{equation}\label{eq:(p.28)}
\vw_n(s)\overset{d}{\longrightarrow}\vw_{\infty}(s)\qquad\text{in }D_k[\epsilon,1]
\end{equation}
by Lemma~\ref{lem:Kato2009}.

In the second step of this proof, it remains to simplify the limiting distribution $\vw_{\infty}(s)$.
To that end, we show that
\[
	\widetilde{\mW}(s) := \mJ_{\ast}^{1/2}\mW_{\ast}(s) + \mK\mOmega^{1/2}\mW(s)
\]
is a Brownian motion.
To do so, it is well-known that it suffices to compute the covariance function $\Cov\big(\widetilde{\mW}(s), \widetilde{\mW}(t)\big)$.
Proceeding similarly as in the proofs of Lemmas~\ref{lem:c2 FS} and \ref{lem:ES 1}, we obtain using Assumption~\ref{ass:Heterogeneity ES}*~\eqref{it:(ii) ES} that
\begin{equation}\label{eq:joint con}
	n^{-1/2}\sum_{t=1}^{\lfloor ns\rfloor} \begin{pmatrix}\psi_{\ast}(\varepsilon_t,\xi_t)\mX_t\\  \psi(\varepsilon_t)\mX_t\end{pmatrix}\overset{d}{\longrightarrow} \overline{\mOmega}^{1/2}\overline{\mW}(s)\qquad\text{in }D_{2k}[0,1],
\end{equation}
where $\overline{\mW}(\cdot)=\big(\overline{\mW}_{1}^\prime(\cdot), \overline{\mW}_{2}^\prime(\cdot)\big)^\prime$ is a standard $(2k)$-dimensional Brownian motion and
\begin{align*}
	\overline{\mOmega}&=\lim_{n\to\infty}\frac{1}{n}\E\Bigg[\sum_{t=1}^{n}\begin{pmatrix}\psi_{\ast}(\varepsilon_t,\xi_t)\mX_t\\  \psi(\varepsilon_t)\mX_t\end{pmatrix}\sum_{t=1}^{n}\big(\psi_{\ast}(\varepsilon_t,\xi_t)\mX_t^\prime, \psi(\varepsilon_t)\mX_t^\prime\big)\Bigg]\\
	&= \lim_{n\to\infty}\frac{1}{n}\sum_{t=1}^{n}\sum_{s=1}^{n}\E\Bigg[\begin{pmatrix}\psi_{\ast}(\varepsilon_t,\xi_t)\psi_{\ast}(\varepsilon_s,\xi_s)\mX_t\mX_s^\prime &  \psi_{\ast}(\varepsilon_t,\xi_t)\psi(\varepsilon_s)\mX_t\mX_s^\prime\\
	\psi(\varepsilon_t)\psi_{\ast}(\varepsilon_s,\xi_s)\mX_t\mX_s^\prime & \psi(\varepsilon_t)\psi(\varepsilon_s)\mX_t\mX_s^\prime\end{pmatrix}\\
	&=\begin{pmatrix}\mJ_{\ast} & \mC^\prime\\
	\mC & \mJ\end{pmatrix}
\end{align*}
is positive definite.
Since $\overline{\mOmega}$ is positive definite by Assumption~\ref{ass:Heterogeneity ES}*~\eqref{it:(ii) ES}, there exists a Choleski decomposition of the form $\overline{\mOmega}=\overline{\mOmega}^{1/2}(\overline{\mOmega}^{1/2})^\prime$ for a lower triangular $\overline{\mOmega}^{1/2}$ \citep[Sec.~A.9.3]{Lüt07}.
This lower triangular matrix is then necessarily of the form
\[
	\overline{\mOmega}^{1/2}=\begin{pmatrix}\mJ_{\ast}^{1/2} & \vzero\\
																				  \mC_{21} & \mJ_{22}\end{pmatrix},
\]
where $\mC_{21}$ and $\mJ_{22}$ are some matrices, and $\mJ_{\ast}^{1/2}$ is the lower triangular matrix from the Choleski decomposition $\mJ_{\ast}=\mJ_{\ast}^{1/2}(\mJ_{\ast}^{1/2})^\prime$ (which exists by positive definiteness of $\mJ_{\ast}$ in Assumption~\ref{ass:Heterogeneity ES}*~\eqref{it:(ii) ES}).
The matrices $\mC_{21}$ and $\mJ_{22}$ therefore satisfy that
\begin{align}
	\overline{\mOmega} &= \overline{\mOmega}^{1/2}(\overline{\mOmega}^{1/2})^\prime\notag\\
	&= \begin{pmatrix} 
	\mJ_{\ast}^{1/2} & \vzero\\
	\mC_{21} & \mJ_{22}
	\end{pmatrix}
	\begin{pmatrix} 
	(\mJ_{\ast}^{1/2})^\prime & \mC_{21}^\prime\\
	\vzero & \mJ_{22}^\prime
	\end{pmatrix}\notag\\
	&= \begin{pmatrix} 
	\mJ_{\ast} & \mJ_{\ast}^{1/2}\mC_{21}^\prime\\
	\mC_{21}(\mJ_{\ast}^{1/2})^\prime & \mC_{21}\mC_{21}^\prime + \mJ_{22}\mJ_{22}^\prime
	\end{pmatrix}= \begin{pmatrix}\mJ_{\ast} & \mC^\prime\\
	\mC & \mJ\end{pmatrix}.\label{eq:CD Omega}
\end{align}

From \eqref{eq:joint con} and the marginal convergences in Lemmas~\ref{lem:c2 FS} and \ref{lem:ES 1}, we conclude that
\begin{equation*}
\begin{pmatrix}\mJ_{\ast}^{1/2}\mW_{\ast}(s) \\
	\mJ^{1/2}\mW(s)\end{pmatrix}
	\overset{d}{=}\overline{\mOmega}^{1/2}\overline{\mW}(s)=\begin{pmatrix} 
	\mJ_{\ast}^{1/2} & \vzero\\
	\mC_{21} & \mJ_{22}
	\end{pmatrix}
	\begin{pmatrix}
	\overline{\mW}_1(s)\\
	\overline{\mW}_2(s)
	\end{pmatrix}.
\end{equation*}
Therefore, also using that $\mOmega^{1/2}=(\mD^{-1}\mJ\mD^{-1})^{1/2}=\mD^{-1}\mJ^{1/2}$,
\begin{align}
	\widetilde{\mW}(s) &= \mJ_{\ast}^{1/2}\mW_{\ast}(s) + \mK\mOmega^{1/2}\mJ^{-1/2}\mJ^{1/2}\mW(s)\notag\\
	&= \mJ_{\ast}^{1/2}\mW_{\ast}(s) + \mK\mD^{-1}\mJ^{1/2}\mW(s)\notag\\
	&= \big(\begin{array}{c|c}
		\mI & \mK\mD^{-1}
	\end{array}\big)
	\begin{pmatrix}\mJ_{\ast}^{1/2}\mW_{\ast}(s) \\
	\mJ^{1/2}\mW(s)\end{pmatrix}\notag\\
	&\overset{d}{=}\big(\begin{array}{c|c}
		\mI & \mK\mD^{-1}
	\end{array}\big)
	\begin{pmatrix} 
	\mJ_{\ast}^{1/2} & \vzero\\
	\mC_{21} & \mJ_{22}
	\end{pmatrix}
	\begin{pmatrix}
	\overline{\mW}_1(s)\\
	\overline{\mW}_2(s)
	\end{pmatrix}\notag\\
	&= \big(\begin{array}{c|c}
		\mJ_{\ast}^{1/2} + \mK\mD^{-1}\mC_{21} & \mK\mD^{-1}\mJ_{22}
	\end{array}\big)
	\begin{pmatrix}
	\overline{\mW}_1(s)\\
	\overline{\mW}_2(s)
	\end{pmatrix}\notag\\
	&= \big(\begin{array}{c|c}
		\mJ_{\ast}^{1/2} + \mK\mD^{-1}\mC(\mJ_{\ast}^{-1/2})^\prime & \mK\mD^{-1}\mJ_{22}\end{array}\big)
	\begin{pmatrix}
	\overline{\mW}_1(s)\\
	\overline{\mW}_2(s)
	\end{pmatrix},\label{eq:(Wtilde)}
\end{align}
where the last step exploits that $\mC_{21}=\mC(\mJ_{\ast}^{-1/2})^\prime$ from \eqref{eq:CD Omega}.
With this and (without loss of generality) $t<s$,
\begin{align*}
	\Cov\big(\widetilde{\mW}(s), \widetilde{\mW}(t)\big) &= \Cov\big(\widetilde{\mW}(s) - \widetilde{\mW}(t), \widetilde{\mW}(t)\big) + \Cov\big(\widetilde{\mW}(t), \widetilde{\mW}(t)\big)\\
	&= \Cov\big(\widetilde{\mW}(t), \widetilde{\mW}(t)\big)\\
	&= t\big(\begin{array}{c|c}
		\mJ_{\ast}^{1/2} + \mK\mD^{-1}\mC(\mJ_{\ast}^{-1/2})^\prime & \mK\mD^{-1}\mJ_{22}\end{array}\big)\times\\
		&\hspace{4cm}\times
		\begin{pmatrix}
		(\mJ_{\ast}^{1/2})^\prime + \big[\mK\mD^{-1}\mC(\mJ_{\ast}^{-1/2})^\prime\big]^\prime \\
		\big[\mK\mD^{-1}\mJ_{22}\big]^\prime\end{pmatrix}\\
		&=t\Big\{\mJ_{\ast} + \mK\mD^{-1}\mC + (\mK\mD^{-1}\mC)^\prime\\
		&\hspace{4cm} + \mK\mD^{-1}\mC\mJ_{\ast}^{-1}\mC^\prime(\mK\mD^{-1})^\prime\\
		&\hspace{4cm} + \mK\mD^{-1}\mJ_{22}\mJ_{22}^\prime(\mK\mD^{-1})^\prime\Big\}\\
		&=t\Big\{\mJ_{\ast} + \mK\mD^{-1}\mC + (\mK\mD^{-1}\mC)^\prime\\
		&\hspace{4cm} + \mK\mD^{-1}\mC\mJ_{\ast}^{-1}\mC^\prime(\mK\mD^{-1})^\prime\\
		&\hspace{4cm} + \mK\mD^{-1}(\mJ-\mC\mJ_{\ast}^{-1}\mC^\prime)(\mK\mD^{-1})^\prime\Big\}\\
		&=t\Big\{\mJ_{\ast} + \mK\mD^{-1}\mC + (\mK\mD^{-1}\mC)^\prime + \mK\mD^{-1}\mJ\mD^{-1}\mK\Big\}\\
		&=t\mSigma_{\ast},
\end{align*}
where the second step follows from \eqref{eq:(Wtilde)} and the independence of the increments of $\big(\overline{\mW}_1(s), \overline{\mW}_2(s)\big)^\prime$, and the fifth step uses that $\mJ_{22}\mJ_{22}^\prime=\mJ - \mC_{21}\mC_{21}^\prime = \mJ - \mC(\mJ_{\ast}^{-1/2})^\prime\mJ_{\ast}^{-1/2}\mC^\prime=\mJ - \mC\mJ_{\ast}^{-1}\mC^\prime$ from \eqref{eq:CD Omega}.
We obtain that $\Cov\big(\widetilde{\mW}(s), \widetilde{\mW}(t)\big)=\min\{s,t\}\mSigma_{\ast}$, which characterizes Brownian motion \citep[e.g.,][Theorem~4.4~II]{JS87}, such that $\widetilde{\mW}(s)\overset{d}{=}\mSigma_{\ast}^{1/2}\mW_k(s)$ for a $k$-dimensional standard Brownian motion $\mW_k(\cdot)$.
We conclude from \eqref{eq:vw char} that
\[
	\vw_{\infty}(s) = \frac{1}{s}\frac{1}{1-\tau}\mD_{\ast}^{-1}\mSigma_{\ast}^{1/2}\mW_k(s).
\]
From this and \eqref{eq:(p.28)}, the desired convergence claimed in Theorem~\ref{thm:ES est} easily follows.
Finally, from \citet[Sec.~A.8.3]{Lüt07} we obtain that
\[
	\mSigma_{\ast}=\big(\begin{array}{c|c}\mI & \mK\mD^{-1}\end{array}\big)\overline{\mOmega}\begin{pmatrix} \mI\\ \mD^{-1}\mK\end{pmatrix}
\]
is positive definite, since $\big(\begin{array}{c|c}\mI & \mK\mD^{-1}\end{array}\big)$ obviously has full rank and $\overline{\mOmega}$ is positive definite.
\end{proof}

%
%
%
%

\section{Proofs of Lemmas~\ref{lem:c2 FS}--\ref{lem:c3 FS}}\label{sec:QRest Lemmas}

\begin{proof}[{\textbf{Proof of Lemma~\ref{lem:c2 FS}:}}]
	We derive the distributional limit of $\frac{1}{\sqrt{n}}\sum_{t=1}^{\lfloor ns\rfloor}\psi(\varepsilon_t)\mX_{t}$ by invoking the $\alpha$-mixing functional central limit theorem (FCLT) of \citet[Corollary 29.19]{Dav94}.
	To do so, we verify the assumptions of this FCLT.
	Note that $\psi(\varepsilon_t)\mX_{t}$ has mean zero because, by the law of iterated expectations and recalling that $\E_t[\cdot]=\E[\,\cdot\mid\mX_t]$,
	\[
	\E[\psi(\varepsilon_t)\mX_{t}] = \E\big[\E_{t}\{\psi(\varepsilon_t)\}\mX_t\big]=\vzero.
	\]
	From Assumption~\ref{ass:Heterogeneity}~\eqref{it:(i)} it follows that
	\[
	\E\big[\psi^r(\varepsilon_t)\Vert\mX_{t}\Vert^r\big]\leq \E\big[\Vert\mX_{t}\Vert^r\big]\leq K<\infty.
	\]
	Moreover, as a function of $(\varepsilon_t, \mX^\prime_t)^\prime$, the $(k\times 1)$-vector sequence $\{\psi(\varepsilon_t)\mX_{t}\}$ is $\alpha$-mixing of size $-r/(r-2)$ by Theorem~14.1 in \citet{Dav94} and Assumption~\ref{ass:Serial Dependence}. 
	Since the long-run variance of $\{\psi(\varepsilon_t)\mX_t\}$ also converges appropriately (see Assumption~\ref{ass:Heterogeneity}~\eqref{it:(ii)}), applying Corollary 29.19 of \citet{Dav94} gives the desired conclusion.
\end{proof}

\begin{proof}[{\textbf{Proof of Lemma~\ref{lem:c3 FS}:}}]
	Define
	\begin{align*}
		\nu_{t}(\vw) &:= (\varepsilon_t - \vw^\prime n^{-1/2}\mX_{t})\big(\1_{\{\vw^\prime n^{-1/2}\mX_{t}< \varepsilon_t<0\}} - \1_{\{0<\varepsilon_t< \vw^\prime n^{-1/2}\mX_{t}\}}\big),\\
		\overline{\nu}_{t}(\vw) &:= \E_{t}\Big[(\varepsilon_t - \vw^\prime n^{-1/2}\mX_{t})\big(\1_{\{\vw^\prime n^{-1/2}\mX_{t}< \varepsilon_t<0\}} - \1_{\{0<\varepsilon_t< \vw^\prime n^{-1/2}\mX_{t}\}}\big)\Big],\\
		V_n(\vw,s) &:= \sum_{t=1}^{\lfloor ns\rfloor}\nu_t(\vw),\\
		\overline{V}_n(\vw,s) &:= \sum_{t=1}^{\lfloor ns\rfloor}\overline{\nu}_t(\vw).
	\end{align*}
	We establish the result by showing that
	\begin{align}
		\sup_{s\in[0, 1]}\big| \overline{V}_n(\vw,s) - \frac{1}{2}s\vw^\prime\mD\vw \big| &= o_{\P}(1), \label{eq:(P.7.0)}\\
		\sup_{s\in[0,1]}\big|V_n(\vw,s) - \overline{V}_n(\vw,s)\big| &= o_{\P}(1).  \label{eq:(P.7.1)}
	\end{align}
	
	We first prove \eqref{eq:(P.7.0)}. 
	Subadditivity, Markov's inequality and the assumption that $\E\big[\Vert\mX_{t}\Vert^{2r}\big]\leq \Delta$ for some $r>2$ (see Assumption~\ref{ass:Heterogeneity}~\eqref{it:(i)}) imply
	\begin{align}
		\P\Big\{\max_{t=1,\ldots,n}n^{-1/2}\Vert\mX_{t}\Vert>\varepsilon\Big\} &\leq \sum_{t=1}^{n}\P\big\{\Vert\mX_{t}\Vert>\varepsilon n^{1/2}\big\}\notag\\
		&\leq \sum_{t=1}^{n}\frac{\E\big[\Vert\mX_{t}\Vert^{2r}\big]}{\varepsilon^{2r}n^{r}}\notag\\
		&=o(1),\notag
	\end{align}
	such that
	\begin{equation}\label{eq:unif Z}
		\max_{t=1,\ldots,n}n^{-1/2}\Vert \mX_t\Vert=o_{\P}(1).
	\end{equation}
	Therefore, the event $\big\{\max_{t=1,\ldots,n}|\vw^\prime n^{-1/2}\mX_{t}|\leq d\big\}$ occurs with probability approaching $1$ (w.p.a.~1), as $n\to\infty$.
	On this set, we may exploit Assumption~\ref{ass:innov} to decompose $\overline{V}_n(\vw,s)$ as follows:
	\begin{align}
		\overline{V}_n(\vw,s) &= \sum_{t=1}^{\lfloor ns\rfloor}\E_{t}\Big[(\varepsilon_t - \vw^\prime n^{-1/2}\mX_{t})\big(\1_{\{\vw^\prime n^{-1/2}\mX_{t}< \varepsilon_t<0\}} - \1_{\{0<\varepsilon_t< \vw^\prime n^{-1/2}\mX_{t}\}}\big)\Big] \notag\\
		&=\sum_{t=1}^{\lfloor ns\rfloor}\1_{\{\vw^\prime n^{-1/2}\mX_{t}<0\}}\int_{\vw^\prime n^{-1/2}\mX_{t}}^{0}(x-\vw^\prime n^{-1/2}\mX_{t})f_{\varepsilon_t\mid\mX_t}(x)\D x\notag\\
		&\hspace{2cm} - \1_{\{\vw^\prime n^{-1/2}\mX_{t}>0\}}\int_{0}^{\vw^\prime n^{-1/2}\mX_{t}}(x-\vw^\prime n^{-1/2}\mX_{t})f_{\varepsilon_t\mid\mX_t}(x)\D x\notag\\
		&=\sum_{t=1}^{\lfloor ns\rfloor}\1_{\{\vw^\prime n^{-1/2}\mX_{t}<0\}}\int_{\vw^\prime n^{-1/2}\mX_{t}}^{0}\big[F_{\varepsilon_t\mid\mX_t}(0) - F_{\varepsilon_t\mid\mX_t}(x)\big]\D x\notag\\
		&\hspace{2cm} - \1_{\{\vw^\prime n^{-1/2}\mX_{t}>0\}}\int_{0}^{\vw^\prime n^{-1/2}\mX_{t}}\big[F_{\varepsilon_t\mid\mX_t}(0) - F_{\varepsilon_t\mid\mX_t}(x)\big]\D x,\label{eq:(P.1)}
	\end{align}
	where the final line follows from integration by parts and $F_{\varepsilon_t\mid\mX_t}(\cdot)$ denotes the cumulative distribution function of $\varepsilon_t\mid\mX_t$. 
	By the mean value theorem it holds for some $x^\ast=x^\ast(x)$ between $0$ and $x$ that
	\begin{align}
		\frac{F_{\varepsilon_t\mid\mX_t}(x)-F_{\varepsilon_t\mid\mX_t}(0)}{x-0} &= f_{\varepsilon_t\mid\mX_t}(x^\ast)\notag\\
		&= f_{\varepsilon_t\mid\mX_t}(0) + \big[f_{\varepsilon_t\mid\mX_t}(x^\ast) - f_{\varepsilon_t\mid\mX_t}(0)\big].\label{eq:(H.MVT)}
	\end{align}
	Plugging this into \eqref{eq:(P.1)} yields that
	\begin{align*}
		\overline{V}_n(\vw,s) &= \sum_{t=1}^{\lfloor ns\rfloor}\bigg\{-\1_{\{\vw^\prime n^{-1/2}\mX_{t}<0\}}\int_{\vw^\prime n^{-1/2}\mX_{t}}^{0}xf_{\varepsilon_t\mid\mX_t}(0)\D x\\
		&\hspace{2cm} + \1_{\{\vw^\prime n^{-1/2}\mX_{t}>0\}}\int_{0}^{\vw^\prime n^{-1/2}\mX_{t}}xf_{\varepsilon_t\mid\mX_t}(0)\D x\bigg\}\\
		&\hspace{1cm} + \sum_{t=1}^{\lfloor ns\rfloor}\bigg\{-\1_{\{\vw^\prime n^{-1/2}\mX_{t}<0\}}\int_{\vw^\prime n^{-1/2}\mX_{t}}^{0}x\big[f_{\varepsilon_t\mid\mX_t}(x^\ast) - f_{\varepsilon_t\mid\mX_t}(0)\big]\D x\\
		&\hspace{2cm} + \1_{\{\vw^\prime n^{-1/2}\mX_{t}>0\}}\int_{0}^{\vw^\prime n^{-1/2}\mX_{t}}x\big[f_{\varepsilon_t\mid\mX_t}(x^\ast) - f_{\varepsilon_t\mid\mX_t}(0)\big]\D x\bigg\}\\
		&=:\overline{V}_{1n}(\vw,s) + \overline{V}_{2n}(\vw,s).
	\end{align*}
	We establish the convergence in \eqref{eq:(P.7.0)} by showing that
	\begin{align}
		\sup_{s\in[0,1]}\big| \overline{V}_{1n}(\vw,s) - \frac{1}{2}s\vw^\prime \mD\vw \big| & = o_{\P}(1),\label{eq:(P.8.0)}\\
		\sup_{s\in[0,1]}\big|\overline{V}_{2n}(\vw,s)\big| &=o_{\P}(1). \label{eq:(P.8.1)}
	\end{align}

	For \eqref{eq:(P.8.1)}, on the set $\{\max_{t=1,\ldots,n}n^{-1/2}\Vert\mX_t\Vert\leq d\}$ (which, by \eqref{eq:unif Z}, occurs w.p.a.~$1$, as $n\to\infty$) we may deduce from Assumption~\ref{ass:innov}~\eqref{it:Lipschitz} that
	\begin{align}
		\big|\overline{V}_{2n}(\vw,s)\big| &\leq 2\sum_{t=1}^{\lfloor ns\rfloor}\int_{0}^{|\vw^\prime n^{-1/2}\mX_{t}|}x L |\vw^\prime n^{-1/2}\mX_{t}|\D x\notag\\
		&\leq 2L\Vert\vw\Vert \sum_{t=1}^{n} \Vert n^{-1/2}\mX_{t}\Vert \frac{1}{2}(\vw^\prime n^{-1/2}\mX_{t})^2\notag\\
		&\leq L\Vert\vw\Vert^3 \max_{t=1,\ldots,n}\big\{n^{-1/2}\Vert\mX_{t}\Vert\big\} \frac{1}{n}\sum_{t=1}^{n}\Vert\mX_{t}\Vert^2.\label{eq:ov W2}
	\end{align}
	Moreover, from Markov's inequality and $\E\big[\Vert\mX_{t}\Vert^2\big]\leq K$, it follows that, as $M\to\infty$,
	\[
	\P\bigg\{\frac{1}{n}\sum_{t=1}^{n} \Vert\mX_{t}\Vert^2>M\bigg\}\leq\frac{1}{M}\frac{1}{n}\sum_{t=1}^{n}\E\big[\Vert\mX_t\Vert^2\big]\leq\frac{K}{M} \longrightarrow0,
	\]
	such that
	\begin{equation}\label{eq:ov W2 h}
		\frac{1}{n}\sum_{t=1}^{n} \Vert\mX_{t}\Vert^2 = O_{\P}(1).
	\end{equation}
	Combining \eqref{eq:unif Z} with \eqref{eq:ov W2} and \eqref{eq:ov W2 h}, we obtain that $\sup_{s\in[0,1]}\big|\overline{V}_{2n}(\vw,s)\big|=o_{\P}(1)$.

	To prove \eqref{eq:(P.8.0)}, write
	\begin{align*}
		\overline{V}_{1n}(\vw,s) &=\sum_{t=1}^{\lfloor ns\rfloor}\bigg\{\1_{\{\vw^\prime n^{-1/2}\mX_{t}<0\}}f_{\varepsilon_t\mid\mX_t}(0)\frac{1}{2}(\vw^\prime n^{-1/2}\mX_{t})^2\\
		&\hspace{2cm} + \1_{\{\vw^\prime n^{-1/2}\mX_{t}>0\}}f_{\varepsilon_t\mid\mX_t}(0)\frac{1}{2}(\vw^\prime n^{-1/2}\mX_{t})^2\bigg\}\\
		&=\frac{1}{2}\vw^\prime\bigg(\frac{1}{n}\sum_{t=1}^{\lfloor ns\rfloor}f_{\varepsilon_t\mid\mX_t}(0)\mX_{t}\mX^\prime_{t}\bigg)\vw.
	\end{align*}
	Since $f_{\varepsilon_t\mid\mX_t}(0)$ is a $\mX_t$-measurable random variable, $\{\vw^\prime f_{\varepsilon_t\mid\mX_t}(0)\mX_t\mX_t^\prime\vw\}$ is mixing of the same size as $\{\mX_t\}$ \citep[Theorem~14.1]{Dav94}.
	This implies that $\big\{\vw^\prime f_{\varepsilon_t\mid\mX_t}(0)\mX_t\mX_t^\prime\vw - \vw^\prime\E[ f_{\varepsilon_t\mid\mX_t}(0)\mX_t\mX_t^\prime]\vw\big\}$ is a (zero mean) $L_1$-mixingale by arguments in \citet[Section~3]{And88}.
	The corollary of \citet[p.~493]{Han92} therefore gives that
	\[
	\sup_{s\in[0,1]}\bigg|\overline{V}_{1n}(\vw,s) - \frac{1}{2}\vw^\prime \Big\{\frac{1}{n}\sum_{t=1}^{\lfloor ns\rfloor}\E\big[f_{\varepsilon_t\mid\mX_t}(0)\mX_{t}\mX^\prime_{t}\big]\Big\}\vw\bigg|\overset{\P}{\longrightarrow}0.
	\]
	With this,
	\begin{align*}
		\sup_{s\in[0,1]}&\big| \overline{V}_{1n}(\vw,s) - \frac{1}{2}s\vw^\prime \mD\vw \big|\\
		&\leq \sup_{s\in[0,1]}\bigg| \overline{V}_{1n}(\vw,s) - \frac{1}{2}\vw^\prime \Big\{\frac{1}{n}\sum_{t=1}^{\lfloor ns\rfloor}\E\big[f_{\varepsilon_t\mid\mX_t}(0)\mX_{t}\mX^\prime_{t}\big]\Big\}\vw\bigg|\\
		&\hspace{1cm} + \sup_{s\in[0,1]}\bigg| \frac{1}{2}\vw^\prime \Big\{\frac{1}{n}\sum_{t=1}^{\lfloor ns\rfloor}\E\big[f_{\varepsilon_t\mid\mX_t}(0)\mX_{t}\mX^\prime_{t}\big]\Big\}\vw - \frac{1}{2}s\vw^\prime \mD\vw \bigg| \\
		&=o_{\P}(1) + \sup_{s\in[0,1]}\bigg| \frac{1}{2}\vw^\prime \frac{\lfloor ns\rfloor}{n}\Big\{\frac{1}{\lfloor ns\rfloor}\sum_{t=1}^{\lfloor ns\rfloor}\E\big[f_{\varepsilon_t\mid\mX_t}(0)\mX_{t}\mX^\prime_{t}\big]\Big\}\vw - \frac{1}{2}s\vw^\prime \mD\vw \bigg|\\
		&=o_{\P}(1) + o(1),
	\end{align*}
	where we used Assumption~\ref{ass:Heterogeneity}~\eqref{it:(iii)} in the last step.
	This proves \eqref{eq:(P.8.0)}, whence \eqref{eq:(P.7.0)} follows.
	
	It remains to show \eqref{eq:(P.7.1)}.
	To this end, write
	\begin{align*}
		V_n(\vw,s) - \overline{V}_n(\vw,s) & = \sum_{t=1}^{\lfloor ns\rfloor}\big[\nu_t(\vw) - \overline{\nu}_t(\vw)\big]\\
		&= \sum_{t=1}^{\lfloor ns\rfloor}\big[\nu_{1t}(\vw) - \overline{\nu}_{1t}(\vw)\big] + \sum_{t=1}^{\lfloor ns\rfloor}\big[\nu_{2t}(\vw) - \overline{\nu}_{2t}(\vw)\big],
	\end{align*}
	where 
	\begin{align*}
		\nu_{1t}(\vw) &= (\varepsilon_t - \vw^\prime n^{-1/2}\mX_{t})\1_{\{\vw^\prime n^{-1/2}\mX_{t}< \varepsilon_t<0\}},&& \overline{\nu}_{1t}(\vw)= \E_{t}\big[\nu_{1t}(\vw)\big],\\
		\nu_{2t}(\vw) &= (\vw^\prime n^{-1/2}\mX_{t} - \varepsilon_t)\1_{\{0<\varepsilon_t< \vw^\prime n^{-1/2}\mX_{t}\}},&& \overline{\nu}_{2t}(\vw)= \E_{t}\big[\nu_{2t}(\vw)\big].
	\end{align*}
	Therefore, it suffices to prove that $\sup_{s\in[0,1]}\big|\sum_{t=1}^{\lfloor ns\rfloor}\big[\nu_{it}(\vw) - \overline{\nu}_{it}(\vw)\big]\big|=o_{\P}(1)$ for $i=1,2$.
	We do so only for $i=2$, as the proof for $i=1$ is very similar.
	
	To task this, we use the Ottaviani-type inequality for $\alpha$-mixing random variables due to \citet[Lemma~3]{Büc15}. 
	This inequality implies that for any $\varepsilon>0$ and $1\leq \ell_n < n$,
	\begin{align}
		\P&\bigg\{ \sup_{s\in[0,1]}\bigg|\sum_{t=1}^{\lfloor ns\rfloor}\big[\nu_{2t}(\vw) - \overline{\nu}_{2t}(\vw)\big]\bigg|> 3\varepsilon \bigg\} \notag\\
		& = \P\bigg\{ \max_{k=1,\ldots,n} \bigg|\sum_{t=1}^{k}\big[\nu_{2t}(\vw) - \overline{\nu}_{2t}(\vw)\big]\bigg|> 3\varepsilon \bigg\}\notag\\
		& \leq \frac{A_{1n} + B_{1n} + \lfloor n/\ell_n\rfloor \alpha(\ell_n)}{1-\max_{k=1,\ldots,n}\P\bigg\{\Big|\sum_{t=k+1}^{n}\big[\nu_{2t}(\vw) - \overline{\nu}_{2t}(\vw)\big]\Big|> \varepsilon\bigg\}},\label{eq:Ottaviani}
	\end{align}
	where 
	\begin{align*}
		A_{1n} &= \P\bigg\{\Big|\sum_{t=1}^{n}\big[\nu_{2t}(\vw) - \overline{\nu}_{2t}(\vw)\big]\Big|> \varepsilon\bigg\},\\
		B_{1n} &= \P\bigg\{\max_{\substack{j<k\in\{1,\ldots,n\}\\ k-j\leq 2 \ell_n}}\Big|\sum_{t=j+1}^{k}\big[\nu_{2t}(\vw) - \overline{\nu}_{2t}(\vw)\big]\Big|> \varepsilon\bigg\}.
	\end{align*}
	Next, we show that the terms in the numerator of \eqref{eq:Ottaviani} converge to zero, and the term in the denominator is bounded away from zero.
	To do so, we choose the integer sequence $\ell_n$ as follows. 
	Since the $\alpha$-mixing coefficients $\alpha(\cdot)$ of $\{(\varepsilon_t,\mX_t^\prime)^\prime\}$ are of size $-r/(r-2)$, this means that $\alpha(n)=O(n^{-\varphi})$ for some $\varphi>r/(r-2)$.
	In particular, there exists some $\overline{\delta}>0$, such that $\alpha(n)=O(n^{-\overline{\varphi}})$ for $\overline{\varphi}=\frac{r+2\overline{\delta}}{r-2}$.
	Now, put $\xi:=\frac{1}{2}\frac{r-2}{r-1+\delta}$ for some $\delta\in(0,\overline{\delta})$ and set $\ell_n:=\lfloor n^\xi\rfloor$.

	First, we deal with the terms in the numerator of \eqref{eq:Ottaviani}.
	By Chebyshev's inequality \citep[e.g.,][Corollary 3.1.3]{AL06},
	\begin{align}
		A_{1n} &\leq \varepsilon^{-2}\Var\bigg(\sum_{t=1}^{n}\big[\nu_{2t}(\vw) - \overline{\nu}_{2t}(\vw)\big]\bigg)\notag\\
		&= \varepsilon^{-2}\bigg\{ \sum_{t=1}^{n}\Var\big(\nu_{2t}(\vw) - \overline{\nu}_{2t}(\vw)\big)\notag\\
		&\hspace{2cm} + 2\sum_{t=1}^{n-1}\sum_{h=1}^{n-t}\Cov\big(\nu_{2t}(\vw) - \overline{\nu}_{2t}(\vw), \nu_{2,t+h}(\vw) - \overline{\nu}_{2,t+h}(\vw)\big) \bigg\}.\label{eq:(1.eck)}
	\end{align}
	To bound the variance and covariance terms in \eqref{eq:(1.eck)}, we use the $c_r$-inequality \citep[e.g.,][Theorem~9.28]{Dav94} to obtain that
	\begin{equation}\label{eq:(3.1)}
		\E\big[|\nu_{2t}(\vw) - \overline{\nu}_{2t}(\vw)|^r\big] \leq 2^{r-1}\Big\{\E\big[\nu_{2t}(\vw)^r\big] + \E\big[\overline{\nu}_{2t}(\vw)^r\big]\Big\}.
	\end{equation}
	Use the law of iterated expectations \citep[LIE; e.g.,][Theorem~10.8]{Dav94} and Assumption~\ref{ass:innov}~\eqref{it:dens bound} to arrive at
	\begin{align}
		\E\big[\nu_{2t}(\vw)^r\big] &= \E\big[(\vw^\prime n^{-1/2}\mX_t-\varepsilon_t)^r\1_{\{0<\varepsilon_t< \vw^\prime n^{-1/2}\mX_t\}}\big]\notag\\
		&= \E\Big[\E_{t}\big\{(\vw^\prime n^{-1/2}\mX_t-\varepsilon_t)^r\1_{\{0<\varepsilon_t< \vw^\prime n^{-1/2}\mX_t\}}\big\}\Big]\notag\\
		&= \E\Big[\1_{\{0< \vw^\prime n^{-1/2}\mX_t\}} \int_0^{\vw^\prime n^{-1/2}\mX_t} (\vw^\prime n^{-1/2}\mX_t - x)^rf_{\varepsilon_t\mid\mX_t}(x)\D x\Big]\notag\\
		&\leq \E\Big[\1_{\{0< \vw^\prime n^{-1/2}\mX_t\}}\overline{f} \int_0^{\vw^\prime n^{-1/2}\mX_t} (\vw^\prime n^{-1/2}\mX_t - x)^r\D x\Big]\notag\\
		&\leq K \E\big[\1_{\{0< \vw^\prime n^{-1/2}\mX_t\}} (\vw^\prime n^{-1/2}\mX_t)^{r+1}\big]\notag\\
		&\leq K n^{-(r+1)/2}\E\big[\Vert\mX_t\Vert^{r+1}\big].\label{eq:(3.eck)}
	\end{align}
	By similar arguments,
	\begin{align}
		\E\big[\overline{\nu}_{2t}(\vw)^r\big] &= \E\Big[\E_{t}^r\big\{(\vw^\prime n^{-1/2}\mX_t-\varepsilon_t)\1_{\{0<\varepsilon_t< \vw^\prime n^{-1/2}\mX_t\}}\big\}\Big]\notag\\
		&= \E\bigg[\1_{\{0< \vw^\prime n^{-1/2}\mX_t\}} \bigg\{\int_0^{\vw^\prime n^{-1/2}\mX_t} (\vw^\prime n^{-1/2}\mX_t - x)f_{\varepsilon_t\mid\mX_t}(x)\D x\bigg\}^{r}\bigg]\notag\\
		&\leq \E\bigg[\1_{\{0<\vw^\prime n^{-1/2}\mX_t\}}\bigg\{\overline{f} \int_0^{\vw^\prime n^{-1/2}\mX_t} (\vw^\prime n^{-1/2}\mX_t - x)\D x\bigg\}^r\bigg]\notag\\
		&\leq K \E\Big[\1_{\{0< \vw^\prime n^{-1/2}\mX_t\}} \big\{(\vw^\prime n^{-1/2}\mX_t)^{2}\big\}^r\Big]\notag\\
		&\leq K n^{-r}\E\big[\Vert\mX_t\Vert^{2r}\big].\label{eq:(4.eck)}
	\end{align}
	Plugging \eqref{eq:(3.eck)}--\eqref{eq:(4.eck)} into \eqref{eq:(3.1)} yields that
	\begin{equation}\label{eq:conclusio}
		\E\big[|\nu_{2t}(\vw) - \overline{\nu}_{2t}(\vw)|^r\big] \leq K\Big\{n^{-\frac{r+1}{2}}\E\big[\Vert\mX_t\Vert^{r+1}\big] + n^{-r}\E\big[\Vert\mX_t\Vert^{2r}\big]\Big\}\leq K n^{-\frac{r+1}{2}},
	\end{equation}
	where we used Assumption~\ref{ass:Heterogeneity}~\eqref{it:(i)} for the final inequality.
	In particular,
	\begin{align}
		\Vert\nu_{2t}(\vw) - \overline{\nu}_{2t}(\vw)\Vert_{r} &\leq K n^{-\frac{1+1/r}{2}},\label{eq:(4.1.eck)}\\
		\E\big[|\nu_{2t}(\vw) - \overline{\nu}_{2t}(\vw)|^2\big] &\leq K n^{-3/2},\label{eq:(4.0.eck)}
	\end{align}
	where the latter inequality follows because \eqref{eq:conclusio} holds not just for $r>2$ from Assumption~\ref{ass:Serial Dependence}, but also for $r=2$.
	For the variance terms in \eqref{eq:(1.eck)}, we get from \eqref{eq:(4.0.eck)} that
	\[
	\Var\big(\nu_{2t}(\vw) - \overline{\nu}_{2t}(\vw)\big) = \E\big[|\nu_{2t}(\vw) - \overline{\nu}_{2t}(\vw)|^2\big] \leq K n^{-3/2}.
	\]
	For the covariance terms in \eqref{eq:(1.eck)}, Davydov's inequality \citep[e.g.,][Corollary~16.2.4]{AL06} and \eqref{eq:(4.1.eck)} imply for $\widetilde{r}=r/(r-2)$ that
	\begin{align*}
		\Big|&\Cov\big(\nu_{2t}(\vw) - \overline{\nu}_{2t}(\vw), \nu_{2,t+h}(\vw) - \overline{\nu}_{2,t+h}(\vw)\big)\Big|\\
		&\leq 2\widetilde{r}[2\alpha(h)]^{1/\widetilde{r}}\Vert\nu_{2t}(\vw) - \overline{\nu}_{2t}(\vw)\Vert_{r} \Vert\nu_{2,t+h}(\vw) - \overline{\nu}_{2,t+h}(\vw)\Vert_{r}\\
		&\leq K[\alpha(h)]^{1/\widetilde{r}} \big(n^{-\frac{1+1/r}{2}}\big)^2\\
		&\leq \frac{1}{n^{1/r}}\frac{K}{n}[\alpha(h)]^{(r-2)/r}.
	\end{align*}
	Plugging these variance and covariance bounds into \eqref{eq:(1.eck)} gives that
	\begin{equation*}
		A_{1n} \leq \frac{K}{\sqrt{n}} + 2\frac{1}{n^{1/r}}\frac{K}{n} \sum_{t=1}^{n-1}\sum_{h=1}^{n-t}[\alpha(h)]^{\frac{r-2}{r}}.
	\end{equation*}
	By Assumption~\ref{ass:Serial Dependence} there exists some $\varphi>r/(r-2)$, such that
	\begin{equation*}
		\sum_{h=1}^{n-t}[\alpha(h)]^{(r-2)/r} \leq \sum_{h=1}^{n}[\alpha(h)]^{\frac{r-2}{r}}\leq K\sum_{h=1}^{\infty}h^{-\varphi\frac{r-2}{r}}\leq K,
	\end{equation*}
	where the last inequality follows from \citet[Theorem~2.27~(iii)]{Dav94}.
	Therefore, 
	\begin{equation}\label{eq:A 1n}
		A_{1n} \leq \frac{K}{\sqrt{n}} + 2\frac{1}{n^{1/r}}\frac{K}{n} \sum_{t=1}^{n-1}K= O(n^{-1/2} + n^{-1/r})=o(1).
	\end{equation}

	For $B_{1n}$, we use the bound $\big|\nu_{2t}(\vw) - \overline{\nu}_{2t}(\vw)\big|\leq K n^{-1/2} \Vert\mX_t\Vert$ to deduce that
	\begin{align}
		B_{1n} & \leq \P\bigg\{Kn^{-1/2}\max_{\substack{j<k\in\{1,\ldots,n\}\\ k-j\leq 2 \ell_n}}\sum_{t=j+1}^{k}\Vert\mX_t\Vert > \varepsilon\bigg\}\notag\\
		& \leq \P\bigg\{  \max_{m=0,\ldots,\lfloor n/(2\ell_n)\rfloor} \sum_{t=\lfloor m2\ell_n\rfloor +1}^{\lfloor (m+2)2\ell_n\rfloor} \Vert\mX_t\Vert>\varepsilon n^{1/2}/K\bigg\} \notag\\
		&\leq \sum_{m=0}^{\lfloor n/(2\ell_n)\rfloor} \P\bigg\{ \sum_{t=\lfloor m2\ell_n\rfloor +1}^{\lfloor (m+2)2\ell_n\rfloor} \Vert\mX_t\Vert>\varepsilon n^{1/2}/K\bigg\} \notag\\
		&\leq \sum_{m=0}^{\lfloor n/(2\ell_n)\rfloor} \frac{K^r}{\varepsilon^r n^{r/2}}\E\Bigg[\bigg|\sum_{t=\lfloor m2\ell_n\rfloor +1}^{\lfloor (m+2)2\ell_n\rfloor} \Vert\mX_t\Vert\bigg|^r\Bigg]\notag\\
		& \leq  \sum_{m=0}^{\lfloor n/(2\ell_n)\rfloor} \frac{K^r}{\varepsilon^r n^{r/2}} (4\ell_n)^{r-1}\sum_{t=\lfloor m2\ell_n\rfloor +1}^{\lfloor (m+2)2\ell_n\rfloor}\E\big[\Vert\mX_t\Vert^r\big]\notag\\
		& \leq K\lfloor n/(2\ell_n)\rfloor\frac{\ell_n^{r-1}}{n^{r/2}}\ell_n\notag\\
		& =O\big(n^{1-r/2}\ell_n^{r-1}\big),\label{eq:(H.A)}
	\end{align}
	where the third line follows from subadditivity, the fourth from Markov's inequality, and the fifth from the $c_r$-inequality \citep[Theorem~9.28]{Dav94}.
	By our choice $\ell_n=\lfloor n^\xi\rfloor$ it holds that
	\[
	n^{1-r/2}\ell_n^{r-1} = O\big(n^{1-r/2}n^{\xi(r-1)}\big) = O\Big(n^{1-r/2}n^{\frac{1}{2}\frac{r-2}{r-1+\delta}(r-1)}\Big) = O\Big(n^{1-r/2} n^{(r/2-1) \frac{r-1}{r-1+\delta}} \Big)=o(1),
	\]
	such that $B_{1n}=o(1)$.
	
	For the term involving the mixing coefficients in the numerator of \eqref{eq:Ottaviani}, we also obtain by our choice of $\ell_n=\lfloor n^\xi\rfloor$ that
	\begin{align}
		\lfloor n/\ell_n\rfloor \alpha(\ell_n) &=O\Big(\frac{n}{\ell_n} \ell_n^{-\overline{\varphi}}\Big)=O\Big(\frac{n}{\ell_n} \ell_n^{-\frac{r+2\overline{\delta}}{r-2}}\Big) = O\Big(n \ell_n^{-2\frac{r-1+\overline{\delta}}{r-2}}\Big)\notag\\
		&= O\Big(n\cdot n^{-2\xi\frac{r-1+\overline{\delta}}{r-2}}\Big) = O\Big(n\cdot n^{-2\frac{1}{2}\frac{r-2}{r-1+\delta}\frac{r-1+\overline{\delta}}{r-2}}\Big)=o(1),\label{eq:(H.mix)}
	\end{align}
	because $0<\delta<\overline{\delta}$.
	
	Now turn to the denominator of \eqref{eq:Ottaviani}. 
	The triangle inequality implies that
	\begin{align}
		\bigg|\sum_{t=k+1}^{n}\big[\nu_{2t}(\vw) - \overline{\nu}_{2t}(\vw)\big]\bigg| &= \bigg|\sum_{t=1}^{n}\big[\nu_{2t}(\vw) - \overline{\nu}_{2t}(\vw)\big] - \sum_{t=1}^{k}\big[\nu_{2t}(\vw) - \overline{\nu}_{2t}(\vw)\big]\bigg|\notag\\
		&\leq \bigg|\sum_{t=1}^{n}\big[\nu_{2t}(\vw) - \overline{\nu}_{2t}(\vw)\big]\bigg| + \bigg|\sum_{t=1}^{k}\big[\nu_{2t}(\vw) - \overline{\nu}_{2t}(\vw)\big]\bigg|.\label{eq:decomp nu2t}
	\end{align}
	Since $A_{1n}=o(1)$ it follows that $\big|\sum_{t=1}^{n}\big[\nu_{2t}(\vw) - \overline{\nu}_{2t}(\vw)\big]\big|=o_{\P}(1)$. 
	Therefore, for arbitrary $\iota\in(0,1)$ and $\varepsilon>0$ there exists some $n_0\in\mathbb{N}$, such that
	\begin{equation}\label{eq:nu large k}
		\P\bigg\{\bigg|\sum_{t=1}^{m}\big[\nu_{2t}(\vw) - \overline{\nu}_{2t}(\vw)\big]\bigg|>\frac{\varepsilon}{2}\bigg\}\leq\frac{\iota}{2}
	\end{equation}
	for all $m\geq n_0$.
	Therefore, for $k,n\geq n_0$,
	\begin{align*}
		\P&\bigg\{\bigg|\sum_{t=k+1}^{n}\big[\nu_{2t}(\vw) - \overline{\nu}_{2t}(\vw)\big]\bigg|>\varepsilon\bigg\}\\
		&\leq \P\bigg\{\bigg|\sum_{t=1}^{n}\big[\nu_{2t}(\vw) - \overline{\nu}_{2t}(\vw)\big]\bigg|>\frac{\varepsilon}{2}\bigg\} + \P\bigg\{\bigg|\sum_{t=1}^{k}\big[\nu_{2t}(\vw) - \overline{\nu}_{2t}(\vw)\big]\bigg|>\frac{\varepsilon}{2}\bigg\}\\
		&\leq\frac{\iota}{2} + \frac{\iota}{2}=\iota.
	\end{align*}
	In particular,
	\begin{equation}\label{eq:(C.22p)}
		\max_{k=n_0,\ldots,n}\P\bigg\{\bigg|\sum_{t=k+1}^{n}\big[\nu_{2t}(\vw) - \overline{\nu}_{2t}(\vw)\big]\bigg|>\varepsilon\bigg\}\leq \iota.
	\end{equation}
	For $k < n_0$
	\[
	\bigg|\sum_{t=1}^{k}\big[\nu_{2t}(\vw) - \overline{\nu}_{2t}(\vw)\big]\bigg|\leq K n_0\max_{t=1,\ldots,k}n^{-1/2}\Vert\mX_t\Vert\leq K n_0\max_{t=1,\ldots,n}n^{-1/2}\Vert\mX_t\Vert =o_{\P}(1)
	\]
	uniformly in $1\leq k <n_0$.
	In particular, for all such $k$,
	\begin{equation*}
		\bigg|\sum_{t=1}^{k}\big[\nu_{2t}(\vw) - \overline{\nu}_{2t}(\vw)\big]\bigg|<\frac{\varepsilon}{2}
	\end{equation*}
	with probability approaching $1$, as $n\to\infty$.
	From this, \eqref{eq:decomp nu2t} and \eqref{eq:nu large k} it follows that for sufficiently large $n$,
	\[
	\max_{k=1,\ldots,n_0-1}\P\bigg\{\Big|\sum_{t=k+1}^{n}\big[\nu_{2t}(\vw) - \overline{\nu}_{2t}(\vw)\big]\Big|> \varepsilon\bigg\}\leq\iota.
	\]
	This and \eqref{eq:(C.22p)} imply that
	\[
	\max_{k=1,\ldots,n}\P\bigg\{\Big|\sum_{t=k+1}^{n}\big[\nu_{2t}(\vw) - \overline{\nu}_{2t}(\vw)\big]\Big|> \varepsilon\bigg\}\leq\iota,
	\]
	such that the denominator remains bounded away from zero.
	
	Combining the results for the numerator and the denominator, we obtain from \eqref{eq:Ottaviani} that $\sup_{s\in[0,1]}\big|\sum_{t=1}^{\lfloor ns\rfloor}\big[\nu_{2t}(\vw) - \overline{\nu}_{2t}(\vw)\big]\big|=o_{\P}(1)$.
	This concludes the proof.
\end{proof}

\section{Proofs of Lemmas~\ref{lem:ES 1}--\ref{lem:ES 4}}\label{sec:ESest Lemmas}

\begin{proof}[{\textbf{Proof of Lemma~\ref{lem:ES 1}:}}]
	As in the proof of Lemma~\ref{lem:c2 FS}, we invoke the $\alpha$-mixing FCLT of \citet[Corollary 29.19]{Dav94}.
	To do so, we verify the assumptions of that corollary.
	First, note that the $\{\1_{\{\varepsilon_t>0\}}\xi_t\mX_t\}$ have zero mean, because
	\begin{eqnarray*}
		\E_t\big[\1_{\{\varepsilon_t>0\}}\xi_t\big] &=& \P_{t}\{\varepsilon_t>0\}\frac{1}{\P_{t}\{\varepsilon_t>0\}}\E_{t}\big[\1_{\{\varepsilon_t>0\}}\xi_t\big]\\
		&=& \P_{t}\{\varepsilon_t>0\}\E_t\big[\xi_t\mid \varepsilon_t>0\big]\\
		&\overset{\eqref{eq:QR and ES errors}}{=}& (1-\tau) \E_t\big[\xi_t\mid \xi_t>Q_{\tau}(\xi_t\mid\mX_t)\big]\\
		&=& (1-\tau)\ES_{\tau}(\xi_t\mid\mX_t)\\
		&=& 0,
	\end{eqnarray*}
	such that by the LIE,
	\begin{equation*}
		\E\big[\1_{\{\varepsilon_t>0\}}\xi_t\mX_t\big] = \E\Big[\E_t\big\{\1_{\{\varepsilon_t>0\}}\xi_t\big\}\mX_t\Big]  = \vzero.
	\end{equation*}
	
	Moreover, Assumption~\ref{ass:Heterogeneity ES}*~\eqref{it:(i) ES} implies
	\[
	\E\big[\1_{\{\varepsilon_t>0\}}^r|\xi_t|^r\Vert\mX_t\Vert^r\big]\leq \E\big[(|\xi_t|\cdot\Vert\mX_t\Vert)^r\big]\leq \Delta<\infty.
	\]
	
	Finally, $\{\1_{\{\varepsilon_t>0\}}\xi_t\mX_t\}$ is $\alpha$-mixing of size $-r/(r-2)$ because $\{(\varepsilon_t,\xi_t,\mX_t^\prime)^\prime\}$ is $\alpha$-mixing of size $-r/(r-2)$ by the arguments below \eqref{eq:QR and ES errors}.
	
	Since also the long-run variance converges appropriately by Assumption~\ref{ass:Heterogeneity ES}*~\eqref{it:(ii) ES}, Corollary~29.19 of \citet{Dav94} gives the desired result.
\end{proof}

\begin{proof}[{\textbf{Proof of Lemma~\ref{lem:ES 2}:}}]
	Write
	\begin{align*}
		\frac{1}{n}\sum_{t=1}^{\lfloor ns\rfloor} \1_{\{\varepsilon_t>0\}}\mX_t\mX_t^\prime - s(1-\tau)\mD_{\ast} &= \frac{1}{n}\sum_{t=1}^{\lfloor ns\rfloor} \Big\{\1_{\{\varepsilon_t>0\}}\mX_t\mX_t^\prime - \E\big[\1_{\{\varepsilon_t>0\}}\mX_t\mX_t^\prime\big]\Big\} \\
		&\hspace{1cm}+ \frac{1}{n}\sum_{t=1}^{\lfloor ns\rfloor}\E\big[\1_{\{\varepsilon_t>0\}}\mX_t\mX_t^\prime\big] - s(1-\tau)\mD_{\ast}\\
		&=:A_{2n} + B_{2n}.
	\end{align*}
	The LIE implies that
	\begin{equation*}
		\E\big[\1_{\{\varepsilon_t>0\}}\mX_t\mX_t^\prime\big] = \E\Big[\E_t\big\{\1_{\{\varepsilon_t>0\}}\big\}\mX_t\mX_t^\prime\Big]=\E\Big[\P_{t}\big\{\varepsilon_t>0\big\}\mX_t\mX_t^\prime\Big] = (1-\tau)\E\big[\mX_t\mX_t^\prime\big].
	\end{equation*}
	In light of this and Assumption~\ref{ass:Heterogeneity ES}*~\eqref{it:(iii) ES} it follows that
	\begin{align*}
		B_{2n} &= \frac{1}{n}\sum_{t=1}^{\lfloor ns\rfloor}(1-\tau)\E\big[\mX_t\mX_t^\prime\big] - s(1-\tau)\mD_{\ast}\\
		&= (1-\tau)\bigg\{\frac{\lfloor ns\rfloor}{n}\frac{1}{\lfloor ns\rfloor}\sum_{t=1}^{\lfloor ns\rfloor}\E\big[\mX_t\mX_t^\prime\big] - s\mD_{\ast}\bigg\}\\
		&= (1-\tau)\bigg\{\frac{\lfloor ns\rfloor}{n}\bigg(\frac{1}{\lfloor ns\rfloor}\sum_{t=1}^{\lfloor ns\rfloor}\E\big[\mX_t\mX_t^\prime\big] - \mD_{\ast}\bigg) - \bigg(s-\frac{\lfloor ns\rfloor}{n}\bigg)\mD_{\ast}\bigg\}\\
		&= o(1)
	\end{align*}
	uniformly in $s\in[0,1]$.
	
	For $A_{2n}$, note that $\big\{\1_{\{\varepsilon_t>0\}}\mX_t\mX_t^\prime - \E[\1_{\{\varepsilon_t>0\}}\mX_t\mX_t^\prime]\big\}$ is a mean-zero, $\alpha$-mixing sequence of size $-r/(r-2)$.
	Thus, by arguments in \citet[Section~3]{And88}, the sequence is an $L_{1}$-mixingale.
	The corollary of \citet[p.~493]{Han92} therefore implies that
	\[
	\sup_{s\in[0,1]}\bigg\Vert\frac{1}{n}\sum_{t=1}^{\lfloor ns\rfloor} \Big\{\1_{\{\varepsilon_t>0\}}\mX_t\mX_t^\prime - \E\big[\1_{\{\varepsilon_t>0\}}\mX_t\mX_t^\prime\big]\Big\}\bigg\Vert = o_{\P}(1),
	\]
	i.e., $A_{2n}=o_{\P}(1)$ uniformly in $s\in[0,1]$.
	
	The desired conclusion of the lemma follows upon combining the results for $A_{2n}$ and $B_{2n}$.
\end{proof}

The proof of Lemma~\ref{lem:ES 3} requires several preliminary results.
To introduce these, define
\begin{align*}
	\vnu_t(\vv) &:=\frac{1}{\sqrt{n}}\big[\1_{\{\vv^\prime n^{-1/2}\mX_t<\varepsilon_t\leq 0 \}} - \1_{\{0<\varepsilon_t\leq \vv^\prime n^{-1/2}\mX_t\}}\big]\xi_t\mX_t,\\
	\overline{\vnu}_t(\vv) &:= \E_{t}\big[\vnu_t(\vv)\big],\\
	\mV_n(\vv, s) &:= \sum_{t=1}^{\lfloor ns\rfloor} \vnu_t(\vv),\\
	\overline{\mV}_n(\vv, s) &:= \sum_{t=1}^{\lfloor ns\rfloor} \overline{\vnu}_t(\vv).
\end{align*}

\setcounter{lem}{7}	

\begin{lem}\label{lem:(D.7)}
	Under the assumptions of Theorem~\ref{thm:ES est} it holds for any $K>0$ that, as $n\to\infty$,
	\[
	\sup_{s\in[0,1]} \sup_{\Vert\vv\Vert\leq K} \big\Vert \overline{\mV}_n(\vv, s) - s\mK\vv\big\Vert=o_{\P}(1).
	\]
\end{lem}

\begin{proof}
	See Section~\ref{sec:prel lem}.
\end{proof}

\begin{lem}\label{lem:(D.8prep)}
	Under the assumptions of Theorem~\ref{thm:ES est} it holds for any fixed $\vv\in\mathbb{R}^{k}$ that, as $n\to\infty$,
	\begin{equation*}
		\sup_{s\in[0,1]}\big\Vert \mV_{n}(\vv,s) - \overline{\mV}_{n}(\vv,s)\big\Vert=o_{\P}(1).
	\end{equation*}
\end{lem}

\begin{proof}
	See Section~\ref{sec:prel lem}.
\end{proof}

\begin{lem}\label{lem:(D.8)}
	Under the assumptions of Theorem~\ref{thm:ES est} it holds for any $K>0$ that, as $n\to\infty$,
	\begin{equation*}
		\sup_{s\in[0,1]}\sup_{\Vert\vv\Vert\leq K}\big\Vert \mV_{n}(\vv,s) - \overline{\mV}_{n}(\vv,s)\big\Vert=o_{\P}(1).
	\end{equation*}
\end{lem}

\begin{proof}
	See Section~\ref{sec:prel lem}.
\end{proof}

\begin{proof}[{\textbf{Proof of Lemma~\ref{lem:ES 3}:}}]
	Plugging in $\widehat{\vv}_n(s)=\sqrt{n}\big[\widehat{\valpha}(s)-\valpha_0\big]$ for $\vv$ in Lemma~\ref{lem:(D.7)} yields by Theorem~\ref{thm:std est} that, as $n\to\infty$,
	\[
	\overline{\mV}_n(\widehat{\vv}_n(s), s)\overset{d}{\longrightarrow}\mK \mOmega^{1/2}\mW(s)\qquad\text{in}\ D_k[\epsilon,1],
	\]
	where $\mW(\cdot)$ is a $k$-variate standard Brownian motion.
	It also holds that
	\begin{align*}
		\P&\Big\{\sup_{s\in[0,1]}\big\Vert \mV_n(\widehat{\vv}_n(s), s) - \overline{\mV}_n(\widehat{\vv}_n(s), s) \big\Vert>\varepsilon\Big\} \\
		&= \P\Big\{\sup_{s\in[0,1]}\big\Vert \mV_n(\widehat{\vv}_n(s), s) - \overline{\mV}_n(\widehat{\vv}_n(s), s) \big\Vert>\varepsilon, \sup_{s\in[0,1]}\Vert\widehat{\vv}_n(s)\Vert\leq K\Big\}\\
		&\hspace{3.0cm} + \P\Big\{\sup_{s\in[0,1]}\big\Vert \mV_n(\widehat{\vv}_n(s), s) - \overline{\mV}_n(\widehat{\vv}_n(s), s) \big\Vert>\varepsilon, \sup_{s\in[0,1]}\Vert\widehat{\vv}_n(s)\Vert> K\Big\}\\
		&\leq \P\Big\{\sup_{s\in[0,1]}\sup_{\Vert\vv\Vert\leq K}\big\Vert \mV_n(\vv, s) - \overline{\mV}_n(\vv, s) \big\Vert>\varepsilon, \sup_{s\in[0,1]}\Vert\widehat{\vv}_n(s)\Vert\leq K\Big\} + \P\Big\{\sup_{s\in[0,1]} \Vert\widehat{\vv}_n(s)\Vert> K\Big\}\\
		& \leq \P\Big\{\sup_{s\in[0,1]}\sup_{\Vert\vv\Vert\leq K}\big\Vert \mV_n(\vv, s) - \overline{\mV}_n(\vv, s) \big\Vert>\varepsilon\Big\}+ \P\Big\{ \sup_{s\in[0,1]}\Vert\widehat{\vv}_n(s)\Vert> K\Big\}\\
		&=o(1)+o(1)
	\end{align*}
	as $n\to\infty$, followed by $K\to\infty$, from Lemma~\ref{lem:(D.8)} and Theorem~\ref{thm:std est}.
	Combining the above two displays implies that
	\begin{align*}
		\mV_n(\widehat{\vv}_n(s), s) &= \overline{\mV}_n(\widehat{\vv}_n(s), s) + \mV_n(\widehat{\vv}_n(s), s) - \overline{\mV}_n(\widehat{\vv}_n(s)\\
		&=\overline{\mV}_n(\widehat{\vv}_n(s), s) + o_{\P}(1)\\
		&\overset{d}{\longrightarrow}\mK \mOmega^{1/2}\mW(s)\qquad\text{in}\ D_k[\epsilon,1],
	\end{align*}
	which is the claimed result.
\end{proof}

\begin{proof}[{\textbf{Proof of Lemma~\ref{lem:ES 4}:}}]
	We establish the lemma by showing that
	\begin{equation*}
		\sup_{s\in[\epsilon,1]}\bigg\Vert\frac{1}{n}\sum_{t=1}^{\lfloor ns\rfloor} \1_{\{\mX_t^\prime(\widehat{\valpha}(s) - \valpha_0)<\varepsilon_t\leq 0 \}} \mX_t\mX_t^\prime\bigg\Vert\overset{\P}{\longrightarrow}0
	\end{equation*}
	and
	\begin{equation*}
		\sup_{s\in[\epsilon,1]}\bigg\Vert\frac{1}{n}\sum_{t=1}^{\lfloor ns\rfloor}  \1_{\{0<\varepsilon_t\leq \mX_t^\prime(\widehat{\valpha}(s) - \valpha_0)\}}\mX_t\mX_t^\prime\bigg\Vert\overset{\P}{\longrightarrow}0.
	\end{equation*}
	For brevity, we only prove the latter convergence.
	Write
	\begin{align*}
		\P&\bigg\{\sup_{s\in[\epsilon,1]}\bigg\Vert\frac{1}{n}\sum_{t=1}^{\lfloor ns\rfloor}  \1_{\{0<\varepsilon_t\leq \mX_t^\prime(\widehat{\valpha}(s) - \valpha_0)\}}\mX_t\mX_t^\prime\bigg\Vert>\varepsilon\bigg\} \\
		& = \P\bigg\{\sup_{s\in[\epsilon,1]}\bigg\Vert\frac{1}{n}\sum_{t=1}^{\lfloor ns\rfloor}  \1_{\{0<\varepsilon_t\leq n^{-1/2}\mX_t^\prime\sqrt{n}(\widehat{\valpha}(s) - \valpha_0)\}}\mX_t\mX_t^\prime\bigg\Vert>\varepsilon,\ \sup_{s\in[\epsilon,1]}\big\Vert\sqrt{n}(\widehat{\valpha}(s) - \valpha_0)\big\Vert\leq K\bigg\} \\
		& \hspace{0.5cm}+  \P\bigg\{\sup_{s\in[\epsilon,1]}\bigg\Vert\frac{1}{n}\sum_{t=1}^{\lfloor ns\rfloor}  \1_{\{0<\varepsilon_t\leq n^{-1/2}\mX_t^\prime\sqrt{n}(\widehat{\valpha}(s) - \valpha_0)\}}\mX_t\mX_t^\prime\bigg\Vert>\varepsilon,\ \sup_{s\in[\epsilon,1]}\big\Vert\sqrt{n}(\widehat{\valpha}(s) - \valpha_0)\big\Vert>K\bigg\}\\
		&\leq \P\bigg\{\sup_{s\in[\epsilon,1]}\bigg\Vert\frac{1}{n}\sum_{t=1}^{\lfloor ns\rfloor}  \1_{\{0<\varepsilon_t\leq Kn^{-1/2}\Vert\mX_t\Vert\}}\mX_t\mX_t^\prime\bigg\Vert>\varepsilon\bigg\} + \P\bigg\{\sup_{s\in[\epsilon,1]}\big\Vert\sqrt{n}(\widehat{\valpha}(s) - \valpha_0)\big\Vert>K\bigg\}\\
		&\leq \P\bigg\{\frac{1}{n}\sum_{t=1}^{n}  \1_{\{0<\varepsilon_t\leq Kn^{-1/2}\Vert\mX_t\Vert\}}\Vert\mX_t\Vert^2>\varepsilon\bigg\} + o(1)
	\end{align*}
	as $n\to\infty$, followed by $K\to\infty$, where the final step follows from $\sup_{s\in[\epsilon,1]}\big\Vert\sqrt{n}(\widehat{\valpha}(s) - \valpha_0)\big\Vert=O_{\P}(1)$, which---in turn---follows from Theorem~\ref{thm:std est} and the continuous mapping theorem \citep[e.g.,][Theorem~26.13]{Dav94}.
	In light of this, we only have to show that
	\[
	\frac{1}{n}\sum_{t=1}^{n}  \1_{\{0<\varepsilon_t\leq Kn^{-1/2}\Vert\mX_t\Vert\}}\Vert\mX_t\Vert^2=o_{\P}(1).
	\]
	
	To do so, observe that by the LIE and Assumption~\ref{ass:innov}~\eqref{it:dens bound},
	\begin{align*}
		\E\big[\1_{\{0<\varepsilon_t\leq Kn^{-1/2}\Vert\mX_t\Vert\}}\Vert\mX_t\Vert^2\big] &=\E\Big[\E_t\big\{\1_{\{0<\varepsilon_t\leq Kn^{-1/2}\Vert\mX_t\Vert\}}\big\} \Vert\mX_t\Vert^2\Big] \\
		&=\E\bigg[\int_{0}^{Kn^{-1/2}\Vert\mX_t\Vert}f_{\varepsilon_t\mid\mX_t}(x)\D x \cdot\Vert\mX_t\Vert^2\bigg] \\
		&\leq \E\big[Kn^{-1/2}\Vert\mX_t\Vert \cdot\overline{f}\cdot \Vert\mX_t\Vert^2\big]\\
		&\leq Kn^{-1/2} \E\big[\Vert\mX_t\Vert^3\big].
	\end{align*}
	Markov's inequality now implies
	\begin{align*}
		\P\bigg\{\frac{1}{n}\sum_{t=1}^{n}\1_{\{0<\varepsilon_t\leq Kn^{-1/2}\Vert\mX_t\Vert\}}\Vert\mX_t\Vert^2>\varepsilon\bigg\} &\leq \frac{1}{\varepsilon} \frac{1}{n}\sum_{t=1}^{n} \E\big[ \1_{\{0<\varepsilon_t\leq Kn^{-1/2}\Vert\mX_t\Vert\}}\Vert\mX_t\Vert^2\big] \\
		&\leq Kn^{-1/2} \frac{1}{n}\sum_{t=1}^{n}\E\big[\Vert\mX_t\Vert^3\big]\\
		&=o(1),
	\end{align*}
	such that the conclusion follows.
\end{proof}

\subsection{Proofs of Lemmas~\ref{lem:(D.7)}--\ref{lem:(D.8)}}\label{sec:prel lem}

\begin{proof}[{\textbf{Proof of Lemma~\ref{lem:(D.7)}:}}]
	Note that
	\begin{equation}\label{eq:(6.1)}
		\overline{\mV}_n(\vv, s) = \frac{1}{\sqrt{n}} \sum_{t=1}^{\lfloor ns\rfloor} \E_{t}\Big[\big(\1_{\{\vv^\prime n^{-1/2}\mX_t<\varepsilon_t\leq 0 \}} - \1_{\{0<\varepsilon_t\leq \vv^\prime n^{-1/2}\mX_t\}}\big)\xi_t\Big]\mX_t.
	\end{equation}
	Use \eqref{eq:QR and ES errors} to rewrite the conditional expectation as
	\begin{align}
		\E_{t}&\Big[\big(\1_{\{\vv^\prime n^{-1/2}\mX_t<\varepsilon_t\leq 0 \}} - \1_{\{0<\varepsilon_t\leq \vv^\prime n^{-1/2}\mX_t\}}\big)\xi_t\Big]\notag\\
		&= \E_{t}\Big[\big(\1_{\{\vv^\prime n^{-1/2}\mX_t<\varepsilon_t\leq 0 \}} - \1_{\{0<\varepsilon_t\leq \vv^\prime n^{-1/2}\mX_t\}}\big)\big(\mX_t^\prime(\valpha_0-\vbeta_0) + \varepsilon_t\big)\Big]\notag\\
		&= \1_{\{\vv^\prime n^{-1/2}\mX_t<0\}}\int_{\vv^\prime n^{-1/2}\mX_t}^{0} \big[\mX_t^\prime(\valpha_0-\vbeta_0) + x\big] f_{\varepsilon_t\mid\mX_t}(x)\D x\notag\\
		&\hspace{2cm}- \1_{\{\vv^\prime n^{-1/2}\mX_t>0\}}\int_{0}^{\vv^\prime n^{-1/2}\mX_t}\big[\mX_t^\prime(\valpha_0-\vbeta_0) + x\big] f_{\varepsilon_t\mid\mX_t}(x)\D x.\label{eq:(7.0)}
	\end{align}
	Integration by parts implies that
	\begin{align}
		\int_{0}^{\vv^\prime n^{-1/2}\mX_t}&\big[\mX_t^\prime(\valpha_0-\vbeta_0) + x\big] f_{\varepsilon_t\mid\mX_t}(x)\D x\notag\\
		&=\Big[\big(\mX_t^\prime(\valpha_0-\vbeta_0) + x\big) F_{\varepsilon_t\mid\mX_t}(x)\Big]_{0}^{\vv^\prime n^{-1/2}\mX_t} - \int_0^{\vv^\prime n^{-1/2}\mX_t}F_{\varepsilon_t\mid\mX_t}(x)\D x\notag\\
		&= \big[\mX_t^\prime(\valpha_0-\vbeta_0) + \vv^\prime n^{-1/2}\mX_t\big] F_{\varepsilon_t\mid\mX_t}(\vv^\prime n^{-1/2}\mX_t) - \mX_t^\prime(\valpha_0-\vbeta_0) F_{\varepsilon_t\mid\mX_t}(0)\notag\\
		&\hspace{9cm}- \int_0^{\vv^\prime n^{-1/2}\mX_t}F_{\varepsilon_t\mid\mX_t}(x)\D x\notag\\
		&= \mX_t^\prime(\valpha_0-\vbeta_0)\big[F_{\varepsilon_t\mid\mX_t}(\vv^\prime n^{-1/2}\mX_t)-F_{\varepsilon_t\mid\mX_t}(0)\big] \notag\\
		&\hspace{5.3cm}- \int_0^{\vv^\prime n^{-1/2}\mX_t}\big[F_{\varepsilon_t\mid\mX_t}(x)-F_{\varepsilon_t\mid\mX_t}(\vv^\prime n^{-1/2}\mX_t)\big]\D x\notag\\
		&= \mX_t^\prime(\valpha_0-\vbeta_0)(\vv^\prime n^{-1/2}\mX_t)\Big\{f_{\varepsilon_t\mid\mX_t}(0)+\big[f_{\varepsilon_t\mid\mX_t}(x^\ast) - f_{\varepsilon_t\mid\mX_t}(0)\big]\Big\} \notag\\
		&\hspace{1.7cm}- \int_0^{\vv^\prime n^{-1/2}\mX_t}\Big\{f_{\varepsilon_t\mid\mX_t}(0)+\big[f_{\varepsilon_t\mid\mX_t}(x^\ast) - f_{\varepsilon_t\mid\mX_t}(0)\big]\Big\}(x-\vv^\prime n^{-1/2}\mX_t)\D x\notag\\
		&= \mX_t^\prime(\valpha_0-\vbeta_0)(\vv^\prime n^{-1/2}\mX_t)f_{\varepsilon_t\mid\mX_t}(0)\notag\\
		&\hspace{1.7cm} + \int_0^{\vv^\prime n^{-1/2}\mX_t}f_{\varepsilon_t\mid\mX_t}(0)(\vv^\prime n^{-1/2}\mX_t-x)\D x\notag\\
		&\hspace{1.7cm} +\mX_t^\prime(\valpha_0-\vbeta_0)(\vv^\prime n^{-1/2}\mX_t)\big[f_{\varepsilon_t\mid\mX_t}(x^\ast)-f_{\varepsilon_t\mid\mX_t}(0)\big]\notag\\
		&\hspace{1.7cm} + \int_0^{\vv^\prime n^{-1/2}\mX_t}\big[f_{\varepsilon_t\mid\mX_t}(x^\ast)-f_{\varepsilon_t\mid\mX_t}(0)\big](\vv^\prime n^{-1/2}\mX_t-x)\D x\notag\\
		&= \mX_t^\prime(\valpha_0-\vbeta_0)(\vv^\prime n^{-1/2}\mX_t)f_{\varepsilon_t\mid\mX_t}(0)\notag\\
		&\hspace{1.7cm} + \frac{1}{2}f_{\varepsilon_t\mid\mX_t}(0)(\vv^\prime n^{-1/2}\mX_t)^2\notag\\
		&\hspace{1.7cm} +\mX_t^\prime(\valpha_0-\vbeta_0)(\vv^\prime n^{-1/2}\mX_t)\big[f_{\varepsilon_t\mid\mX_t}(x^\ast)-f_{\varepsilon_t\mid\mX_t}(0)\big]\notag\\
		&\hspace{1.7cm} + \int_0^{\vv^\prime n^{-1/2}\mX_t}\big[f_{\varepsilon_t\mid\mX_t}(x^\ast)-f_{\varepsilon_t\mid\mX_t}(0)\big](\vv^\prime n^{-1/2}\mX_t-x)\D x,\label{eq:(7.1)}
	\end{align}
	where the fourth step follows from \eqref{eq:(H.MVT)} and $x^\ast=x^\ast(x)$ is some mean value between $0$ and $x$ that may change from line to line.
	For the other integral in \eqref{eq:(7.0)}, we obtain similarly that
	\begin{align}
		\int_{\vv^\prime n^{-1/2}\mX_t}^{0}&\big[\mX_t^\prime(\valpha_0-\vbeta_0) + x\big] f_{\varepsilon_t\mid\mX_t}(x)\D x\notag\\
		&=\Big[\big(\mX_t^\prime(\valpha_0-\vbeta_0) + x\big) F_{\varepsilon_t\mid\mX_t}(x)\Big]_{\vv^\prime n^{-1/2}\mX_t}^{0} - \int_{\vv^\prime n^{-1/2}\mX_t}^{0}F_{\varepsilon_t\mid\mX_t}(x)\D x\notag\\
		&= \mX_t^\prime(\valpha_0-\vbeta_0) F_{\varepsilon_t\mid\mX_t}(0) - \big[\mX_t^\prime(\valpha_0-\vbeta_0) + \vv^\prime n^{-1/2}\mX_t\big] F_{\varepsilon_t\mid\mX_t}(\vv^\prime n^{-1/2}\mX_t) \notag\\
		&\hspace{9cm}- \int_{\vv^\prime n^{-1/2}\mX_t}^{0}F_{\varepsilon_t\mid\mX_t}(x)\D x\notag\\
		&= -\mX_t^\prime(\valpha_0-\vbeta_0)\big[F_{\varepsilon_t\mid\mX_t}(\vv^\prime n^{-1/2}\mX_t)-F_{\varepsilon_t\mid\mX_t}(0)\big] \notag\\
		&\hspace{5.3cm}- \int_{\vv^\prime n^{-1/2}\mX_t}^{0}\big[F_{\varepsilon_t\mid\mX_t}(x)-F_{\varepsilon_t\mid\mX_t}(\vv^\prime n^{-1/2}\mX_t)\big]\D x\notag\\
		&= -\mX_t^\prime(\valpha_0-\vbeta_0)(\vv^\prime n^{-1/2}\mX_t)\Big\{f_{\varepsilon_t\mid\mX_t}(0)+\big[f_{\varepsilon_t\mid\mX_t}(x^\ast) - f_{\varepsilon_t\mid\mX_t}(0)\big]\Big\} \notag\\
		&\hspace{1.7cm}- \int_{\vv^\prime n^{-1/2}\mX_t}^{0}\Big\{f_{\varepsilon_t\mid\mX_t}(0)+\big[f_{\varepsilon_t\mid\mX_t}(x^\ast) - f_{\varepsilon_t\mid\mX_t}(0)\big]\Big\}(x-\vv^\prime n^{-1/2}\mX_t)\D x\notag\\
		&= -\mX_t^\prime(\valpha_0-\vbeta_0)(\vv^\prime n^{-1/2}\mX_t)f_{\varepsilon_t\mid\mX_t}(0)\notag\\
		&\hspace{1.7cm} - \int_{\vv^\prime n^{-1/2}\mX_t}^{0}f_{\varepsilon_t\mid\mX_t}(0)(x-\vv^\prime n^{-1/2}\mX_t)\D x\notag\\
		&\hspace{1.7cm} -\mX_t^\prime(\valpha_0-\vbeta_0)(\vv^\prime n^{-1/2}\mX_t)\big[f_{\varepsilon_t\mid\mX_t}(x^\ast)-f_{\varepsilon_t\mid\mX_t}(0)\big]\notag\\
		&\hspace{1.7cm} - \int_{\vv^\prime n^{-1/2}\mX_t}^{0}\big[f_{\varepsilon_t\mid\mX_t}(x^\ast)-f_{\varepsilon_t\mid\mX_t}(0)\big](x-\vv^\prime n^{-1/2}\mX_t)\D x\notag\\
		&= -\mX_t^\prime(\valpha_0-\vbeta_0)(\vv^\prime n^{-1/2}\mX_t)f_{\varepsilon_t\mid\mX_t}(0)\notag\\
		&\hspace{1.7cm} - \frac{1}{2}f_{\varepsilon_t\mid\mX_t}(0)(\vv^\prime n^{-1/2}\mX_t)^2\notag\\
		&\hspace{1.7cm} -\mX_t^\prime(\valpha_0-\vbeta_0)(\vv^\prime n^{-1/2}\mX_t)\big[f_{\varepsilon_t\mid\mX_t}(x^\ast)-f_{\varepsilon_t\mid\mX_t}(0)\big]\notag\\
		&\hspace{1.7cm} - \int_{\vv^\prime n^{-1/2}\mX_t}^{0}\big[f_{\varepsilon_t\mid\mX_t}(x^\ast)-f_{\varepsilon_t\mid\mX_t}(0)\big](x-\vv^\prime n^{-1/2}\mX_t)\D x.\label{eq:(9.1)}
	\end{align}
	Plugging \eqref{eq:(7.0)}--\eqref{eq:(9.1)} into \eqref{eq:(6.1)} gives
	\begin{align}
		\overline{\mV}_n(\vv, s) &= \frac{1}{\sqrt{n}} \sum_{t=1}^{\lfloor ns\rfloor} \mX_t^\prime(\valpha_0-\vbeta_0)(\vv^\prime n^{-1/2}\mX_t)f_{\varepsilon_t\mid\mX_t}(0)\mX_t\notag\\
		&\hspace{1cm} + \frac{1}{\sqrt{n}} \sum_{t=1}^{\lfloor ns\rfloor} \frac{1}{2}f_{\varepsilon_t\mid\mX_t}(0)(\vv^\prime n^{-1/2}\mX_t)^2\mX_t\notag\\
		&\hspace{1cm} + \frac{1}{\sqrt{n}} \sum_{t=1}^{\lfloor ns\rfloor}\Big\{\1_{\{\vv^\prime n^{-1/2}\mX_t>0\}}\mX_t^\prime(\valpha_0-\vbeta_0)(\vv^\prime n^{-1/2}\mX_t)\big[f_{\varepsilon_t\mid\mX_t}(x^\ast)-f_{\varepsilon_t\mid\mX_t}(0)\big]\mX_t\notag\\
		&\hspace{2.5cm} + \1_{\{\vv^\prime n^{-1/2}\mX_t<0\}}\mX_t^\prime(\valpha_0-\vbeta_0)(\vv^\prime n^{-1/2}\mX_t)\big[f_{\varepsilon_t\mid\mX_t}(x^\ast)-f_{\varepsilon_t\mid\mX_t}(0)\big]\mX_t\Big\}\notag\\
		&\hspace{1cm} + \frac{1}{\sqrt{n}} \sum_{t=1}^{\lfloor ns\rfloor}\Big\{\1_{\{\vv^\prime n^{-1/2}\mX_t>0\}}\int_0^{\vv^\prime n^{-1/2}\mX_t}\big[f_{\varepsilon_t\mid\mX_t}(x^\ast)-f_{\varepsilon_t\mid\mX_t}(0)\big](\vv^\prime n^{-1/2}\mX_t-x)\D x\mX_t\notag\\
		&\hspace{2.5cm} + \1_{\{\vv^\prime n^{-1/2}\mX_t<0\}}\int_{\vv^\prime n^{-1/2}\mX_t}^{0}\big[f_{\varepsilon_t\mid\mX_t}(x^\ast)-f_{\varepsilon_t\mid\mX_t}(0)\big](x-\vv^\prime n^{-1/2}\mX_t)\D x\mX_t\Big\}\notag\\
		&=:\overline{\mV}_{1n}(\vv,s) + \overline{\mV}_{2n}(\vv,s) + \overline{\mV}_{3n}(\vv,s) + \overline{\mV}_{4n}(\vv,s),\label{eq:V gesammelt}
	\end{align}
	where $x^\ast$ is again some mean value that may change from appearance to appearance.
	We consider each term in \eqref{eq:V gesammelt} in turn.
	
	The first term simplifies to
	\[
	\overline{\mV}_{1n}(\vv,s) = \frac{1}{n} \sum_{t=1}^{\lfloor ns\rfloor} \mX_t\mX_t^\prime(\valpha_0-\vbeta_0)\mX_t^{\prime}f_{\varepsilon_t\mid\mX_t}(0)\vv.
	\]
	As for $A_{2n}$ in the proof of Lemma~\ref{lem:ES 2}, we obtain from the corollary in \citet[p.~493]{Han92} that
	\[
	\sup_{s\in[0,1]}\bigg\Vert\frac{1}{n} \sum_{t=1}^{\lfloor ns\rfloor} \Big\{\mX_t\mX_t^\prime(\valpha_0-\vbeta_0)\mX_t^{\prime}f_{\varepsilon_t\mid\mX_t}(0) - \E\big[\mX_t\mX_t^\prime(\valpha_0-\vbeta_0)\mX_t^{\prime}f_{\varepsilon_t\mid\mX_t}(0)\big]\Big\} \bigg\Vert=o_{\P}(1).
	\]
	With this and the observation that
	\[
	\E\big[\mX_t\mX_t^\prime(\valpha_0-\vbeta_0)\mX_t^{\prime}f_{\varepsilon_t\mid\mX_t}(0)\big] \overset{\eqref{eq:QR and ES errors}}{=} \E\big[(\xi_t-\varepsilon_t)f_{\varepsilon_t\mid\mX_t}(0)\mX_t\mX_t^{\prime}\big],
	\]
	we obtain using Assumption~\ref{ass:Heterogeneity ES}*~\eqref{it:(iv) ES} that
	\begin{align*}
		\sup_{s\in[0,1]}\sup_{\Vert\vv\Vert\leq K}\big\Vert\overline{\mV}_{1n}(\vv,s) - s \mK\vv \big\Vert & \leq \sup_{s\in[0,1]}\sup_{\Vert\vv\Vert\leq K}\bigg\Vert\overline{\mV}_{1n}(\vv,s) - \frac{1}{n} \sum_{t=1}^{\lfloor ns\rfloor} \E\big[\mX_t\mX_t^\prime(\valpha_0-\vbeta_0)\mX_t^{\prime}f_{\varepsilon_t\mid\mX_t}(0)\big]\vv \bigg\Vert \\
		&\hspace{1cm} + \sup_{s\in[0,1]}\sup_{\Vert\vv\Vert\leq K}\bigg\Vert\frac{1}{n} \sum_{t=1}^{\lfloor ns\rfloor} \E\big[(\xi_t-\varepsilon_t)f_{\varepsilon_t\mid\mX_t}(0)\mX_t\mX_t^{\prime}\big]\vv - s \mK\vv \bigg\Vert\\
		&=o_{\P}(1) + o(1)\\
		&=o_{\P}(1).
	\end{align*}
	
	For $\overline{\mV}_{2n}(\vv,s)$ note that by Markov's inequality and Assumption~\ref{ass:Heterogeneity ES}*~\eqref{it:(i) ES},
	\begin{equation*}
		\P\bigg\{\frac{1}{n}\sum_{t=1}^{n} \Vert\mX_t\Vert^3>M\bigg\}\leq \frac{1}{M}\E\bigg[\frac{1}{n}\sum_{t=1}^{n} \Vert\mX_t\Vert^3\bigg]\leq \frac{1}{M}\frac{1}{n}\sum_{t=1}^{n} \E\big[\Vert\mX_t\Vert^3\big]\leq \frac{K}{M},
	\end{equation*}
	which tends to zero as $M\to\infty$, such that $\frac{1}{n}\sum_{t=1}^{n} \Vert\mX_t\Vert^3=O_{\P}(1)$.
	Therefore, also using Assumption~\ref{ass:innov}~\eqref{it:dens bound},
	\begin{align*}
		\sup_{s\in[0,1]}\sup_{\Vert\vv\Vert\leq K}\big\Vert\overline{\mV}_{2n}(\vv,s)\big\Vert &\leq \sup_{s\in[0,1]}\frac{K}{\sqrt{n}}\frac{1}{n}\sum_{t=1}^{\lfloor ns\rfloor} f_{\varepsilon_t\mid\mX_t}(0)\Vert\mX_t\Vert^3\\
		&\leq \frac{K}{\sqrt{n}}\frac{1}{n}\sum_{t=1}^{n} \Vert\mX_t\Vert^3\\
		&=\frac{K}{\sqrt{n}}O_{\P}(1)\\
		&=o_{\P}(1).
	\end{align*}
	
	Similarly, exploiting Assumption~\ref{ass:innov}~\eqref{it:Lipschitz} now,
	\begin{align*}
		\sup_{s\in[0,1]}\sup_{\Vert\vv\Vert\leq K}\big\Vert\overline{\mV}_{3n}(\vv,s)\big\Vert &\leq \sup_{s\in[0,1]}\frac{K}{n}\sum_{t=1}^{\lfloor ns\rfloor} \Vert\mX_t\Vert^3L|x^{\ast}-0|\\
		&\leq \frac{K}{\sqrt{n}}\frac{1}{n}\sum_{t=1}^{n} \Vert\mX_t\Vert^4\\
		&=\frac{K}{\sqrt{n}}O_{\P}(1)\\
		&=o_{\P}(1).
	\end{align*}
	
	Finally, from once more similar arguments,
	\begin{align*}
		\sup_{s\in[0,1]}\sup_{\Vert\vv\Vert\leq K}\big\Vert\overline{\mV}_{4n}(\vv,s)\big\Vert &\leq \sup_{s\in[0,1]}\frac{K}{\sqrt{n}}\sum_{t=1}^{\lfloor ns\rfloor} \Vert\mX_t\Vert \bigg\{\int_{0}^{\vv^\prime n^{-1/2}\mX_t} L|x^{\ast}-0|\cdot|\vv^\prime n^{-1/2}\mX_t|\D x \\
		&\hspace{4cm}+ \int_{\vv^\prime n^{-1/2}\mX_t}^{0} L|x^{\ast}-0|\cdot|\vv^\prime n^{-1/2}\mX_t|\D x\bigg\}\\
		&\leq \frac{K}{n}\frac{1}{n}\sum_{t=1}^{n} \Vert\mX_t\Vert^4\\
		&=\frac{K}{n}O_{\P}(1)\\
		&=o_{\P}(1).
	\end{align*}
	
	Combining the results for the $\overline{\mV}_{in}(\vv,s)$'s, we obtain from \eqref{eq:V gesammelt} that
	\begin{equation*}
		\sup_{s\in[0,1]}\sup_{\Vert\vv\Vert\leq K}\big\Vert \overline{\mV}_{n}(\vv,s) - s\mK\vv\big\Vert=o_{\P}(1),
	\end{equation*}
	which is the desired result.
\end{proof}

\begin{proof}[{\textbf{Proof of Lemma~\ref{lem:(D.8prep)}:}}]
	For convenience, we overload notation here and redefine $\overline{\mV}_{1n}(\vv,s)$ and $\overline{\mV}_{2n}(\vv,s)$ from \eqref{eq:V gesammelt} in this proof as follows:
	\begin{align*}
		\mV_{n}(\vv,s) - \overline{\mV}_{n}(\vv,s) &= \big[\mV_{1n}(\vv,s) - \overline{\mV}_{1n}(\vv,s)\big] - \big[\mV_{2n}(\vv,s) - \overline{\mV}_{2n}(\vv,s)\big]\\
		&:=\sum_{t=1}^{\lfloor ns\rfloor}\big[\vnu_{1t}(\vv)- \overline{\vnu}_{1t}(\vv)\big] - \sum_{t=1}^{\lfloor ns\rfloor}\big[\vnu_{2t}(\vv)- \overline{\vnu}_{2t}(\vv)\big],
	\end{align*}
	where 
	\begin{align*}
		\vnu_{1t}(\vv) &:=\frac{1}{\sqrt{n}}\1_{\{\vv^\prime n^{-1/2}\mX_t<\varepsilon_t\leq 0\}}\xi_t\mX_t, && \overline{\vnu}_{1t}(\vv):=\E_{t}\big[\vnu_{1t}(\vv)\big],\\
		\vnu_{2t}(\vv) &:=\frac{1}{\sqrt{n}}\1_{\{0<\varepsilon_t\leq \vv^\prime n^{-1/2}\mX_t\}}\xi_t\mX_t, && \overline{\vnu}_{2t}(\vv):=\E_{t}\big[\vnu_{2t}(\vv)\big].
	\end{align*}
	In view of this decomposition, it suffices to show that for $i=1,2$,
	\[
	\sup_{s\in[0,1]}\big\Vert \mV_{in}(\vv,s) - \overline{\mV}_{in}(\vv,s)\big\Vert=o_{\P}(1).
	\]
	We do so only for $i=2$, as the proof for $i=1$ is almost identical.
	
	Apply Lemma~3 of \citet{Büc15} once more to deduce that
	\begin{align}
		\P&\Big\{\sup_{s\in[0,1]}\big\Vert \mV_{2n}(\vv,s) - \overline{\mV}_{2n}(\vv,s)\big\Vert>3\varepsilon\Big\}\notag\\
		&= \P\Big\{\max_{k=1,\ldots,n}\Big\Vert \sum_{t=1}^{k}\big[\vnu_{2t}(\vv)-\overline{\vnu}_{2t}(\vv)\big]\Big\Vert>3\varepsilon\Big\}\notag\\
		&\leq \frac{A_{3n} + B_{3n} + \lfloor n/\ell_n\rfloor\alpha(\ell_n)}{1-\max_{k=1,\ldots,n}\P\bigg\{\Big\Vert \sum_{t=k+1}^{n}\big[\vnu_{2t}(\vv)-\overline{\vnu}_{2t}(\vv)\big] \Big\Vert>\varepsilon\bigg\}},\label{eq:(13.1)}
	\end{align}
	where
	\begin{align*}
		A_{3n} &:= \P\Big\{\big\Vert\mV_{2n}(\vv, 1) - \overline{\mV}_{2n}(\vv, 1)\big\Vert>\varepsilon\Big\},\\
		B_{3n} &:= \P\Bigg\{\max_{\substack{j<k\in\{1,\ldots,n\}\\ k-j\leq 2\ell_n}}\Big\Vert\sum_{t=j+1}^{k}\big[\vnu_{2t}(\vv)-\overline{\vnu}_{2t}(\vv)\big]\Big\Vert>\varepsilon\Bigg\}.
	\end{align*}
	
	First, we consider the terms in the numerator of \eqref{eq:(13.1)}.
	To establish that $A_{3n}=o(1)$, we show that
	\begin{equation}\label{eq:(14.1)}
		\big\Vert\mV_{2n}(\vv, 1) - \overline{\mV}_{2n}(\vv, 1)\big\Vert=o_{\P}(1).
	\end{equation}
	We do this by proving that each component of $\mV_{2n}(\vv, 1) - \overline{\mV}_{2n}(\vv, 1)$ is $o_{\P}(1)$.
	Therefore, if we denote by $X_t$ a generic component of $\mX_t$ and (slightly overloading notation introduced in the proof of Lemma~\ref{lem:c3 FS}) define
	\[
	\nu_{2t}(\vv):=n^{-1/2}\1_{\{0<\varepsilon_t\leq \vv^\prime n^{-1/2}\mX_t\}}\xi_t X_t,\qquad \overline{\nu}_{2t}(\vv):=\E_{t}\big[\nu_{2t}(\vv)\big],
	\]
	then we must show that
	\[
	\sum_{t=1}^{n}\big[\nu_{2t}(\vv) - \overline{\nu}_{2t}(\vv)\big]=o_{\P}(1).
	\]
	For this, we follow the proof strategy used to deal with $A_{1n}$ in \eqref{eq:Ottaviani}.
	In fact, the arguments leading to \eqref{eq:A 1n} go through verbatim if we can show that a similar bound as in \eqref{eq:conclusio} is in force.
	To establish such a bound, apply the $c_r$-inequality to deduce that
	\begin{equation}\label{eq:B15 redo}
		\E\big[|\nu_{2t}(\vv) - \overline{\nu}_{2t}(\vv)|^r\big] \leq 2^{r-1}\Big\{\E\big[|\nu_{2t}(\vv)|^r\big] + \E\big[|\overline{\nu}_{2t}(\vv)|^r\big]\Big\}.
	\end{equation}
	For $\E\big[|\nu_{2t}(\vv)|^r\big]$ we obtain that
	\begin{align*}
		\E\big[|\nu_{2t}(\vv)|^r\big] &= n^{-r/2}\E\big[\1_{\{0<\varepsilon_t\leq \vv^\prime n^{-1/2}\mX_t\}}|\xi_t|^r |X_t|^r\big]\\
		&= n^{-r/2}\E\Big[|X_t|^r \E_t\big\{\1_{\{0<\varepsilon_t\leq \vv^\prime n^{-1/2}\mX_t\}}|\xi_t|^r\big\}\Big]\\
		&= n^{-r/2}\E\Big[|X_t|^r \E_t\big\{\1_{\{0<\varepsilon_t\leq \vv^\prime n^{-1/2}\mX_t\}}|\mX_t^\prime(\valpha_0-\vbeta_0) + \varepsilon_t|^r\big\}\Big]\\
		&\leq n^{-r/2}\E\Big[|X_t|^r \E_t\big\{\1_{\{0<\varepsilon_t\leq \vv^\prime n^{-1/2}\mX_t\}}2^{r-1}(|\mX_t^\prime(\valpha_0-\vbeta_0)|^r + |\varepsilon_t|^r)\big\}\Big]\\
		&\leq K n^{-r/2}\bigg\{ \E\Big[|X_t|^{2r} \E_t\big\{\1_{\{0<\varepsilon_t\leq \vv^\prime n^{-1/2}\mX_t\}}\big\}\Big]\\
		&\hspace{4cm} + \E\Big[|X_t|^r\E_t\big\{\1_{\{0<\varepsilon_t\leq \vv^\prime n^{-1/2}\mX_t\}}|\varepsilon_t|^r\big\}\Big] \bigg\}\\
		&=K n^{-r/2}\bigg\{ \E\Big[|X_t|^{2r} \1_{\{0<\vv^\prime n^{-1/2}\mX_t\}}\int_{0}^{\vv^\prime n^{-1/2}\mX_t} f_{\varepsilon_t\mid\mX_t}(x)\D x\Big]\\
		&\hspace{4cm} + \E\Big[|X_t|^r\1_{\{0<\vv^\prime n^{-1/2}\mX_t\}}\int_{0}^{\vv^\prime n^{-1/2}\mX_t}|x|^r f_{\varepsilon_t\mid\mX_t}(x)\D x\Big] \bigg\}\\
		&\leq K n^{-r/2}\bigg\{ \E\Big[|X_t|^{2r} \overline{f}\big(\vv^\prime n^{-1/2}\mX_t\big)\Big]\\
		&\hspace{4cm} + \E\Big[|X_t|^r\overline{f}(r+1)^{-1}\big|\vv^\prime n^{-1/2}\mX_t\big|^{r+1}\Big] \bigg\}\\
		&\leq K n^{-r/2}\Big\{K n^{-1/2}\E\big[\Vert\mX_t\Vert^{2r+1}\big] + Kn^{-(r+1)/2}\E\big[\Vert\mX_t\Vert^{2r+1}\big]\Big\}\\
		&\leq K n^{-(r+1)/2},
	\end{align*}
	where the second step follows from the LIE, the third step from \eqref{eq:QR and ES errors}, the fourth step from the $c_r$-inequality, the third-to-last step from Assumption~\ref{ass:innov}~\eqref{it:dens bound} and the last step from Assumption~\ref{ass:Heterogeneity ES}*~\eqref{it:(i) ES}.
	For the other expectation on the right-hand side of \eqref{eq:B15 redo}, similar arguments show that
	\begin{align*}
		\E\big[|\overline{\nu}_{2t}(\vv)|^r\big] &= n^{-r/2}\E\Big[\big|\E_t\{\1_{\{0<\varepsilon_t\leq \vv^\prime n^{-1/2}\mX_t\}}\xi_t X_t\}\big|^r\Big]\\
		&\leq n^{-r/2}\E\Big[|X_t|^r\big|\E_t\{\1_{\{0<\varepsilon_t\leq \vv^\prime n^{-1/2}\mX_t\}}\xi_t\}\big|^r\Big]\\
		&= n^{-r/2}\E\Big[|X_t|^r\big|\E_t\{\1_{\{0<\varepsilon_t\leq \vv^\prime n^{-1/2}\mX_t\}}(\mX_t^\prime(\valpha_0-\vbeta_0) + \varepsilon_t)\}\big|^r\Big]\\
		&= n^{-r/2}\E\Big[|X_t|^r\big| \mX_t^\prime(\valpha_0-\vbeta_0) \E_t\{\1_{\{0<\varepsilon_t\leq \vv^\prime n^{-1/2}\mX_t\}}\} +  \E_t\{\1_{\{0<\varepsilon_t\leq \vv^\prime n^{-1/2}\mX_t\}}\varepsilon_t\}\big|^r\Big]\\
		&= n^{-r/2}\E\Big[|X_t|^r\big| \mX_t^\prime(\valpha_0-\vbeta_0) \1_{\{0< \vv^\prime n^{-1/2}\mX_t\}}\int_{0}^{\vv^\prime n^{-1/2}\mX_t}f_{\varepsilon_t\mid\mX_t}(x)\D x\\
		&\hspace{2cm}+  \1_{\{0< \vv^\prime n^{-1/2}\mX_t\}}\int_{0}^{\vv^\prime n^{-1/2}\mX_t}x f_{\varepsilon_t\mid\mX_t}(x)\D x\big|^r\Big]\\
		&\leq n^{-r/2}\E\Big[|X_t|^r\Big| \big|\mX_t^\prime(\valpha_0-\vbeta_0)\big| \overline{f} |\vv^\prime n^{-1/2}\mX_t|+  \overline{f}(1/2)(\vv^\prime n^{-1/2}\mX_t)^2 \Big|^r\Big]\\
		&\leq  K n^{-r/2}\E\Big[|X_t|^r \Big|K n^{-1/2}\Vert\mX_t\Vert^{2} +  Kn^{-1}\Vert\mX_t\Vert^2 \Big|^r\Big]\\
		&\leq K n^{-r/2}\E\big[|X_t|^r K n^{-r/2}\Vert\mX_t\Vert^{2r}\big]\\
		&\leq K n^{-r}\E\big[\Vert\mX_t\Vert^{3r} \big]\\
		&\leq K n^{-r}.
	\end{align*}
	Plugging these two results into \eqref{eq:B15 redo}, we obtain the analog of \eqref{eq:conclusio}, i.e.,
	\[
	\E\big[|\nu_{2t}(\vv) - \overline{\nu}_{2t}(\vv)|^r\big]\leq K n^{-(r+1)/2}.
	\]
	Therefore, $A_{3n}=o(1)$ follows as $A_{1n}=o(1)$ in \eqref{eq:A 1n}.
	
	Turning to $B_{3n}$, note that from \eqref{eq:QR and ES errors},
	\begin{align*}
		\big\Vert\vnu_{2t}(\vv)\big\Vert &\leq n^{-1/2} \1_{\{0<\varepsilon_t\leq \vv^\prime n^{-1/2}\mX_t\}}|\xi_t| \cdot\Vert\mX_t \Vert\\
		&= n^{-1/2} \1_{\{0<\varepsilon_t\leq \vv^\prime n^{-1/2}\mX_t\}}|\mX_t^\prime(\valpha_0-\vbeta_0) + \varepsilon_t|\cdot \Vert\mX_t \Vert\\
		&\leq n^{-1/2} \1_{\{0<\varepsilon_t\leq \vv^\prime n^{-1/2}\mX_t\}}\big(K\Vert\mX_t\Vert + |\vv^\prime n^{-1/2}\mX_t|\big) \Vert\mX_t \Vert\\
		&\leq Kn^{-1/2} \Vert\mX_t\Vert^2
	\end{align*}
	and, similarly, 
	\begin{align*}
		\big\Vert\overline{\vnu}_{2t}(\vv)\big\Vert &\leq n^{-1/2} \E_t\big[\1_{\{0<\varepsilon_t\leq \vv^\prime n^{-1/2}\mX_t\}}|\xi_t|\cdot \Vert\mX_t \Vert\big]\\
		&= n^{-1/2} \Vert\mX_t \Vert \E_t\big[\1_{\{0<\varepsilon_t\leq \vv^\prime n^{-1/2}\mX_t\}}|\mX_t^\prime(\valpha_0-\vbeta_0) + \varepsilon_t| \big]\\
		&\leq n^{-1/2} \Vert\mX_t \Vert \E_t\big[\1_{\{0<\varepsilon_t\leq \vv^\prime n^{-1/2}\mX_t\}}\big(K\Vert\mX_t\Vert + |\vv^\prime n^{-1/2}\mX_t|\big) \big]\\
		&\leq Kn^{-1/2} \Vert\mX_t\Vert^2.
	\end{align*}
	Thus, by the triangle inequality,
	\[
	\big\Vert \vnu_{2t}(\vv) - \overline{\vnu}_{2t}(\vv)\big\Vert \leq K n^{-1/2}\Vert\mX_t\Vert^2.
	\]
	Then, $B_{3n}=o(1)$ follows as in \eqref{eq:(H.A)}, with the only difference that we now exploit $\E\big[\Vert\mX_t\Vert^{2r}\big]\leq K$ from Assumption~\ref{ass:Heterogeneity ES}*~\eqref{it:(i) ES} (instead of $\E\big[\Vert\mX_t\Vert^{r}\big]\leq K$ as in \eqref{eq:(H.A)}).
	
	That $\lfloor n/\ell_n\rfloor \alpha(\ell_n)=o(1)$ follows as in \eqref{eq:(H.mix)}.
	
	In sum, the numerator of \eqref{eq:(13.1)} converges to zero.
	
	That the denominator of \eqref{eq:(13.1)} remains bounded away from zero, follows similarly as in the proof of Lemma~\ref{lem:c3 FS}.
	We omit details for brevity.
	
	This establishes the lemma.
\end{proof}

\begin{proof}[{\textbf{Proof of Lemma~\ref{lem:(D.8)}:}}]
	The proof is most transparent for scalar-valued $\mX_t=X_t$.
	To reflect this in our notation, we now write $V_n(v,s)-\overline{V}_n(v,s)$ in place of $\mV_n(\vv,s)-\overline{\mV}_n(\vv,s)$ and so on.
	We omit the extension to vector-valued $\mX_t$, as it is only notationally more complicated.
	Write
	\begin{align*}
		V_n(v,s)-\overline{V}_n(v,s) &= \big[V_{1n}(v,s)-\overline{V}_{1n}(v,s)\big] - \big[V_{2n}(v,s)-\overline{V}_{2n}(v,s)\big]\\
		&:=\sum_{t=1}^{\lfloor ns\rfloor} \big[\nu_{1t}(v)-\overline{\nu}_{1t}(v)\big] - \sum_{t=1}^{\lfloor ns\rfloor} \big[\nu_{2t}(v)-\overline{\nu}_{2t}(v)\big],
	\end{align*}
	where, similarly as in the proof of Lemma~\ref{lem:(D.8prep)},
	\begin{align*}
		\nu_{1t}(v)&:= n^{-1/2}\1_{\{vn^{-1/2}X_t<\varepsilon_t\leq 0\}}\xi_tX_t,& \overline{\nu}_{1t}(v)&:=\E_{t}[\nu_{1t}(v)],\\
		\nu_{2t}(v)&:= n^{-1/2}\1_{\{0<\varepsilon_t\leq vn^{-1/2}X_t\}}\xi_tX_t,& \overline{\nu}_{2t}(v)&:=\E_{t}[\nu_{2t}(v)].
	\end{align*}
	Therefore, it suffices to show that
	\[
	\sup_{|v|\leq K}\sup_{s\in[0,1]}\bigg|\sum_{t=1}^{\lfloor ns\rfloor} \big[\nu_{it}(v)-\overline{\nu}_{it}(v)\big]\bigg|=o_{\P}(1)
	\]
	for $i=1,2$.
	Once again, we only do so for $i=2$, as the proof for $i=1$ is virtually identical.
	Fix $\rho>0$ and assume for ease of presentation that $2K/\rho$ is an integer.
	Then,
	\begin{align*}
		&\sup_{|v|\leq K}\sup_{s\in[0,1]}\bigg|\sum_{t=1}^{\lfloor ns\rfloor} \big[\nu_{2t}(v)-\overline{\nu}_{2t}(v)\big]\bigg|\\
		&\hspace{1cm} \leq \max_{j\in\mathbb{Z}}\sup_{s\in[0,1]}\bigg|\sum_{t=1}^{\lfloor ns\rfloor} \big[\nu_{2t}(j\rho)-\overline{\nu}_{2t}(j\rho)\big]\bigg|\\
		&\hspace{2cm}+ \sup_{|v_1-v_2|<\rho}\sup_{s\in[0,1]}\bigg|\sum_{t=1}^{\lfloor ns\rfloor} \big[\nu_{2t}(v_1)-\overline{\nu}_{2t}(v_1)\big] - \sum_{t=1}^{\lfloor ns\rfloor} \big[\nu_{2t}(v_2)-\overline{\nu}_{2t}(v_2)\big]\bigg|,
	\end{align*}
	where the maximum is taken over the $j$'s with $j\rho\in[-K,K]$.
	This implies that
	\begin{align}
		&\P\bigg\{\sup_{|v|\leq K}\sup_{s\in[0,1]}\bigg|\sum_{t=1}^{\lfloor ns\rfloor} \big[\nu_{2t}(v)-\overline{\nu}_{2t}(v)\big]\bigg|>\varepsilon\bigg\}\notag\\
		&\hspace{1cm} \leq \P\bigg\{\max_{j\in\mathbb{Z}}\sup_{s\in[0,1]}\bigg|\sum_{t=1}^{\lfloor ns\rfloor} \big[\nu_{2t}(j\rho)-\overline{\nu}_{2t}(j\rho)\big]\bigg|>\frac{\varepsilon}{2}\bigg\}\notag\\
		&\hspace{2cm} + \P\bigg\{\sup_{|v_1-v_2|<\rho}\sup_{s\in[0,1]}\bigg|\sum_{t=1}^{\lfloor ns\rfloor} \big[\nu_{2t}(v_1)-\overline{\nu}_{2t}(v_1)\big] - \sum_{t=1}^{\lfloor ns\rfloor} \big[\nu_{2t}(v_2)-\overline{\nu}_{2t}(v_2)\big]\bigg|>\frac{\varepsilon}{2}\bigg\}.\label{eq:decomp nu2}
	\end{align}
	By subadditivity and Lemma~\ref{lem:(D.8prep)},
	\begin{multline*}
		P\bigg\{\max_{j\in\mathbb{Z}}\sup_{s\in[0,1]}\bigg|\sum_{t=1}^{\lfloor ns\rfloor} \big[\nu_{2t}(j\rho)-\overline{\nu}_{2t}(j\rho)\big]\bigg|>\frac{\varepsilon}{2}\bigg\}\\
		\leq \sum_{j:\, j\rho\in[-K,K]}P\bigg\{\sup_{s\in[0,1]}\bigg|\sum_{t=1}^{\lfloor ns\rfloor} \big[\nu_{2t}(j\rho)-\overline{\nu}_{2t}(j\rho)\big]\bigg|>\frac{\varepsilon}{2}\bigg\}=o(1).
	\end{multline*}
	
	Therefore, we only have to show that the final right-hand side term in \eqref{eq:decomp nu2} converges to zero.
	Note that for $-K\leq v_1,v_2\leq K$,
	\begin{align}
		\sup_{|v_1-v_2|<\rho}&\sup_{s\in[0,1]}\bigg|\sum_{t=1}^{\lfloor ns\rfloor} \big[\nu_{2t}(v_1)-\overline{\nu}_{2t}(v_1)\big] - \sum_{t=1}^{\lfloor ns\rfloor} \big[\nu_{2t}(v_2)-\overline{\nu}_{2t}(v_2)\big]\bigg|\notag\\
		&=\sup_{|v_1-v_2|<\rho}\sup_{s\in[0,1]}\bigg|\sum_{t=1}^{\lfloor ns\rfloor} \big[\nu_{2t}(v_1)-\nu_{2t}(v_2)\big] - \sum_{t=1}^{\lfloor ns\rfloor} \big[\overline{\nu}_{2t}(v_1)-\overline{\nu}_{2t}(v_2)\big]\bigg|\notag\\
		&\leq \sup_{|v_1-v_2|<\rho}\sum_{t=1}^{n} \big|\nu_{2t}(v_1)-\nu_{2t}(v_2)\big| + \sup_{|v_1-v_2|<\rho}\sum_{t=1}^{n} \big|\overline{\nu}_{2t}(v_1)-\overline{\nu}_{2t}(v_2)\big|\notag\\
		&=:A_{4n} + B_{4n}.\label{eq:A4n B4n}
	\end{align}
	For $B_{4n}$, we obtain the bound
	\begin{equation}\label{eq:B4n}
		B_{4n} \leq \max_{\ell}\frac{1}{\sqrt{n}}\sum_{t=1}^{n}|X_t|\E_{t}\big[(\1_{\{0<\varepsilon_t\leq (\ell+2)\rho n^{-1/2}X_t\}} - \1_{\{0<\varepsilon_t\leq \ell\rho n^{-1/2}X_t\}})|\xi_t|\big],
	\end{equation}	
	where the maximum is taken over the integers $\ell$, such that $[\ell\rho,(l+2)\rho]\subset[-K,K]$.
	Now,
	\begin{align*}
		\E_{t}&\big[(\1_{\{0<\varepsilon_t\leq (\ell+2)\rho n^{-1/2}X_t\}} - \1_{\{0<\varepsilon_t\leq \ell\rho n^{-1/2}X_t\}})|\xi_t|\big]\\
		&\leq\E_{t}\big[\1_{\{\ell\rho n^{-1/2}X_t<\varepsilon_t\leq (\ell+2)\rho n^{-1/2}X_t\}} |\xi_t|\big]\\
		&= \E_{t}\big[\1_{\{\ell\rho n^{-1/2}X_t<\varepsilon_t\leq (\ell+2)\rho n^{-1/2}X_t\}} |X_t(\alpha_0-\beta_0) + \varepsilon_t|\big]\\
		&\leq |\alpha_0-\beta_0|\cdot|X_t|\E_{t}\big[\1_{\{\ell\rho n^{-1/2}X_t<\varepsilon_t\leq (\ell+2)\rho n^{-1/2}X_t\}}\big]\\
		&\hspace{1cm} +\E_{t}\big[\1_{\{\ell\rho n^{-1/2}X_t<\varepsilon_t\leq (\ell+2)\rho n^{-1/2}X_t\}} |\varepsilon_t|\big]\\
		&= |\alpha_0-\beta_0|\cdot|X_t| \int_{\ell\rho n^{-1/2}X_t}^{(\ell+2)\rho n^{-1/2}X_t} f_{\varepsilon_t\mid X_t}(x)\D x\\
		&\hspace{1cm} + \int_{\ell\rho n^{-1/2}X_t}^{(\ell+2)\rho n^{-1/2}X_t} |x| f_{\varepsilon_t\mid X_t}(x)\D x\\
		& \leq K|X_t|\overline{f}2\rho n^{-1/2}|X_t| + \overline{f}\big[2^{-1}x^2\big]_{\ell\rho n^{-1/2}X_t}^{(\ell+2)\rho n^{-1/2}X_t}\\
		&\leq  K|X_t|^2 \rho n^{-1/2} + 2^{-1}\overline{f}\big[(\ell+2)^2-\ell^2\big]\rho^2n^{-1} X_t^2\\
		&\leq K|X_t|^2 \rho n^{-1/2},
	\end{align*}
	which is independent of $\ell$.
	Plugging this into \eqref{eq:B4n} yields that
	\begin{equation}\label{eq:(D.11+)}
		B_{4n} \leq K\rho \frac{1}{n}\sum_{t=1}^{n}|X_t|^3=\rho O_{\P}(1),
	\end{equation}
	where $\frac{1}{n}\sum_{t=1}^{n}|X_t|^3=O_{\P}(1)$ holds because of $\E|X_t|^3\leq K<\infty$ and Markov's inequality.
	Therefore,
	\begin{equation}\label{eq:(4.12)}
		\lim_{\rho\downarrow0}\limsup_{n\to\infty}\P\bigg\{\sup_{|v_1-v_2|<\rho}\sum_{t=1}^{n} \big|\overline{\nu}_{2t}(v_1)-\overline{\nu}_{2t}(v_2)\big|>\frac{\varepsilon}{2}\bigg\}=0.
	\end{equation}
	
	For $A_{4n}$, we obtain the bound
	\begin{equation}\label{eq:(3.1new)}
		A_{4n} \leq \max_{\ell}\frac{1}{\sqrt{n}}\sum_{t=1}^{n}\big[\1_{\{0<\varepsilon_t\leq (\ell+2)\rho n^{-1/2}X_t\}}|\xi_t X_t| - \1_{\{0<\varepsilon_t\leq \ell\rho n^{-1/2}X_t\}})|\xi_t X_t|\big],
	\end{equation}	
	where the maximum is likewise taken over the integers $\ell$, such that $[\ell\rho,(l+2)\rho]\subset[-K,K]$.
	Similar arguments used to show $A_{3n}=o(1)$ in the proof of Lemma~\ref{lem:(D.8prep)} also give that for each such $\ell$,
	\begin{multline}\label{eq:(3.2)}
		\frac{1}{\sqrt{n}}\sum_{t=1}^{n}\big[\1_{\{0<\varepsilon_t\leq (\ell+2)\rho n^{-1/2}X_t\}}|\xi_t X_t| - \1_{\{0<\varepsilon_t\leq \ell\rho n^{-1/2}X_t\}})|\xi_t X_t|\big]\\
		- \Big\{\E_{t}\big[\1_{\{0<\varepsilon_t\leq (\ell+2)\rho n^{-1/2}X_t\}}|\xi_t X_t|\big] - \E_{t}\big[\1_{\{0<\varepsilon_t\leq \ell\rho n^{-1/2}X_t\}}|\xi_t X_t|\big]\Big\}=o_{\P}(1).
	\end{multline}
	By \eqref{eq:B4n} and \eqref{eq:(D.11+)},
	\[
	\frac{1}{\sqrt{n}}\sum_{t=1}^{n}\Big\{\E_{t}\big[\1_{\{0<\varepsilon_t\leq (\ell+2)\rho n^{-1/2}X_t\}}|\xi_t X_t|\big] - \E_{t}\big[\1_{\{0<\varepsilon_t\leq \ell\rho n^{-1/2}X_t\}}|\xi_t X_t|\big]\Big\}=o_{\P}(1),
	\]
	as $n\to\infty$, followed by $\rho\downarrow$.
	From this and \eqref{eq:(3.2)},
	\[
	\frac{1}{\sqrt{n}}\sum_{t=1}^{n}\big[\1_{\{0<\varepsilon_t\leq (\ell+2)\rho n^{-1/2}X_t\}}|\xi_t X_t| - \1_{\{0<\varepsilon_t\leq \ell\rho n^{-1/2}X_t\}})|\xi_t X_t|\big]=o_{\P}(1),
	\]
	as $n\to\infty$, followed by $\rho\downarrow$, such that by \eqref{eq:(3.1new)},
	\begin{equation*}
		\lim_{\rho\downarrow0}\limsup_{n\to\infty}\P\bigg\{\sup_{|v_1-v_2|<\rho}\sum_{t=1}^{n} \big|\nu_{2t}(v_1)-\nu_{2t}(v_2)\big|>\frac{\varepsilon}{2}\bigg\}=0.
	\end{equation*}
	From this, \eqref{eq:A4n B4n} and \eqref{eq:(4.12)}, 
	\[
	\sup_{|v_1-v_2|<\rho}\sup_{s\in[0,1]}\bigg|\sum_{t=1}^{\lfloor ns\rfloor} \big[\nu_{2t}(v_1)-\overline{\nu}_{2t}(v_1)\big] - \sum_{t=1}^{\lfloor ns\rfloor} \big[\nu_{2t}(v_2)-\overline{\nu}_{2t}(v_2)\big]\bigg|=o_{\P}(1)
	\] 
	as $n\to\infty$, followed by $\rho\downarrow0$.
	Overall, the conclusion follows.
\end{proof}

\singlespacing

\bibliographystyle{APA}
\bibliography{thebib}

\end{document}